\documentclass[twocolumn,showpacs,amsfonts,amsmath,floatfix,superscriptaddress]{revtex4-2}

\usepackage{amsmath, amsfonts, amssymb, amsthm, bbm}

\usepackage{multirow}
\usepackage{dsfont}
\usepackage[T1]{fontenc}
\usepackage{enumitem}
\usepackage{bm}
\usepackage{mathtools}
\usepackage{times}
\usepackage{physics}
\usepackage[caption=false]{subfig}
\usepackage{listings}

\usepackage{makecell}

\DeclarePairedDelimiter\floor{\lfloor}{\rfloor}
\DeclarePairedDelimiter\cfloor{\lfloor}{\rceil}

\input{Qcircuit}

\newcommand{\targMinus}{*+<.02em,.02em>{\xy ="i","i"-<.39em,0em>;"i"+<.39em,0em> **\dir{-}, "i"-<0em,.39em>;"i"+<0em,.39em> **\dir{},"i"*\xycircle<.4em>{} \endxy} \qw}

\usepackage{orcidlink}
\usepackage{hyperref}
\begin{document}

\title{Sufficient condition for universal quantum computation using bosonic circuits}

\author{Cameron Calcluth\,\orcidlink{0000-0001-7654-9356}}
\email{calcluth@gmail.com}
\affiliation{Department of Microtechnology and Nanoscience (MC2), Chalmers University of Technology, SE-412 96 G\"{o}teborg, Sweden}
\author{Nicolas Reichel}
\affiliation{Department of Microtechnology and Nanoscience (MC2), Chalmers University of Technology, SE-412 96 G\"{o}teborg, Sweden}
\author{Alessandro Ferraro\,\orcidlink{0000-0002-7579-6336}}
\affiliation{Centre for Theoretical Atomic, Molecular and Optical Physics, Queen's University Belfast, Belfast BT7 1NN, United Kingdom}
\affiliation{Dipartimento di Fisica ``Aldo Pontremoli,''
Università degli Studi di Milano, I-20133 Milano, Italy}
\author{Giulia Ferrini\,\orcidlink{0000-0002-7130-6723}}
\affiliation{Department of Microtechnology and Nanoscience (MC2), Chalmers University of Technology, SE-412 96 G\"{o}teborg, Sweden}

\begin{abstract}
Continuous-variable bosonic systems stand as prominent candidates for implementing quantum computational tasks. While various necessary criteria have been established to assess their resourcefulness, sufficient conditions have remained elusive. We address this gap by focusing on promoting circuits that are otherwise simulatable to computational universality. The class of simulatable, albeit non-Gaussian, circuits that we consider is composed of Gottesman-Kitaev-Preskill (GKP) states, Gaussian operations, and homodyne measurements.
Based on these circuits, we first introduce a general framework for mapping a continuous-variable state into a qubit state. Subsequently, we cast existing maps into this framework, including the modular and stabilizer subsystem decompositions. By combining these findings with established results for discrete-variable systems, we formulate a sufficient condition for achieving universal quantum computation. Leveraging this, we evaluate the computational resourcefulness of a variety of states, including Gaussian states, finite-squeezing GKP states, and cat states. Furthermore, our framework reveals that both the stabilizer subsystem decomposition and the modular subsystem decomposition (of position-symmetric states) can be constructed in terms of simulatable operations. This establishes a robust resource-theoretical foundation for employing these techniques to evaluate the logical content of a generic continuous-variable state, which can be of independent interest.
\end{abstract}

\maketitle

\section{Introduction}
Despite recent progress in understanding the relationship between genuine quantum properties and quantum computation \cite{horodecki2003, brunner2014bell, streltsov2017, chitambar2019, budroni2022kochen}, unraveling the origin of quantum computational power remains a challenging task. Adopting insight from the framework of resource theories \cite{chitambar2019}, one approach to develop our understanding consists of breaking down the design of quantum computing architectures into two sub-parts: \textit{(i)} the implementation of a restricted class of circuits, which can be efficiently simulated with a classical device and therefore are deemed as \textit{free} or \textit{allowed}; \textit{(ii)} the preparation of specific states which are able to promote the restricted class to a universal model~\cite{bravyi2005} and are therefore deemed as {\it resources}. The latter identifies key properties that enable quantum advantage --- namely the ability to solve certain computational problems exponentially faster than classical computers~\cite{harrow2017}.

The choice of the restricted class depends on the model of quantum computation (QC).
In discrete-variable (DV) qubit-based QC, the restricted class most commonly considered is the set of Clifford circuits acting on stabilizer states~\cite{gottesman1997,nielsen2000}. Clifford circuits alone are incapable of achieving universality, and consequently quantum advantage. 
However certain states, such as the ``magic'' $T$-state, are capable of promoting these circuits to universality~\cite{bravyi2005}. States that have a fidelity to the $T$-state beyond a certain threshold also fulfill this scope by means of magic state distillation, whereby a large number of low-quality magic states can be converted to a smaller number of nearly ideal ones~\cite{bravyi2005,bravyi2012,litinski2019,reichardt2009,campbell2017}. Hence the fidelity to the closest ideal magic state yields a \textit{sufficient} criterion for universality.

In continuous-variable (CV) quantum computing, Gaussian quantum circuits ~\cite{ferraro2005, menicucci2006,weedbrook2012, adesso2014continuous} are commonly chosen as the counterpart to Clifford circuits. In fact, it is known that Gaussian circuits are efficiently simulatable and therefore incapable of performing universal QC~\cite{bartlett2002}. Adding access to certain CV resource states, such as the cubic phase state~\cite{gottesman2001,gu2009,ghose2007,miyata2016}, Gottesman-Kitaev-Preskill (GKP) states~\cite{baragiola2019} or cat state~\cite{vasconcelos2010, Weigand2018} promotes these circuits to universality. More broadly, a significant effort has been devoted to identifying efficiently simulatable circuits  \cite{bartlett2002, mari2012, veitch2013, rahimi-keshari2015}, and therefore the requisite properties for a state to act as a resource in the CV setting. In particular, \textit{necessary} conditions have been provided in terms of the Wigner logarithmic negativity (WLN)~\cite{albarelli2018, takagi2018} and the stellar rank~\cite{chabaud2020}, which quantify the degree of non-Gaussian features of a state. 
However, in contrast to the DV case, no sufficient criterion exists which can identify whether an arbitrary CV state is capable of promoting an otherwise simulatable architecture to universality.

In this work, we establish a sufficient criterion for a CV state to promote an otherwise simulatable class of circuits to universality. To accomplish this, we consider a distinct class, different from Gaussian circuits, as resourceless.
Specifically, we choose circuits composed of ideal GKP stabilizer states, acted on by Gaussian operations~\footnote{We restrict to Gaussian operations parameterized by rational symplectic matrices $\text{Sp}(2n,\mathbb Q)$ and all real phase space displacements $\text{HW}(n)$, which were shown to be simulatable in Ref.~\cite{calcluth2023}.} and measured with homodyne detection. These circuits have been shown to be efficiently simulatable~\cite{bermejo-vega2016,Juani-thesis,calcluth2022,calcluth2023}. As such, throughout this work, we refer to these types of circuits as simulatable GKP (SGKP) circuits.

Leveraging on this criterion, we assess the resourcefulness of generic Gaussian states in this model, thereby extending  the set of previously known resourceful Gaussian states which only included the vacuum and thermal states~\cite{baragiola2019}.
Our approach can be applied to generic states, and in particular we also investigate highly non-Gaussian states, such as realistic GKP states, cat states, and cubic phase states. We identify parameter regimes where they can be considered as resources in this framework, and where they exceed the resourcefulness of the Gaussian states.   

Our approach comprises two steps. We first map the CV state of interest into a two-dimensional space, effectively associating a qubit state to it. The maps that we define are inspired by some subsystem decompositions (SSDs) recently introduced in order to extract the (qubit-like) logical content of a generic CV state \cite{pantaleoni2020, pantaleoni2021, shaw2022}. However, crucially, 
we identify and focus on maps that 
can be implemented using solely SGKP circuits. Therefore they are free maps in a rigorous resource-theoretical sense, ensuring they do not artificially add any resource to the original CV state. In the second step, we apply known results in DV systems to evaluate the resourcefulness of the mapped qubit state, therefore establishing a sufficient condition for the original CV state to be resourceful.

As a byproduct, by establishing which SSD can be obtained using SGKP circuits, we are able to establish whether known SSDs can be grounded in a rigorous resource-theoretic framework. Considering that these SSDs play a pivotal role in extracting logical information from both theoretically proposed \cite{tzitrin2020,konno2021,fukui2022} and experimentally generated states \cite{matsos2023robust}, we anticipate that this result will be of independent interest. 

The subsequent sections are organized as follows. In Sec.~\ref{sec:main-results} we present an overview of the main results of our work. In Sec.~\ref{sec:background} we present previous methods for understanding the resourcefulness of quantum states for universal QC, along with an overview of existing methods for mapping CV states to DV states. In Sec.~\ref{sec:unified} we introduce a unified approach for mapping CV states to DV states and demonstrate how existing maps can be expressed in this framework. We then introduce a new map which is implementable using SGKP circuits. In Sec.~\ref{sec:resourcefulness} we present our technique for quantifying the resourcefulness of CV states for quantum advantage by interpreting the CV state as an encoded DV state, and present results quantifying the resourcefulness of a range of different CV states using our technique. In Sec.~\ref{sec:conclusion} we present the conclusions of our work and provide some open questions. In the Appendix, we provide a physical interpretation of the various maps in terms of circuits, and we demonstrate that the modular subsystem decomposition admits a physical interpretation in terms of SGKP circuits for states symmetric in position.

\section{Main results}
\label{sec:main-results}
To enhance readability, we report in this section a summary of the main results of this work. Comprehensive details and proofs are deferred to subsequent sections.

\begin{figure}[ht]
    \centering
    $$
    \Qcircuit @C=2em @R=1em {
    &&\qw & \multigate{5}{\text{Gaussian}}&\measureD{\hat q_1} \\
    \lstick{\ket{0_{\text{GKP}}}^{\otimes m}}&\vdots&&\ghost{\text{Gaussian}}&\vdots\\
    && \qw & \ghost{\text{Gaussian}}&\measureD{\hat q_n} \\
    &&\qw & \ghost{\text{Gaussian}}&\measureD{\hat q_{n+1}} \\
    \lstick{\hat \rho^{\otimes n}\quad}&\vdots&&\ghost{\text{Gaussian}}&\vdots\\
    && \qw & \ghost{\text{Gaussian}}&\measureD{\hat q_{n+m}} \\
        {\gategroup{1}{2}{3}{2}{1.75em}{\{}}
        {\gategroup{4}{2}{6}{2}{1.75em}{\{}}
        }
        $$
         \caption{A circuit diagram displaying the broad class of circuits that we consider in this work. In input, there are $m$ stabilizer ideal GKP states (in the diagram these are indicated as $0$-logical states without loss of generality) and $n$ arbitrary CV states $\hat \rho$. These states are acted on by Gaussian operations and measured with homodyne measurement. When also $\hat \rho$ are stabilizer GKP states, these circuits are efficiently simulatable yielding SGKP circuits ~\cite{calcluth2022, calcluth2023} (see text for details).}
         \label{fig:circuit-class}
    \end{figure}
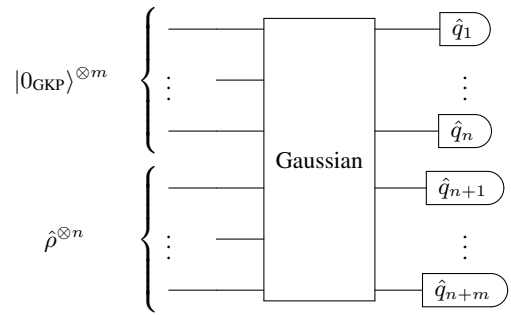

As said, we introduce a framework to address the resourcefulness of generic CV states when combined with the otherwise simulatable class of SGKP circuits. The general type of circuits considered is of the form depicted in Fig.~\ref{fig:circuit-class}, where one can also assume to have access to adaptive operations. As proven in Ref.~\cite{calcluth2023}, these circuits are efficiently simulatable on a classical device when employing only ideal stabilizer GKP states as input. Drawing on insights from qubit magic state distillation \cite{bravyi2005} and GKP error correction~\cite{gottesman2001}, we will prove that the circuit in Fig.~\ref{fig:circuit-class} attains instead universality when the input CV states $\hat \rho$ can be mapped into resourceful encoded qubits state via SGKP circuits alone. This approach therefore establishes a sufficient criterion for determining the resourcefulness of $\hat \rho$.

In more detail, we first introduce a method to systematically map arbitrary single-mode CV states to qubit states. This method unifies previously defined mappings, specifically subsystem decompositions (SSDs), which have recently been introduced for evaluating the qubit-like logical content of CV states. The approach involves transforming a CV state into an encoded GKP state and subsequently analyzing the resulting encoded qubit state. Depending on the choice of mapping, different qubit states will emerge. Of particular relevance for our objectives are those mappings implementable exclusively using SGKP circuits. This is crucial, since it guarantees that the associated SSDs do not introduce additional resources beyond those present in the original CV state. For this reason, following standard resource theory nomenclature, we term them \textit{allowed} mappings. 

We review the existing mappings of \textit{stabilizer} SSD~\cite{shaw2022} and \textit{modular} SSD~\cite{pantaleoni2020,pantaleoni2021}, expressing them within the presented general formalism. By leveraging on the connection between the stabilizer SSD and GKP error correction~\cite{shaw2022}, we show that stabilizer SSD can be constructed in terms of SGKP circuits, therefore it is an allowed mapping. In contrast, the modular SSD lacks this interpretative advantage in general. To address this, we introduce a new map termed \textit{Gaussian modular} SSD, and prove its equivalence to the modular SSD when the input CV state exhibits symmetry in the position representation. Crucially, like the stabilizer SSD, the Gaussian modular SSD can be understood in terms of allowed maps. This implies that the modular SSD, too, is a resource-theoretically grounded mapping for analyzing relevant position symmetric states such as finitely squeezed GKP states, as well as cat states, among others. However, for non-symmetric states, the equivalence breaks down, and the implementation of modular SSD necessitates operations beyond SGKP circuits, thus losing its interpretative status as an allowed map. These three mappings are summarized in Table \ref{tab:summary}, accompanied by their circuit diagrams. As mentioned earlier, while we have considered these mappings here for instrumental reasons to prove our main results, the revealed connection between SSDs and allowed maps is a result of intrinsic interest. In fact, this connection provides a rigorous resource-theoretical basis for recently introduced SSDs.

\begin{table*}[htbp]
\begingroup
\setlength{\tabcolsep}{0.3pt}
\renewcommand{\arraystretch}{2}
\begin{ruledtabular}
 \begin{tabular}{c c c c c l}
\textbf{SSD Type}            & \textbf{Gates} & \textbf{Kraus Operators}  & \textbf{\makecell{Logical\\State}}  & \quad\quad\quad\quad & \multicolumn{1}{c}{\textbf{Circuit}}  \\
& & & && \multirow{3}{*}{$$
        \Qcircuit @C=0.2em @R=.4em {
        \lstick{\hat\rho}              & \qw & \ctrl{1}    & \qw               &  \gate{\hat X(-\{t_q\}_{\sqrt\pi})}      & \qw           & \targMinus & \qw               & \qw                      &  \gate{\hat Z(-\{t_p\}_{\sqrt\pi})}   & \qw  & \qw \\
        \lstick{\ket{0_{\text{GKP}}}}   & \qw & \control\qw & \measureD{\hat p} & \control \cw \cwx   & \lstick{t_q} \\
        \lstick{\ket{0_{\text{GKP}}}}   & \qw & \qw         &\qw                &\qw  & \qw                 & \ctrl{-2}\qw  & \qw   & \measureD{\hat p} & \control \cw \cwx[-2]    & \lstick{t_p} 
        }
        $$}   \\
Stabilizer & Gaussian    &             $\{\hat K(\mathbf t)\}$  & $\hat \rho_{\Pi}$   & & \\
        & & & & & \\
& & & &&\multirow{3}{*}{$$
        \Qcircuit @C=0.2em @R=.4em {
        \lstick{\hat \rho}              & \qw & \ctrl{1}    & \qw               &  \gate{\hat X(-\{t_q\}{\sqrt\pi} )}      & \qw           & \targMinus & \qw               & \qw                      &  \gate{\hat Z(-\{t_q\}{\sqrt\pi})}   & \qw & \gate{\hat R_Z(-t_p\sqrt\pi)} & \qw \\
        \lstick{\ket{0_{\text{GKP}}}}   & \qw & \control\qw & \measureD{\hat p} & \control \cw \cwx   & \lstick{t_q} \\
        \lstick{\ket{0_{\text{GKP}}}}   & \qw & \qw         &\qw                &\qw  & \qw                 & \ctrl{-2}\qw  & \qw   & \measureD{\hat p} & \control \cw \cwx[-2]& \cw& \control \cw \cwx[-2]    & \lstick{t_p} 
        }
        $$}              \\
Modular    & Non-Gaussian   &       $\{\hat R_Z(\sqrt\pi t_p)\hat K(\mathbf t)\}$ & $\hat \rho_L$  & & \\
& & & & & \\
        & & & & & \multirow{3}{*}{$$
        \Qcircuit @C=0.2em @R=.4em {
        \lstick{\hat \rho}              & \qw & \ctrl{1}    & \qw               &  \gate{\hat X(-\{t_q\}_{\sqrt\pi})}      & \qw           & \targMinus & \qw               & \qw                      &  \gate{\hat Z(-\{t_p\}_{\sqrt\pi})}   & \qw & \gate{\mathcal{E}_{t_p}} & \qw \\
        \lstick{\ket{0_{\text{GKP}}}}   & \qw & \control\qw & \measureD{\hat p} & \control \cw \cwx   & \lstick{t_q} \\
        \lstick{\ket{0_{\text{GKP}}}}   & \qw & \qw         &\qw                &\qw  & \qw                 & \ctrl{-2}\qw  & \qw   & \measureD{\hat p} & \control \cw \cwx[-2]   &\cw & \control \cw \cwx[-2]    & \rstick{\, \, t_p} 
        }
        $$} \\ 
\makecell{Gaussian \\Modular}  & Gaussian    &      $\{\Gamma(\mathbf t)\hat Z_L\hat K(\mathbf t),\bar\Gamma(\mathbf t)\hat K(\mathbf t)\}$   & $\hat \rho_L^{\text{G}}$  & &  \\
& & & & &
 \end{tabular}
\end{ruledtabular}
\endgroup
\caption{A summary of the three types of maps considered in this work. The operator $\hat K(\mathbf t)=\hat \Pi \hat V(-\mathbf t)$ is the Kraus operator which is implemented by GKP error correction~\cite{baragiola2019}, where $\mathbf t=(t_q,t_p)$ contains the measurement results. The modular SSD and Gaussian modular SSD are equivalent for CV states that are symmetric in position. Note that the implementation of all of these maps requires access to ideal GKP stabilizer states. The operation $\hat R_Z(\theta)$ is a GKP-encoded rotation around the $Z$-axis on the Bloch sphere, it is therefore non-Gaussian. The functions $\Gamma(\mathbf t)$ and $\bar\Gamma(\mathbf t)=1-\Gamma(\mathbf t)$, defined later in Eq.~(\ref{eq:main-prob-z}), correspond to the probability of implementing each operation. Finally, $\hat Z_L$ is the GKP-encoded Pauli $\hat Z$ operator. The probabilistic implementation of the $\hat Z_L$ operator can equivalently be expressed as a Gaussian channel $
    \varepsilon_{t_p}(\hat\rho)=\bar\Gamma(\mathbf{t})\hat \rho+\Gamma(\mathbf{t})\hat Z\hat \rho\hat Z^\dagger$. The controlled gate with the symbol $\ominus$, shown in each circuit, denotes the inverse of the SUM gate, namely $e^{i\hat q_3 \hat p_1}$~\cite{noh2022low}. Each SSD can be implemented by the circuit shown in this table, whereby the outcome over the measurement results are averaged to produce a mixed state. \label{tab:summary}}
\end{table*}

Using this framework, we provide our main result. Namely, given a generic CV state $\hat \rho$, we define a sufficient condition for promoting the circuits presented in Fig.~\ref{fig:circuit-class} to universality. This condition entails identifying an allowed map that converts $\hat \rho$ into an encoded qubit state sufficiently close to an ideal magic GKP state. In particular, owing to the correspondence between SGKP circuits and Clifford circuits, it sufficies for the mapped qubit state to exhibit fidelity to an ideal magic state surpassing the corresponding known distillation threshold (identified in the context of Clifford quantum computation via state injection). Furthermore, beyond the fidelity, the resourcefulness of qubit states can be quantified by using various magic measures --- such as the robustness of magic (ROM)~\cite{howard2017}, relative entropy of magic~\cite{veitch2014}, GKP magic~\cite{hahn2022} and stabilizer R\'enyi entropy~\cite{leone2022}. 
For certain measures, such as the ROM, there exists a threshold $\mathcal R^*$ above which a state is guaranteed to have a fidelity greater than the threshold for magic state distillation. Therefore, access to a set of qubit states with a value of ROM beyond this threshold is sufficient to promote Clifford circuits to universality. 

In practical terms, given a generic single-mode CV state $\hat \rho$, our sufficient criterion involves the following steps: (i) employ a chosen allowed mapping to transform $\hat \rho$ into a qubit state $\hat \rho^{(P)}_L$ (details on the nomenclature will be provided later), and (ii) determine its corresponding ROM. If the latter exceeds the threshold $\mathcal R^*$, than the state $\hat \rho$ is a resource for universal quantum computation.

We apply the above criterion by analyzing the ROM of the logical states obtained by the two resource-theoretically motivated mappings, i.e., stabilizer SSD and Gaussian modular SSD. In particular, we analyze the set of Gaussian states and three classes of non-Gaussian CV states, namely finitely-squeezed GKP states, cat states, and cubic phase states. This allows us to identify states able to promote SGKP circuits to universal quantum computation which extend what previously reported in the literature. We stress that all pure Gaussian states are equally resourceful for promoting SGKP circuits to universality, since they can all be generated via SGKP circuits from vacuum.  Furthermore, we find that certain non-Gaussian states, albeit not necessarily all, have a value of ROM higher than the set of Gaussian states upon considered allowed mappings. Finally, notice that the fact that Gaussian states can be considered resourceful in this model implies that the resourcefulness for SGKP circuits is independent of the notion of resourcefulness in all-Gaussian circuits.

\section{Background}
\label{sec:background}
A resource is a component of a quantum circuit that promotes an otherwise simulatable model to universality. 
In this section, we review both DV and CV QC and their existing known measures of resourcefulness. We also introduce the families of quantum states which we analyze in the later Sec.~\ref{sec:resourcefulness}. Finally, we also recall existing methods to map CV states to DV, before introducing our unified approach for this type of mapping, in the next Section.

\subsection{Universal quantum computation and measures of resourcefulness in discrete variables}
Quantum computation over DVs involves quantum states defined over a discrete finite eigenspectrum. For example, qubit-based quantum computation involves qubits that are expressed in terms of the eigenstates of Pauli operators. A complete basis can be defined in terms of the eigenstates of the $\hat Z$ Pauli operator, $\hat Z\ket{0}=\ket 0$ and $\hat Z\ket{1}=-\ket 1$.

It is possible to simulate DV quantum circuits under certain conditions. For example, the Gottesman-Knill theorem~\cite{gottesman1999} provides a method to simulate circuits with input Pauli eigenstates, Clifford group operations (i.e., those which map Pauli operators to Pauli operators), and measurements in the Pauli basis. If we introduce access to a distillable magic state, then the circuit can perform universal quantum computation. For example, an ideal magic state such as the $T$ state, defined as~\cite{bravyi2005}
\begin{align}
    \ket{T}= \cos \beta \ket 0 +e^{i\pi /4}\sin\beta \ket 1, \quad \cos(2\beta)=\frac{1}{\sqrt 3}
\end{align}
can be combined with Clifford circuits to produce the full span of qubit circuits~\cite{kitaev1997,nielsen2000}. Furthermore, a supply of states sufficiently close to this state can be converted to a smaller number of higher-quality versions of this state via magic state distillation~\cite{bravyi2005}. The resourcefulness of a single qubit state $\hat \rho$ can be therefore quantified as the fidelity of the state with its closest T-type magic state, i.e.,
\begin{align}
\label{eq:fidelity-to-T}
    F_T^{\text{max}}(\hat \rho)=&\max_{\hat U\in \mathcal C} \bra{T}\hat U^\dagger \hat \rho \hat U \ket{T},
\end{align}
where the set $\mathcal C$ is the set of single qubit Clifford operations. 

\subsubsection{Resourcefulness of DV states: Robustness of magic}

Magic measures, such as the robustness of magic (ROM), also  provide a method to quantify the resourcefulness of a DV state. First note that by defining $\mathcal S_n$ as the set of all pure stabilizer states over $n$ qubits, any non-stabilizer state can be expressed as a sum of such states --- i.e., $\hat\rho=\sum_ix_i\hat\sigma_i$ for $\sigma_i \in  \mathcal S_n$. In general, there may be many different choices of $\{x_i\}$ which give the same $\hat\rho$.
The ROM of the qubit state $\hat \rho$ is defined as the minimal $1$-norm among all those possible choices of $\{x_i\}$. Formally, its expression is given by~\cite{howard2017}
\begin{align}
    \label{eq:rom-general}
    \mathcal R(\hat \rho)=\min_{\{x_i\}} \left\{\sum_i |x_i| \, ; \quad \hat \rho =\sum_{i} x_i\hat \sigma_i\right\}.
\end{align}

If the qubit state $\hat \rho$ is a stabilizer state then the ROM is equal to $1$. For non-stabilizer states, the ROM of a single qubit state can be simplified to the convenient expression~\cite{seddon2021}
\begin{align}
    \label{eq:rom-single}
    \mathcal R^{(1)}(\hat \rho)=\abs{\Tr(\hat \rho \hat X)}+|\Tr(\hat \rho \hat Y)|+|\Tr(\hat \rho \hat Z)|
\end{align}
where $\hat X,\hat Y,\hat Z$ are the Pauli operators. To avoid confusion, we have denoted the ROM of a single qubit as $\mathcal R^{(1)}(\hat \rho)$, where $\hat \rho$ must be a single qubit state. Note that we can also express this value in terms of the coefficients of the qubit density matrix $\hat \rho$,
\begin{align}
    \label{eq:rom-single-coefs}
    \mathcal R^{(1)}(\hat \rho)=2|\Re \rho_{01}|+2|\Im \rho_{01}|+|\rho_{00}-\rho_{11}|.
\end{align}

  For single qubit states, the ROM is directly related to the fidelity to the closest $T$-state in Eq.~(\ref{eq:fidelity-to-T}) by
\begin{align}
        F_T^{\text{max}}(\hat \rho)=\frac{1}{2\sqrt 3} \mathcal R^{(1)}(\hat \rho)+\frac 1 2.
\end{align}
The proof of this relation is given in Appendix \ref{sec:appendix-measure}.
It is known~\cite{bravyi2005} that single qubit states are distillable to $T$-type magic states if they have fidelity ${F_T^{\text{max}}(\hat \rho)>F^*=\frac{1}{2}(1+\sqrt{3/7})}$.
We can express this condition in terms of the ROM as
\begin{align}
    \mathcal R^{(1)}(\hat \rho)>\mathcal R^*=\frac{3}{\sqrt 7}\approx 1.134.
    \label{eq:rom-threshold}
\end{align}
Therefore, to perform magic state distillation, a value of ROM greater than $\mathcal R^*$ is sufficient for universality, in combination with Clifford circuits~\cite{reichardt2009,howard2017,seddon2021}. In addition, the larger the ROM, the more resourceful the state, in the sense that fewer copies of the state are needed for magic state distillation~\cite{bravyi2005}.

\subsection{Universal quantum computation and measures of resourcefulness in continuous variables}

CV QC involves quantum states defined over a continuous eigenspectrum of relevant observables, such as the position $\hat q$ and momentum $\hat p$ quadratures of the electromagnetic field, satisfying the commutation relations $[\hat q,\hat p]=i$. A complete basis in CV can be defined in terms of the eigenvectors of the position operator, $\hat q\ket{\hat q=s}=s\ket{\hat q=s}$.

In CV quantum systems --- as in the case of DV QC --- there exist simulatable models that have no exponential computational advantage over a classical computer. For example, circuits involving all Gaussian input states, Gaussian operations, and Gaussian measurements, such as homodyne measurements, are efficiently simulatable~\cite{bartlett2002, mari2012, veitch2013, rahimi-keshari2015}. Although it is not possible to achieve quantum advantage with this restricted class of circuits, it is known that adding access to specific resource states, such as the cubic phase state or GKP stabilizer states, will promote this model to universality~\cite{gottesman2001, baragiola2019}.

\subsubsection{Resourcefulness of CV states: Wigner logarithmic negativity}
Considering states displaying negative regions in their Wigner function as resources, a rigorous monotone has been introduced within a resource-theoretic framework. Such monotone is dubbed Wigner logarithmic negativity (WLN)~\cite{albarelli2018}, and it is defined as
\begin{align}
    W_{\text{neg}}(\hat \rho)=\log(\int \dd \mathbf q \int \dd \mathbf p |W_{\hat\rho}(\mathbf q,\mathbf p)|),
\end{align}
where the Wigner function $W_{\hat\rho}(\mathbf q,\mathbf p)$ is defined as
\begin{align}
    W_{\hat\rho}(\mathbf q,\mathbf p)=\frac{1}{(2 \pi)^n}\int \dd \mathbf x e^{i \mathbf p \cdot \mathbf x} \bra{\mathbf q+\frac{\mathbf x}{2}}\hat \rho\ket{\mathbf q-\frac{\mathbf x}{2}}_{\hat q}.
\end{align}

All states $\hat \rho$ with a non-negative Wigner function have $W_{\text{neg}}(\hat \rho)=0$. 
Meanwhile, a non-zero value of this quantity is a necessary condition to promote otherwise Gaussian circuits to universality~\cite{veitch2013,mari2012}.
However, it is also known that satisfying this criterion is not sufficient for achieving universality. Namely, circuits with input states that do contain Wigner negativity can be simulatable~\cite{garcia-alvarez2020, calcluth2022, calcluth2023}.

\subsection{Families of CV states}
\label{sec:background-families}
Here we provide a short review  of some families of CV states experimentally relevant to bosonic quantum computation with continuous variables.
We first begin with a quick reminder of Gaussian states \citep{ferraro2005}. Then, we present two types of bosonic code states~\cite{grimsmo2020}, which encode DV quantum information into CV states. Specifically, we present GKP states~\cite{gottesman2001, baragiola2019}  and cat states~\cite{cochrane1999, vasconcelos2010, Weigand2018}. We then recall the cubic phase state~\cite{gottesman2001,lloyd1999,budinger2022}. The last three families are known states able to promote Gaussian circuits to QC universality.

\subsubsection{Gaussian states}
\label{sec:families-gaussian}
Any pure Gaussian state can be produced via a Gaussian unitary operation $\hat U$ acting on the vacuum state. A single-mode Gaussian unitary can be decomposed in terms of a rotation ($\Theta \in [0,2\pi]$),
\begin{align}
    \label{eq:rotation-operator}
    \hat R(\Theta)=e^{i \frac{\Theta}{2} (\hat q^2+\hat p^2)},
\end{align}
squeezing,
\begin{equation}
\label{eq:squeezing}
 \hat S(\zeta)=e^{-\frac i 2 \zeta(\hat q \hat p+\hat p\hat q)},
  \end{equation}
where $\zeta>0$ represents squeezing in the position basis, while $\zeta<0$ represents squeezing in the momentum basis,
and displacement operations~\cite{arvind1995}
\begin{align}
    \label{eq:displacement-op}
    \hat V(\mathbf s)=e^{i\hat q s_p}e^{-i\hat p s_q},
\end{align}
parameterized by $\mathbf s=(s_q,s_p)^T$, where $s_q \in \mathbb{R}$ is the displacement in position while $s_p \in \mathbb{R}$ is the displacement in momentum.

Therefore, we can define any pure single-mode Gaussian state in terms of these operations as
\begin{align}
    \label{eq:single-gaussian-state}
    \ket{\zeta,\Theta,\mathbf s}=\hat V(\mathbf s) \hat R(\Theta)\hat S(\zeta)\ket{0}
\end{align}
where $\ket 0$ is the vacuum state. 

General Gaussian states can then be constructed out of pure Gaussian states by considering convex mixtures of pure states.

\subsubsection{GKP states}
The GKP encoding encodes DV quantum information using grid states~\cite{gottesman2001}. For qubits, the $0$-logical state and the $1$-logical state are defined as
\begin{align}
    \ket{0_{\text{GKP}}}&=\sum_n \ket{\hat q=2n \sqrt\pi},\\
    \ket{1_{\text{GKP}}}&=\sum_n \ket{\hat q=(2n+1) \sqrt\pi}.
\end{align}
Using these two basis states, it is possible to define arbitrary qubit states encoded as logical GKP states. For pure single-qubit states, we have
\begin{align}
    \label{eq:ideal-gkp-def}
    \ket{\psi_{\text{GKP}}}=\cos(\theta/2)\ket{0_{\text{GKP}}}+\sin(\theta/2)e^{i\phi}\ket{1_{\text{GKP}}}.
\end{align}
However, these ideal states are not normalizable and hence are not physically implementable. By using a wavefunction with Gaussian peaks and a Gaussian envelope parameterized by a squeezing parameter $\Delta$, instead of Dirac delta peaks which extend infinitely in position, it is possible to define realistic GKP states in terms of the unnormalized \footnote{Note that the basis states are given in their unnormalized form for two reasons. First, except for the finite squeezing limit of $\Delta \to 0$, the normalization constants do not have a closed analytic form~\cite{gottesman2001}. Second, the encoded logical state is defined in terms of the unnormalized basis states and then the encoded state is normalized.} basis states as~\cite{gottesman2001,matsuura2020,pantaleoni2021}
\begin{align}
    \ket{\bar 0_{\text{GKP}}^\Delta}=&\int \dd x e^{-x^2\Delta^2/2}\vartheta\left(\frac{x}{2\sqrt\pi},\frac{i\pi\Delta^2}{2\pi}\right)\ket{\hat q=x},\nonumber\\
    \ket{\bar 1_{\text{GKP}}^\Delta}=&\int \dd x e^{-x^2\Delta^2/2}\vartheta\left(\frac{x}{2\sqrt\pi}-\frac{1}{2},\frac{i\pi\Delta^2}{2\pi}\right)\ket{\hat q=x},
\end{align}
where $\vartheta(z,\tau)$ is the Jacobi theta function,
\begin{align}
    \vartheta(z,\tau)=\sum_m e^{i \pi m^2 \tau}e^{2\pi imz}.
\end{align}
Combining these states allows us to encode any pure (and hence also mixed) single qubit state as
\begin{align}
\label{eq:realistic-gkp-def}
\ket{ \psi_{\text{GKP}}^{\Delta}}=\frac{1}{\sqrt{\mathcal{N}_{\text{GKP}}}}\left(\cos(\theta/2)\ket{\bar 0_{\text{GKP}}^{\Delta}}+\sin(\theta/2)e^{i\phi}\ket{\bar 1_{\text{GKP}}^{\Delta}}\right),
\end{align}
whereby $\mathcal N_{\text{GKP}}$ is a normalization constant, specific to the squeezing and the parameters of the encoded state.

These states are physically implementable, however, the logical basis states are no longer orthogonal. This introduces errors in the encoding that can be interpreted as qubit errors~\cite{gottesman2001}. Furthermore, while for large squeezing, i.e., $\Delta \ll 1$, the norm of both unnormalized basis states are approximately equal~\cite{gottesman2001}, for larger values of $\Delta$ the normalization factors differ and can introduce an asymmetry in the encoded states~\cite{pantaleoni2021}.

GKP states have been physically implemented in a variety of experimental setups~\cite{fluhmann2019,campagne-ibarcq2020,kudra2022,konno2023} and are known to promote all-Gaussian circuits to universality~\cite{baragiola2019}.

\subsubsection{Cat states}
The second type of non-Gaussian states that we analyze in this work are cat states~\cite{cochrane1999,mirrahimi2014}. Cat states with even symmetry can be used to encode the $0$-logical state of a qubit, while cat states with odd symmetry encode the $1$-logical state of a qubit. The code space is defined in terms of the unnormalized \footnote{As is the case for GKP states, we provide the unnormalized basis states because the encoded qubit state is defined in terms of the two unnormalized basis states.} basis states~\cite{cochrane1999,lund2008,ralph2003}
\begin{align}
    \label{eq:even-cat-def}
    \ket{\bar 0_{\text{cat}}^{\alpha}}=&\left(\ket{\alpha}+\ket{-\alpha}\right),\\
    \ket{\bar 1_{\text{cat}}^{\alpha}}=&\left(\ket{\alpha}-\ket{-\alpha}\right),
\end{align}
where $\ket{\alpha}$ is a coherent state parameterized by the complex number $\alpha\in\mathbb C$, which can equivalently be expressed as ${\alpha=re^{i\phi}}$. The wavefunction of a coherent state $\ket{\alpha}$ in the position basis is given by
\begin{align}
\bra{\hat q=x}\ket{\alpha}= \pi^{-1/4}e^{-\frac 1 2 (x-\sqrt 2 r \cos\phi)^2+i\sqrt 2 r x \sin\phi}.
\end{align}
Any pure (and hence also mixed) qubit state can be encoded using these basis states. 
In what follows, we do not focus on the code aspect of cat states but rather analyze  the ability of the state $\ket{\bar 0_{\text{cat}}^{\alpha}}$ to promote SGKP circuits to universality.

Cat states have been successfully experimentally produced in a variety of different CV architectures~\cite{touzard2018,lewenstein2021,grimm2020,leghtas2015,ofek2016,gertler2021}. These states can also be used to produce GKP states using only Gaussian operations~\cite{vasconcelos2010, Weigand2018}. Therefore, like GKP states, they can also be considered a resource for quantum advantage in Gaussian circuits.

\subsubsection{Cubic phase state}
The final type of state that we analyze is the cubic phase state~\cite{gottesman2001}. This is defined as
\begin{align}
    \label{eq:cubicphase}
    \ket{\gamma,\zeta}=e^{i\gamma \hat q^3}\hat S(\zeta)\ket{0},
\end{align}
where $\ket{0} $ is the vacuum state and the squeezing operator is defined as in Eq.~(\ref{eq:squeezing}).

The cubic phase state can be used to produce both a $T$ gate in the GKP encoding~\cite{gottesman2001} and the CV cubic phase gate, which promotes all-Gaussian circuits to universality~\cite{braunstein05}. Cubic phase states have recently been successfully produced in a microwave cavity~\cite{kudra2022} and in an optical system~\cite{sakaguchi2023}. Theoretical prosposals have been put forward to generate them also in other platforms \cite{houhou2022unconditional} or by Gaussian
conversion from other non-Gaussian states \cite{zheng2021gaussian, hahn2022deterministic}.

\subsection{Existing methods to map CV states to DV states}
\label{sec:background-existing-maps}
There exist different methods~\cite{baragiola2019,pantaleoni2020,shaw2022} to analyze the logical content of a CV state.
GKP states offer a natural analogy to DV quantum states because they specifically encode DV quantum information into a CV state.
Furthermore, the logical action of Clifford operations in DV circuits is obtained by Gaussian operations when acting on GKP states~\cite{gottesman2001}.

Although the mapping from DV states to CV states through the GKP encoding is clear and well-defined~\cite{garcia-alvarez2020,grimsmo2020,gottesman2001}, understanding general CV states in terms of DV states is more challenging. This is due to the fact that the Hilbert space of CV states is infinite and therefore there is an infinite number of possible mappings. However, by grounding our choice of mapping in terms of the information we wish to extract from the CV state, and by using only resourceless states and operations in our mapping, we can define criteria for maps which are appropriate to the situation at hand.

Specifically, in this work, we are interested in maps which inform us of the resourcefulness of CV states to promote otherwise resourceless SGKP circuits to universality.

Here we review two existing methods of SSD. Namely, the stabilizer SSD, which effectively implements ideal GKP error correction on the CV state, and modular SSD, which has a convenient mathematical form. Notice that, prior to this work, neither of them had received a resource-theoretical interpretation.
\subsubsection{Stabilizer subsystem decomposition}
The result of the projection of a CV state $\hat \rho$ into the GKP encoded subspace, due to GKP error correction, gives a state of the form~\cite{baragiola2019}
\begin{align}
    \label{eq:def-stab-ssd-t}
    \hat \rho_{\Pi}(\mathbf t)=\hat \Pi\hat V (-\mathbf t) \hat \rho \hat V^\dagger(-\mathbf t)\hat \Pi,
\end{align}
where $\hat \Pi$ is the GKP projector defined as
\begin{align}
    \label{eq:gkp-projector}
    \hat \Pi=\ket{0_{\text{GKP}}}\bra{0_{\text{GKP}}}+\ket{1_{\text{GKP}}}\bra{1_{\text{GKP}}},
\end{align}
and $\hat V(-\mathbf t)$ is the displacement operator in both position and momentum, given in Eq.~(\ref{eq:displacement-op}). The circuit for implementing the stabilizer SSD is given in Table \ref{tab:summary}.
The output of such a circuit depends on the values $\mathbf t=(t_q,t_p)$. By disposing of these measurement outcomes, after the corrective displacements, we are left with a mixed state.
This state is a GKP-encoded qubit state which encodes the result of stabilizer SSD~\cite{shaw2022}. By a slight abuse of notation, we express the result of the stabilizer SSD as
\begin{align}
    \label{eq:def-stab-ssd}
    \hat \rho_\Pi =\frac{1}{\sqrt\pi} \int^{\sqrt\pi/2}_{-\sqrt\pi/2} \dd t_q\int^{\sqrt\pi/2}_{-\sqrt\pi/2} \dd t_p \hat \rho_{\Pi}(\mathbf t).
\end{align}
Note that the right-hand side of this equation is defined over the continuous-variable Hilbert space, while the left-hand side is defined over the qubit Hilbert space. However, this can be resolved by considering the implicit change of the basis states $\ket l$ to $\ket{l_{\text{GKP}}}$.
We provide further details on this notation in Appendix \ref{sec:appendix-stab}.

\subsubsection{Modular bosonic subsystem decomposition}
\label{sec:background-modular-ssd}
The logical content of a general CV state can also be identified using modular analysis. Modular analysis of CV states has a long history in quantum information~\cite{aharonov1969,ketterer2016}. Notably, it was used to first test the Bell inequalities~\cite{clauser1969,freedman1972,aspect1982}, which enabled much higher detection efficiency in comparison with using DV systems. Furthermore, it has recently been realized that modular analysis can be used to reconstruct the logical content of realistic GKP states ~\cite{pantaleoni2020,pantaleoni2021}.

The modular SSD has been introduced in Ref.~\cite{pantaleoni2021} in an abstract context, without reference to a specific circuit. Its primary feature is to decompose a CV state into a logical component and a gauge part. As in Ref.~\cite{pantaleoni2021}, we begin by providing an example of the decomposition for a real number, $s\in \mathbb R$. It is always possible to write the number in terms of an integer part $\floor{s}$ and its remainder $s-\floor{s}$, where $\floor{\cdot}$ is the integer floor function which rounds the number down to the nearest integer. We can consider this decomposition as splitting the number $s$ into different bins on the real number line, whereby each bin has width $1$.
Similarly, we can find a different decomposition of the number $s$ by using a different bin width $\alpha\in\mathbb R$. We can then decompose the number $s$ into the closest integer multiple of $\alpha$ using the centered floor function $\cfloor{s}_\alpha=\alpha \floor{\frac{s}{\alpha}+\frac 1 2}$ and its remainder $\{s\}_\alpha=s-\cfloor{s}_\alpha$.

The position quadrature $\hat q$ can be similarly decomposed. The position eigenstates $\ket{\hat q=s}$ of the position quadrature operator have eigenvalues over the real numbers. The operator can be written as $\hat q=\alpha \hat m+\hat u$ where $\alpha \hat m=\cfloor{\hat q}_{\alpha}$ is the integer part of the operator and $\{\hat u\}$ is the fractional part. This provides a method of writing the position eigenstates as simultaneous eigenstates of $\alpha \hat m$ and $\hat u$. We can express the position eigenstate as $\ket{\hat q=s}=\ket{\alpha \hat m+\hat u=s}$ or $\ket{\hat q=s}=\ket{\hat m=m,\hat u=u}$, with $\alpha m+u=s$. Furthermore, by separating the odd and even integers $m$ we can define a logical subsystem. This can be achieved by expressing $\hat q=\alpha\hat l+2\alpha \hat m_{\mathcal G}+\hat u_{\mathcal G}$ where $\hat l=\hat m \mod 2$, $\hat u_{\mathcal G}=\hat u$ and $\hat m_{\mathcal G}=\frac 1 2 (\hat m-\hat l)$.
We can then write the position basis states in terms of the logical part and gauge parts
\begin{align}
    \ket{\hat q =s}=\ket{m,u}=\ket{l}_L\ket{m_{\mathcal G},u_{\mathcal G}}_{\mathcal G}.
\end{align}
We can therefore describe the complete Hilbert space of a CV state in terms of a logical qubit and a gauge mode, i.e., ${\mathcal H_{\text{CV}}=\mathcal H_{L}\otimes \mathcal H_{\mathcal G}}$. The identity operator $\mathbbm 1_{\text{CV}}$ can be expressed as
\begin{align}
    \label{eq:identity-mod-vars}
    \sum_{l=1}^d \ket{l}_L\prescript{}{L}{\bra{l}}\otimes \sum_{m_{\mathcal G}=-\infty}^\infty \int^{\alpha/d}_{\alpha/d} \dd u_{\mathcal G} \ket{m_{\mathcal G},u_{\mathcal G}}_{\mathcal G}\prescript{}{\mathcal G}{\bra{m_{\mathcal G},u_{\mathcal G}}}.
\end{align}
It is possible to calculate the logical component of the density matrix by tracing out the gauge part of the state. The logical density matrix can be expressed as
\begin{align}
    \label{eq:definition-mssd}
    \hat \rho_L=\Tr_{\mathcal G}(\hat \rho).
\end{align}
While this method has a clear and robust mathematical definition, it was previously unknown whether this partial trace corresponds to implementable operations using physical circuits. In the next section, specifically Sec.~\ref{sec:resourcefulness-mssd}, we demonstrate that in the analysis of the logical content of GKP states, the modular SSD is in fact a well-motivated mapping that can be implemented with SGKP circuits.

\section{Unified approach for mapping CV states to qubits}
\label{sec:unified}

In this section, we establish a general mapping from CV states to logically encoded GKP states using continuous-variable operations. The modular SSD and the stabilizer SSD, introduced in Sec.~\ref{sec:background-existing-maps}, fall within this broad category. However, this general class of maps lacks a clear interpretation in terms of quantum computational resources. It may encompass operations that could potentially artificially enhance the computational capabilities of the original CV state. To address this, we further narrow down the scope to the class of maps implementable solely using SGKP circuits. This ensures that no artificial computational power is introduced during the mapping process. We will demonstrate that this class includes the stabilizer SSD but not the modular SSD. Additionally, we introduce a new map, the Gaussian modular SSD, inspired by the modular SSD but exclusively relying on components from SGKP circuits. Consequently, it also falls within the restricted class of maps.

\subsection{Mapping CV states to DV}

We begin defining a general map $M_{P}$ from an arbitrary CV state $\rho$ to an encoded qubit GKP state as
\begin{align}
\label{eq:general-map-def}
    M_{P}:\quad \hat \rho \to \int_{R} \dd \mathbf s \sum_i \hat P_i(\mathbf s) \hat \rho \hat P_i^\dagger (\mathbf s),
\end{align}
where $\hat P_i(\mathbf s)$ are Kraus operators which --- according to some parameters $\mathbf s$ that may depend on measurement results --- consist of CV operations, and $R$ is some integrable region of the space of the measurement outcomes $\mathbf s$. We denote the set of these Kraus operators as $P$, i.e., $P=\{\hat P_1(\mathbf s),\dots, \hat P_k(\mathbf s)\}$. These Kraus operators must include the GKP projector such that the state is mapped to a perfectly encoded ideal GKP state; i.e., the Kraus operators $\hat P_i=\hat \Pi\hat P_i''$ are expressed as an arbitrary CV operation $\hat P_i''$ followed by the projection $\hat \Pi$ onto the GKP code space. The encoded qubit state achieved as a result of applying the set of Kraus operators $P$ is denoted $\hat \rho_L^{(P)}$, i.e.,
\begin{align}
    \label{eq:general-map}
    \bra{l}\hat \rho^{(P)}_L\ket{l'}=\bra{l_{\text{GKP}}}\int_R \dd \mathbf s \sum_i \hat P_i(\mathbf s) \hat \rho \hat P_i^\dagger (\mathbf s)\ket{l'_{\text{GKP}}}.
\end{align}
The state that arises from the mapping $M_P$ depends on the choice of the Kraus operators $P$.

As said, the crucial point to notice is that, depending on the choice of Kraus operators, this general map may introduce additional resourcefulness to the original CV state. Therefore, in order to quantify the resourcefulness of CV states for SGKP circuits, we 
must restrict the Kraus operators to be chosen from the set of SGKP-type Kraus operators, which we label $\mathcal{P}_{\text{SGKP}}$. 
The corresponding restricted set of maps, i.e., the set of $M_P$ such that $P\in\mathcal{P}_{\text{SGKP}}$, are hence all maps that can be implemented using resourceless operations. 

As said, each Kraus operator in any set $P$ must project onto the GKP basis using the operator $\hat \Pi$. However, this operator is not, by itself, a valid operation in SGKP circuits. Despite this apparent contradiction, it remains possible to perform GKP error correction using SGKP circuits, which effectively introduces a random displacement and projects the CV state onto the GKP code basis, and which can instead be expressed using the Kraus operator~\cite{baragiola2019}
\begin{align}
    \label{eq:kraus-gkp}
    \hat K(\mathbf t)=\hat \Pi\hat V(-\mathbf t).
\end{align}

Therefore, we identify a class of allowed Kraus operators, which are both implementable with SGKP circuits and also project onto the GKP basis as 
\begin{align}
\label{eq:Kraus:projector-QEC}
    \hat P_i(\mathbf s)=\hat P'(\mathbf s)_i\hat \Pi \hat V(-\mathbf s) \hat U_i
\end{align}
where $\hat P'_i(\mathbf s)$ is selected from the set of Kraus operators implementable by probabilistic GKP-encoded Clifford operations and $\hat U_i$ is any unitary Gaussian operation (encompassed in Ref.~\cite{calcluth2023}), which occurs prior to the GKP error correction routine, and therefore does not depend on the measurement outcomes. For simplicity, we choose $\hat U_i=\mathbbm 1$ in our analysis of CV states. A complete characterization of the class of maps $M_P$ such that $P\in\mathcal{P}_{\text{SGKP}}$ is lacking and we leave it for further investigation.

\subsection{Considered maps in terms of the general map}

The two maps introduced in Sec.~\ref{sec:background-existing-maps} can all be expressed in the form given in Eq.~(\ref{eq:general-map}). As we will now see for the stabilizer subsystem decomposition, as well as for the Gaussian modular subsystem decomposition that we will introduce below, the Kraus operators can be further expressed as in Eq.~(\ref{eq:Kraus:projector-QEC}), implying that these maps can be implemented by means of SGKP circuits. However, the Kraus operators implementing modular SSDs are not in the set $\mathcal{P}_{\text{SGKP}}$.

\subsubsection{Stabilizer subsystem decomposition}
\label{sec:unified-considered-stab}
If we consider the set of Kraus operators $P$ in Eq.~(\ref{eq:general-map}) to consist of a single operator $P=K=\{\hat K(\mathbf s)\}$, where $\hat K(\mathbf s)$ is defined in Eq.~(\ref{eq:kraus-gkp}) and $R$ is the interval $[-\sqrt\pi/2,\sqrt\pi/2)$ over both $s_q$ and $s_p$,
\begin{align}
    \label{eq:unified-stab-ssd}
    \hat \rho^{(K)}_L=\int_R \dd \mathbf s\hat \Pi \hat V(-\mathbf s) \hat \rho \hat V^\dagger (-\mathbf s) \hat\Pi,
\end{align}
 then we recover the stabilizer SSD~\cite{shaw2022} as defined in Eq.~(\ref{eq:def-stab-ssd}), i.e., $\hat \rho^{(K)}_L=\hat \rho_\Pi$.

This map can be implemented by performing GKP error correction according to the original proposal provided by Ref.~\cite{gottesman2001}. In turn, it is easy to see that GKP error correction is a SGKP circuit, namely an allowed map. In fact, from the circuit diagram in Table I, it consists of measuring the two GKP stabilizers and displacing the mode in both position and momentum, whereby the corrective displacements are performed modulo $\sqrt\pi$ over the interval $(-\sqrt\pi/2,\sqrt\pi/2]$. Equivalently, this can be implemented by performing the corrective displacements $t_q,t_p$ directly but only accepting the state when the values of the measurement results $t_q,t_p$, modulo $2\sqrt\pi$, are within the acceptable interval $(-\sqrt\pi/2,\sqrt\pi/2]$; otherwise, the state is discarded~\cite{baragiola2019}. In any case, all these elements belong to the class of SGKP circuits, therefore ensuring that the stabilizer SSD map is an allowed map from a resource theory viewpoint, in that it does not add any computational power to the original state $\hat\rho$. This provides a resource-theoretic foundation to the statiblizer SSD therefore strongly grounding its use when one wants to associate a binary (qubit-like) logical content to a generic CV state $\hat\rho$. 

In Appendix \ref{sec:appendix-stab-pos}, we also provide a new alternative form of the stabilizer SSD, namely expressing it in the position basis. This alternative form is useful for comparing the effect of the stabilizer SSD with the modular SSD and also provides a convenient method to calculate the stabilizer SSD of a general CV state. As mentioned, in Appendix \ref{sec:appendix-stab-circuit} we derive the circuit implementation of the stabilizer SSD, also reproduced in Table~\ref{tab:summary}. 
\subsubsection{Modular subsystem decomposition}
\label{sec:resourcefulness-mssd}

The modular SSD is calculated by tracing out the gauge part of a bosonic state, i.e., Eq.~(\ref{eq:definition-mssd}). For a single mode, this can be expressed in a convenient form using the density matrix of the state in the position basis, as we show in Appendix \ref{sec:appendix-mod-pos}. However, we can also interpret this operationally as performing GKP error correction, followed by a logical $\hat Z$ rotation acting on the logical qubit state, as we explicitly show in Appendix~\ref{sec:appendix-mod-circuit}. The resulting interpretation in terms of a quantum circuit  is reproduced in Table~\ref{tab:summary} and makes explicit the connection between modular SSD with GKP error correction that was implicitly established in Ref.~\cite{pantaleoni2022}. Our analysis allows us to express the modular SSD as
\begin{align}
    \label{eq:main-ssd-logical-projector}
    \hat \rho_L
    =&\int_R \dd \mathbf  t \hat R_{Z} (t_p\sqrt\pi) \hat \rho_\Pi(\mathbf t)\hat R_Z^\dagger (t_p\sqrt\pi),
\end{align}
where the logical $\hat Z$ rotation is given by
\begin{align}
    \label{eq:main-z-rot}
    \hat R_{Z}(\theta)=\cos\frac{\theta}{2} \mathbbm 1-i\sin\frac{\theta}{2} \hat Z,
\end{align}
and $\hat \rho_\Pi(\mathbf t)$ is given in Eq.~(\ref{eq:def-stab-ssd-t}).
The set of Kraus operators defining the modular SSD, in terms of the general map defined in Eq.~(\ref{eq:general-map}), therefore consists of a single Kraus operator, i.e., $P=\{\hat R_Z(t_p\sqrt\pi)\hat K(-\mathbf t)\}$.

It is relevant to notice that the logical $\hat Z$ rotation is, for general $\theta$, a non-Clifford operation in the qubit framework and its GKP-encoded operation is accordingly non-Gaussian.

By inserting the rotation operator given in Eq.~(\ref{eq:main-z-rot}) into Eq.~(\ref{eq:main-ssd-logical-projector}), we show that the expression can be interpreted as a summation of a Gaussian $\hat \rho_{L}^{\text{G}}$ term and a non-Gaussian $\hat \rho_{L}^{\text{NG}}$ term, i.e., $\hat \rho_L=\hat \rho_{L}^{\text{G}}+\hat \rho_{L}^{\text{NG}}$. As we explicitly derive in Appendix \ref{sec:appendix-mod-g-and-ng}, these terms are given by
\begin{align}
    \label{eq:main-ssd-g-part}
    \hat \rho_{L}^{\text{G}}=\int_R \dd \mathbf  t \cos^2\left(\frac{t_p\sqrt\pi}{2}\right)\hat \rho_\Pi(\mathbf t)+\sin^2\left(\frac{t_p\sqrt\pi}{2}\right)\hat Z \hat \rho_\Pi(\mathbf t)\hat Z^\dagger
\end{align}
and
\begin{align}
    \label{eq:main-ssd-ng-part}
    \hat \rho_{L}^{\text{NG}}=-i\int_R \dd \mathbf  t \frac{\sin(t_p\sqrt\pi)}{2}\left(\hat Z \hat \rho_\Pi(\mathbf t)-\hat \rho_\Pi(\mathbf t)\hat Z^\dagger\right).
\end{align}

In general, since the logical $\hat Z$ rotation corresponds to a non-Gaussian operation, it is not implementable via an SGKP circuit and
therefore it could add computational power to
the original state $\hat\rho$, as it could increase the magic content of the corresponding qubit state.  However, for certain states $\hat \rho$, the state $\hat \rho_L$ is equivalent to $\hat \rho_L^{\text{G}}$ and can therefore be prepared with only SGKP circuits. In fact, as we explicitly demonstrate in Appendix \ref{sec:appendix-equivalence-symmetric}, when the input state is symmetric in position, the non-Gaussian part of the density matrix,  Eq.~(\ref{eq:main-ssd-ng-part}), evaluates to zero, i.e.,
\begin{align}
\label{eq:symmetry-implication}
&\bra{\hat q=x}\hat \rho \ket{\hat q=x'}=\bra{\hat q=-x}\hat \rho \ket{\hat q=-x'} \text{for all } x,x'\in\mathbb R \nonumber \\
&   \implies \hat \rho_L=\hat \rho_L^{\text{G}}.
\end{align}

For the purpose of analyzing realistic GKP states, as given in Eq.~(\ref{eq:realistic-gkp-def}), which are symmetric in position, we therefore find that the modular SSD can, in fact, be implemented using only components selected from the class of SGKP circuits. This implies that the modular SSD is also endowed with a resource-theoretic fundation, as the stabilizer SSD, when it is applied to the analysis of the logical content of realistic GKP states.

\subsubsection{Gaussian modular subsystem decomposition}
We now introduce a new map that can be performed using only the set of simulatable SGKP circuits. I.e., the set of Kraus operators $P$ is contained within $\mathcal{P}_{\text{SGKP}}$.
This map is the result of performing only the Gaussian part of the modular SSD and, therefore, the resulting state is given by $\hat \rho_L^{\text{G}}$.

To operationally produce this state from the state $\hat \rho$ with the otherwise free resources of SGKP circuits, we perform GKP error correction which gives measurement outcomes $t_q,t_p$ and then randomly apply a logical $Z$ gate with probability 
\begin{align}
\label{eq:main-prob-z}
\Gamma(\mathbf t)=\sin^2\left(\frac{t_p\sqrt\pi}{2}\right)=\frac{1-\cos(t_p\sqrt\pi)}{2}.
\end{align}
The measurement results should then be discarded to produce the statistical mixture over the possible values of $t_q,t_p$. Further details, and a circuit diagram of this procedure, are presented in Appendix \ref{sec:appendix-gauss}, see also Table \ref{tab:summary}.

The Kraus operators that define this map, in terms of Eq.~(\ref{eq:general-map}), are given by ${P=\{\bar\Gamma(\mathbf t)\hat K(\mathbf t),\Gamma(\mathbf t)\hat Z_L \hat K(\mathbf t)\}}$ where $\bar\Gamma(\mathbf t)=1-\Gamma(\mathbf t)$ is the complement probability, i.e., the probability of not implementing a $\hat Z_L$ operation.

This mapping has the benefit of being implementable with the resourceless SGKP operations, while also maintaining part of the structure of the modular SSD. In fact, as a result of Eq.~(\ref{eq:symmetry-implication}), this map is equivalent to the modular SSD when the input state is symmetric in position.

\section{Resourcefulness of CV states for SGKP circuits}
\label{sec:resourcefulness}
We now use the maps described in the previous subsections to analyze the resourcefulness of arbitrary CV states to promote the otherwise simulatable model of SGKP circuits to universality. 

\subsection{Resourcefulness of DV state resulting from general mapping}
\label{sec:resourcefulness-of-state-after-map}

In order to quantify the resourcefulness of generic CV states, we calculate the ROM of its associated qubit:
\begin{align}
    \label{eq:rom-one-mapping}
    \mathcal R(M_P(\hat \rho)).
\end{align}

As said, for this quantity to have a grounded resource-theoretic meaning, we restrict the allowed Kraus operators to those which are included in the simulatable model of simulatable GKP circuits, i.e., ${P\in \mathcal{P}_{\text{SGKP}}}$.

    By this logic, we can quantify the resourcefulness of an arbitrary single-mode CV state to promote SGKP circuits to universality by means of the functional
    \begin{align}
        \label{eq:gkp-rom}
        \mathcal R_{\text{SGKP}}(\hat \rho)=\max_{P\in \mathcal{P}_{\text{SGKP}}}\mathcal R(M_P(\hat \rho)).
    \end{align}
    
    Although a full search over all possible mappings is challenging, for the purpose of a sufficient condition of universality it is only required that there exists some map such that the ROM is above the threshold of distillability $\mathcal R^*$. This is because SGKP circuits contain stabilizer GKP states and Gaussian operations, yielding encoded Clifford circuits. The addition of a supply of GKP-encoded magic states, above the distillation threshold, promotes these circuits to universal QC.
    
    We can therefore inspect the quantity given in Eq.~(\ref{eq:rom-one-mapping}) for different choices of mappings $M_P$, all with $P\in\mathcal{P}_{\text{SGKP}}$. If, for one of these mappings, the ROM is greater than $\mathcal R^*$, then the CV state can clearly be converted to a GKP-encoded distillable magic state by some allowed mapping.
    Hence, the ROM of the logical state found via a specific mapping $M_{P}$ gives a lower bound of $\mathcal R_{\text{SGKP}}$.
    
   In other words, given access to a supply of the CV state $\hat\rho$, if the ROM of a logical state found via a specific mapping $M_P$ is above the distillation threshold, $\mathcal R(M_P(\hat \rho))>\mathcal R^*$, then it is possible to produce a supply of GKP-encoded magic states above the distillation threshold from $\hat\rho$ using only resourceless SGKP operations. Furthermore, since the operations required for magic state distillation consist of only encoded GKP-Clifford operations and adaptive homodyne measurements, it is possible to produce a supply of $T$ states with arbitrarily high quality using a polynomial number of operations~\cite{bravyi2005}, given access to a supply of the CV state $\hat \rho$. In this sense, the value of ROM after an allowed mapping yields an upper bound to the number of copies to be used in the magic state distillation procedure. Therefore, for a given mapping, the larger the ROM, the more resourceful the state. Note however that different mappings can yield different hierarchies between states, as we show in Appendix \ref{sec:appendix-mapping-choice}.

   Finally, we note that the value of ROM of a logical state found via a mapping of the form given in Eqs.~(\ref{eq:general-map-def},\ref{eq:general-map}) is convex. Specifically, when considering a mixed CV state ${\hat \rho=\sum_k p_k\hat\rho_k}$ with ${\sum_k p_k =1}$ consisting of a weighted sum of pure CV states, the corresponding logical state is equal to a weighted sum of the set of logical states found from each of the corresponding CV pure states $\hat \rho_k$, i.e., $\hat \rho_L=\sum_k p_k M_P(\hat \rho_k)$. Since ROM is convex for qubit states~\cite{howard2017}, we therefore must have that 
   \begin{align}
    \label{eq:convexity-rom-ssd}
       \mathcal R(M_P(\hat \rho))= \mathcal R \left(M_P\left(\sum_k p_k\hat\rho_k\right)\right) \le \sum_k p_k \mathcal R ( M_P(\hat \rho_k)).
   \end{align}

   \subsection{Analysis of CV states}
\label{sec:analysis-of-cv-states}
We use the methods described in Sec.~\ref{sec:resourcefulness-of-state-after-map} to analyze the ROM of the mapped CV states selected from the set of Gaussian states and three families of non-Gaussian states introduced in Sec.~\ref{sec:background-families}. Here we present the values of ROM for the resource-theoretically motivated SSDs, i.e., the stabilizer SSD and the Gaussian modular SSD. For the symmetric cat and GKP states, the Gaussian modular SSD is equivalent to the modular SSD, and therefore the values of the ROM for the two decompositions are equal.

\subsubsection{Gaussian states}
\label{sec:analysis-pure-gauss}
We begin with an analysis of pure Gaussian states. We then consider the case of mixed Gaussian states.

\begin{figure}[ht]
    \centering
    \subfloat[ROM of the stabilizer SSD of a rotated-squeezed Gaussian state. \label{subfig:gauss_stab}]{%
      \includegraphics[width=\linewidth]{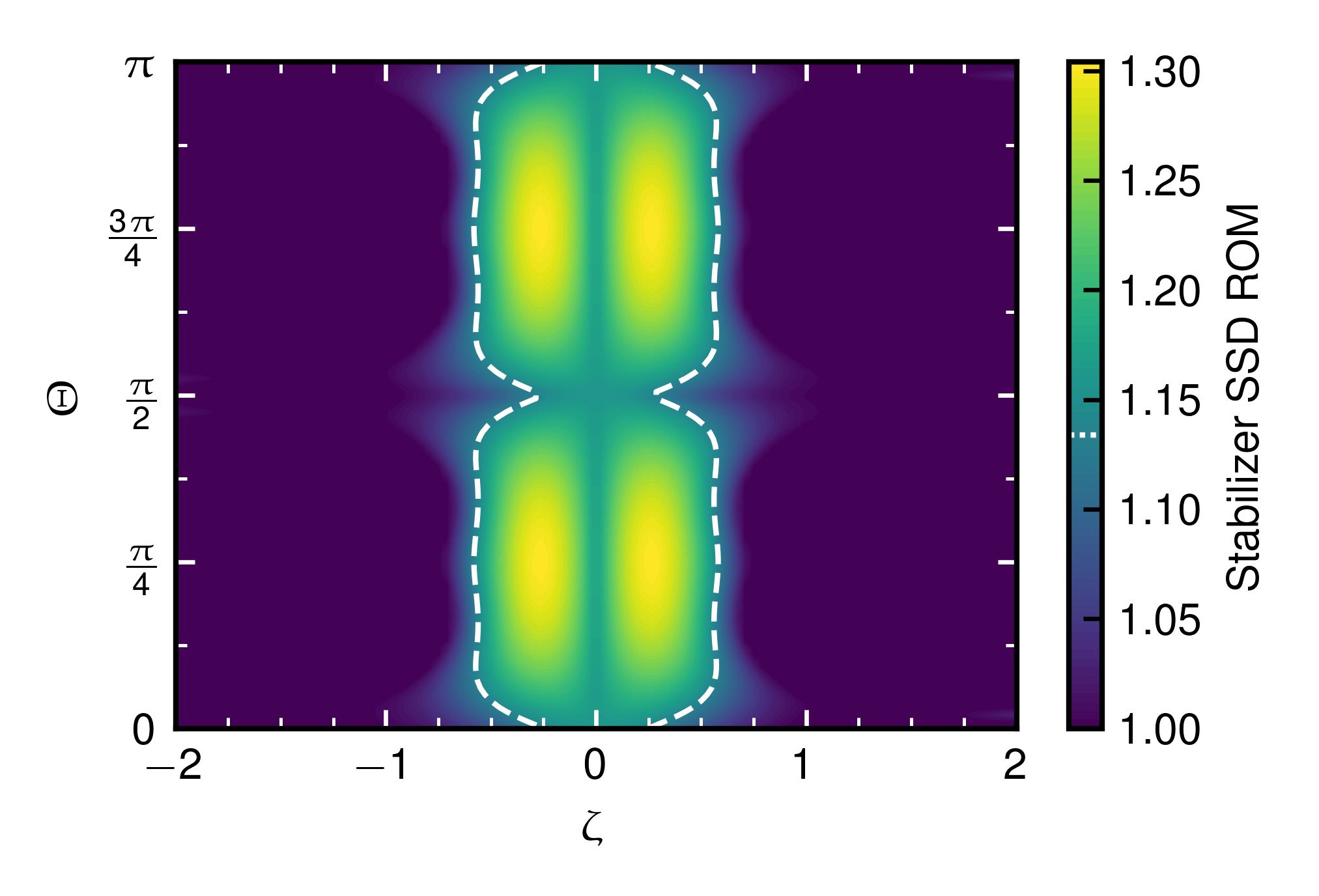}
    }
    \hfill\subfloat[ROM of the (Gaussian) modular SSD of a rotated-squeezed Gaussian state.\label{subfig:gauss_gkpGauss}]{%
      \includegraphics[width=\linewidth]{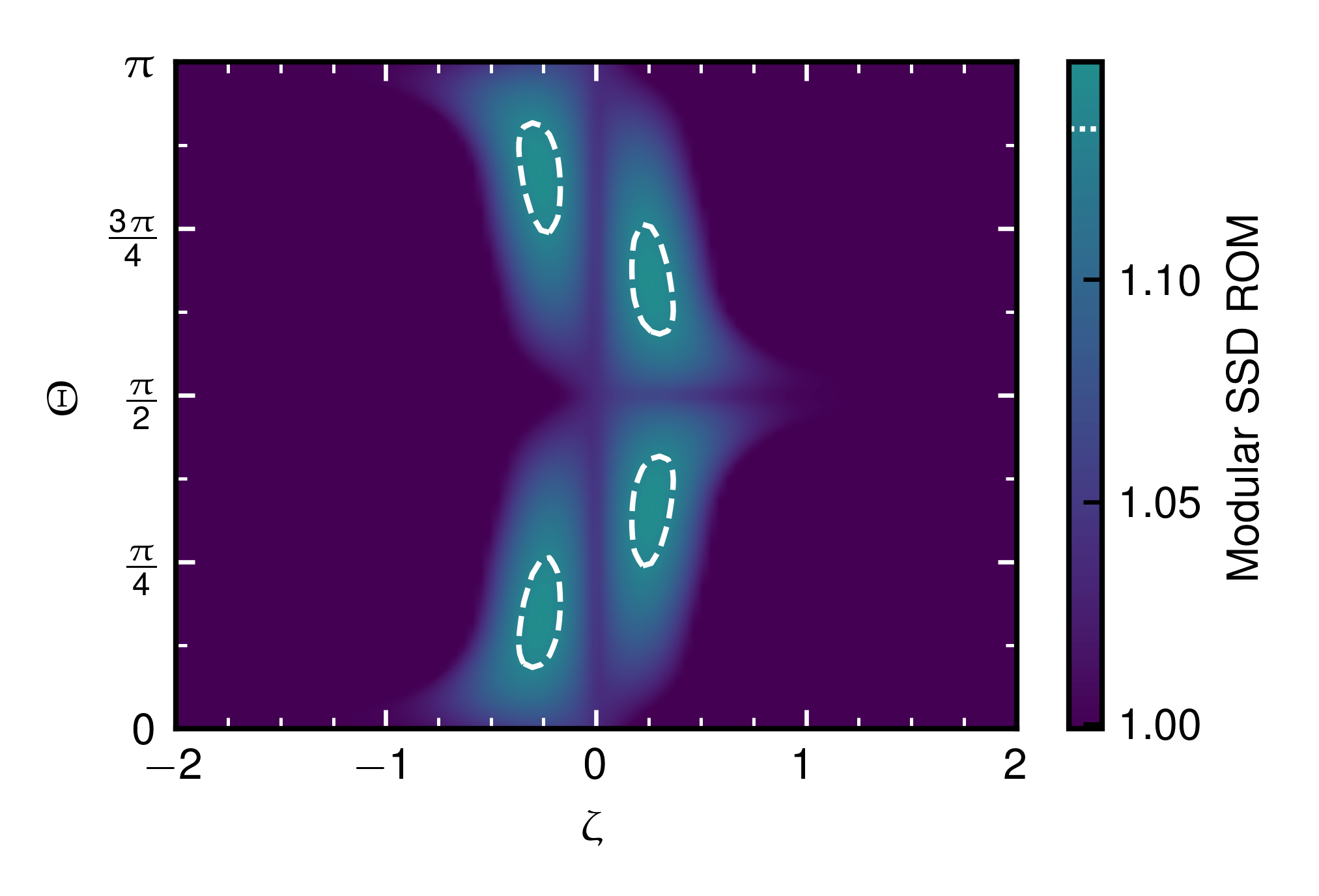}
    }
    \hfill
    
     \caption{The ROM of a decomposed rotated and squeezed Gaussian state, as defined in Eq.~(\ref{eq:realistic-gkp-theta}), for different values of squeezing $\zeta$ and rotation angles $\Theta$. The regions inside the dashed white boundaries in each plot indicate the regions of distillability. }
    \label{fig:roms-gaussian-states}
\end{figure}

We recalled in Sec.~\ref{sec:families-gaussian}, specifically in Eq.~(\ref{eq:single-gaussian-state}) that any pure Gaussian state can be defined via a Gaussian unitary operation --- parameterized in terms of the squeezing parameter $\zeta$, a rotation angle $\Theta$ and a displacement vector $\mathbf s$ --- acting on the vacuum state. 
In Fig.~\ref{fig:roms-gaussian-states} we plot the value of ROM of the qubit state arising from the different choices of SSD of a pure Gaussian state, parameterized by the squeezing parameter $\zeta$ and rotation angle $\Theta$. Specifically, in Fig.~\ref{subfig:gauss_stab} we plot the ROM of the stabilizer SSD of a Gaussian state and in Fig.~\ref{subfig:gauss_gkpGauss} we plot the modular SSD of a Gaussian state. Note that in the figures we choose the value of the displacement vector $\mathbf s$ to be zero in both position and momentum, however when we later optimize to identify the maximally resourceful states, we also optimize over the choice of $\mathbf s$. Also, note that Fig.~\ref{subfig:gauss_gkpGauss} equivalently shows the ROM of the Gaussian modular SSD because the wavefunction of a pure Gaussian state centered in phase space is symmetric in position.

We start by inspecting Fig.~\ref{subfig:gauss_stab}, which shows the ROM of the stabilizer SSD for different pure Gaussian states. As a reminder, values of $\Theta=0$ and $\zeta=0$ correspond to the vacuum state, while non-zero values correspond to a rotated and squeezed state.
We find that the stabilizer SSD of the vacuum state has a value of ROM of $\mathcal R(\hat \rho_\Pi)\approx 1.160$, which is greater than the threshold for $T$-magic state distillation. This result is in line with what reported in Ref.~\cite{baragiola2019}, where it was shown that a supply of vacuum states allows to distill 0-logical GKP states into magic states with Gaussian
operations alone, although in that work  distillation towards H states was rather considered.
  However, the vacuum is not the optimal Gaussian state to achieve a high value of ROM~\footnote{Note that the vacuum state is the optimal Gaussian state for $H$ state distillation~\cite{baragiola2019}.}. Instead, we find that the value of ROM (and hence, fidelity to $T$) is greater when using a rotated squeezed state. Specifically, by numerically optimizing over all the parameters of the Gaussian unitary we find that by choosing rotation angle $\Theta=\pi/4$, squeezing parameter $\zeta\approx 0.26$, and displacement $\mathbf s=(0,0)$, a ROM of $1.303$ can be achieved.  
  Note that the ROM of the stabilizer SSD is symmetric in both $\zeta$ and $\Theta$.

Next, inspecting Fig.~\ref{subfig:gauss_gkpGauss}, we see that the ROM of the modular SSD of the vacuum state is $1$. This is because the modular SSD of the vacuum state evaluates to the maximally mixed state. To achieve a value of ROM above the distillation threshold for the modular SSD, it is necessary to instead use a Gaussian state with both non-zero squeezing and rotation.

Note that, since it is possible to convert between any Gaussian state with only Gaussian operations, in virtue of the possibility of optimizing over Gaussian unitaries in Eq.~(\ref{eq:Kraus:projector-QEC}), all pure Gaussian states should be considered equally resourceful for SGKP circuits.
Furthermore, given that the value of ROM of the SSD of a CV state is convex, as shown in Eq.~(\ref{eq:convexity-rom-ssd}), the value of ROM of the SSD of a mixed Gaussian state can only be less than or equal to the value of ROM of the SSD of a pure Gaussian state. This implies that the optimal values of ROM for the pure single-mode Gaussian states are also optimal over all single-mode Gaussian states, including thermal states.

\subsubsection{GKP states}
\label{sec:analysis-gkp}
\begin{figure}[htb]
    \centering
    \subfloat[ROM of the stabilizer SSD of a realistic GKP state. \label{subfig:gkp_stab}]{%
      \includegraphics[width=\linewidth]{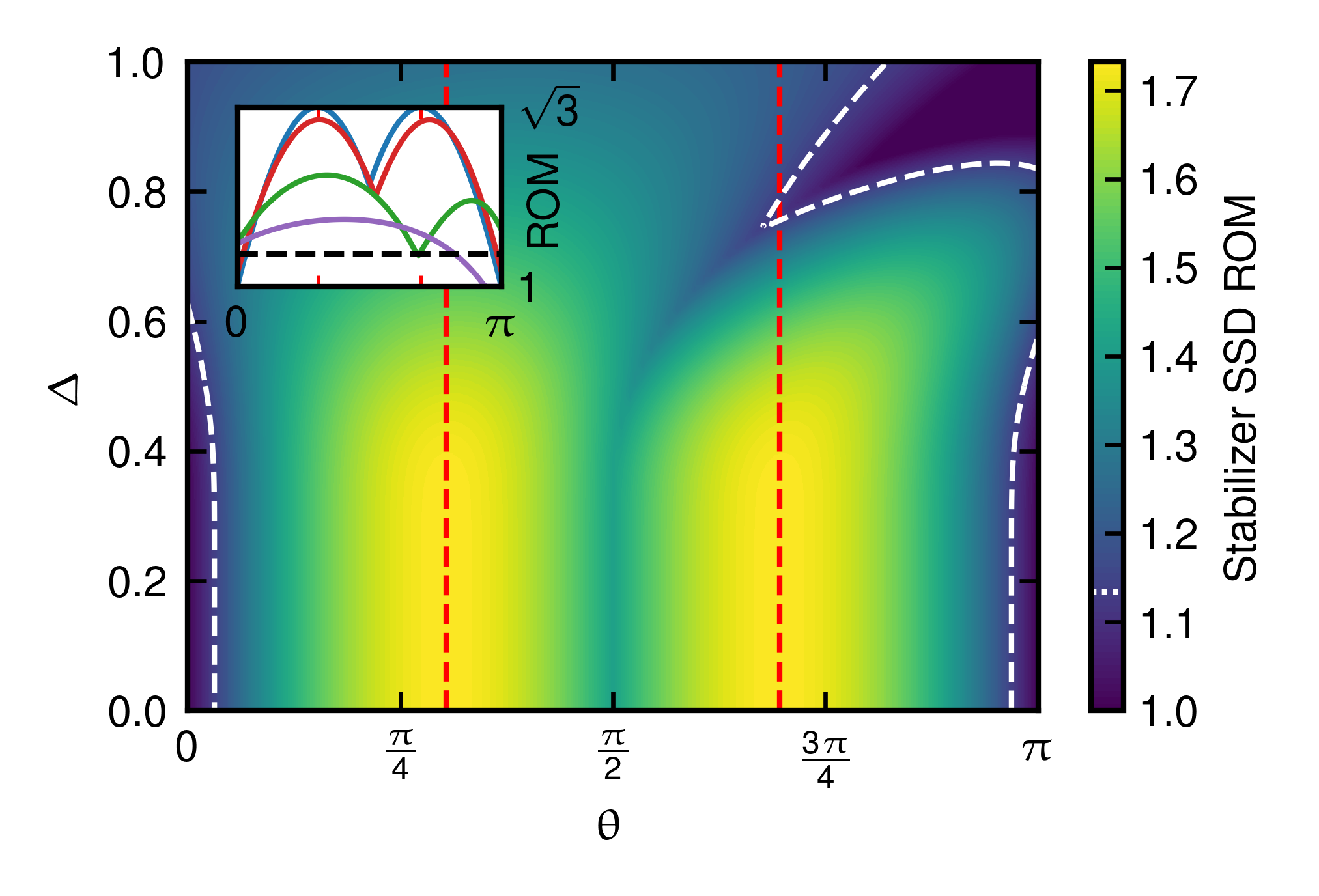}
    }
    \hfill\subfloat[ROM of the (Gaussian) modular SSD of a realistic GKP state.\label{subfig:gkp_gkpGauss}]{%
      \includegraphics[width=\linewidth]{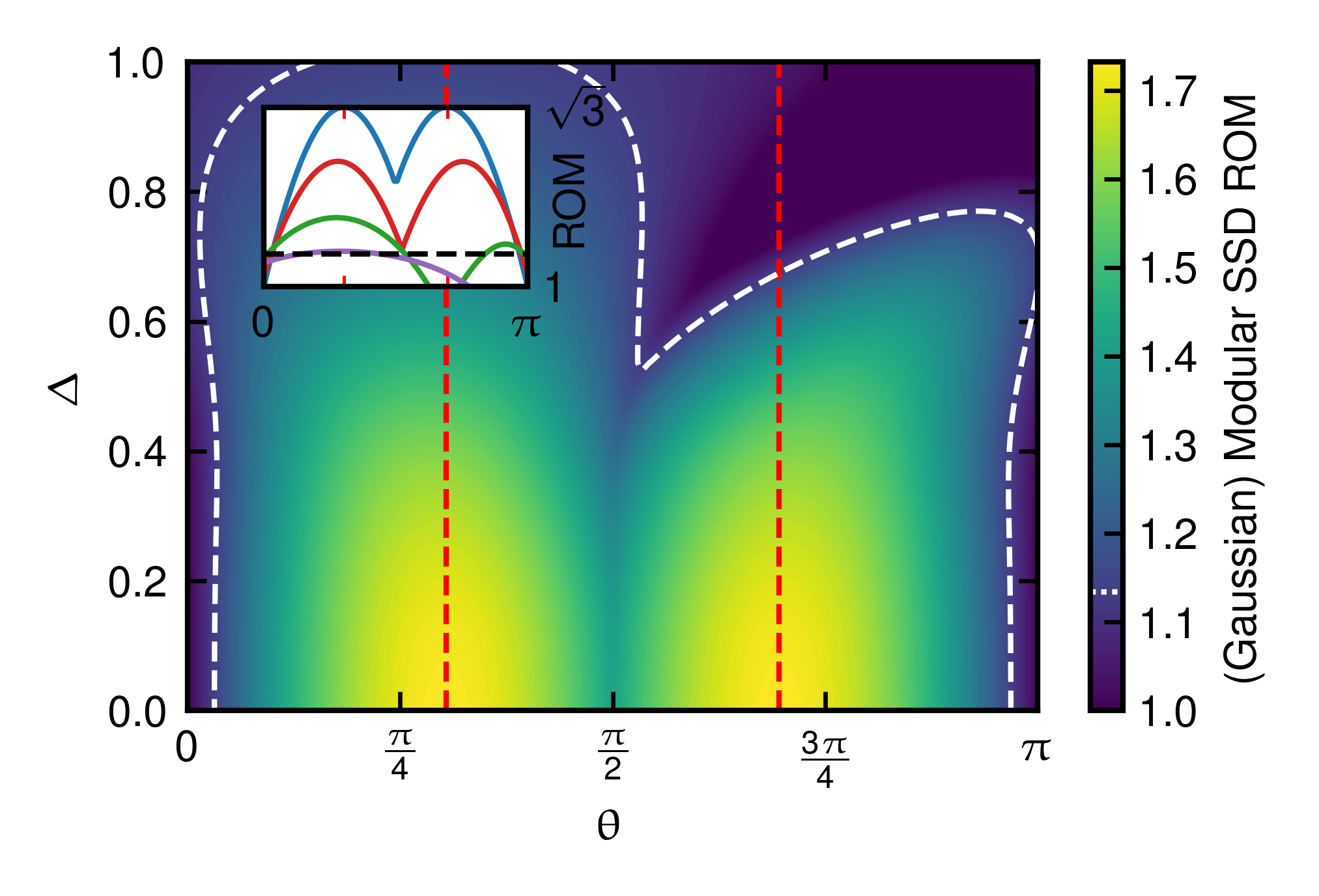}
    }
    \hfill
    
     \caption{The ROM of a decomposed encoded qubit GKP state, as defined in Eq.~(\ref{eq:realistic-gkp-theta}), for different values of squeezing $\Delta$ and rotation angles $\theta$ and a fixed phase of $\phi=\pi/4$. The red dashed lines indicate the values of $\theta$ for which the state is an encoded $T$-state, i.e., ${\theta=\arccos(\pm 1/\sqrt 3)}$.  The large regions inside the dashed white boundaries in each plot indicate the regions of distillability. The inset plots show a subset of the same data plotted with $\theta$ on the $x$-axis and the value of ROM on the $y$-axis. The solid blue, red, green and purple lines correspond to $\Delta=0,1/2,3/4,1$, respectively. Equivalently, the lines for each increasing $\Delta$ have decreasing maxima. Note that the value of ROM in the main figures and the insets is always greater than or equal to $1$.}
    \label{fig:roms-gkp}
\end{figure}

We start by numerically calculating the ROM of an encoded realistic GKP state of the form given in Eq.~(\ref{eq:realistic-gkp-def}), where we fix $\phi=\pi/4$. The state is given by
\begin{align}
\label{eq:realistic-gkp-theta}
&\ket{\psi_{\text{GKP}}^{\Delta}(\theta)} \nonumber \\
=&\frac{1}{\sqrt{\mathcal{N}_{\text{GKP}}}}\left(\cos(\theta/2)\ket{\bar 0_{\text{GKP}}^{\Delta}}+\sin(\theta/2)e^{i\pi/4}\ket{\bar 1_{\text{GKP}}^{\Delta}}\right),
\end{align}
which is only parameterized by the logical encoding angle $\theta$ and the squeezing parameter $\Delta$. Fixing $\phi=\pi/4$ allows us to identify a selection of insightful logical states with only varying $\theta$. For $\theta=0$, this state is simply the $0$-logical GKP state. For $\theta=\pi$, this state is the $1$-logical GKP state. For $\theta=\arccos(1/\sqrt 3)$ and ${\theta=\pi - \arccos(1/\sqrt 3)}$, the state is an encoded magic $\ket{T}$ state and its orthogonal magic state, respectively. Each of these states is an encoded finitely squeezed state, parameterized by $\Delta$. From these states, we evaluate the resulting qubit state from each SSD introduced in the previous Sec.~\ref{sec:unified} and plot the ROM of each state in Fig.~\ref{fig:roms-gkp}. The red vertical lines correspond to the encoded $T$ states and have the highest ROM for any choice of $\Delta$, in fact reaching the maximal achievable ROM (i.e, $\mathcal R = \sqrt 3$) for $\Delta=0$. The plot shows ample regions of distillability in terms of the parameters $\theta$ and $\Delta$, i.e., regions where $\mathcal R(M_P(\hat \rho))>\mathcal R^*$, identified by the shaded white contour. These results are in line with and generalize what is reported in Ref.~\cite{calcluth2023}, 
where it was shown that a supply of realistic GKP states allows to distill magic states from 0-logical GKP states with Gaussian operations alone, although that result referred to  distillation towards H states.

\begin{figure}[ht]
    \centering
    \includegraphics[width=\linewidth]{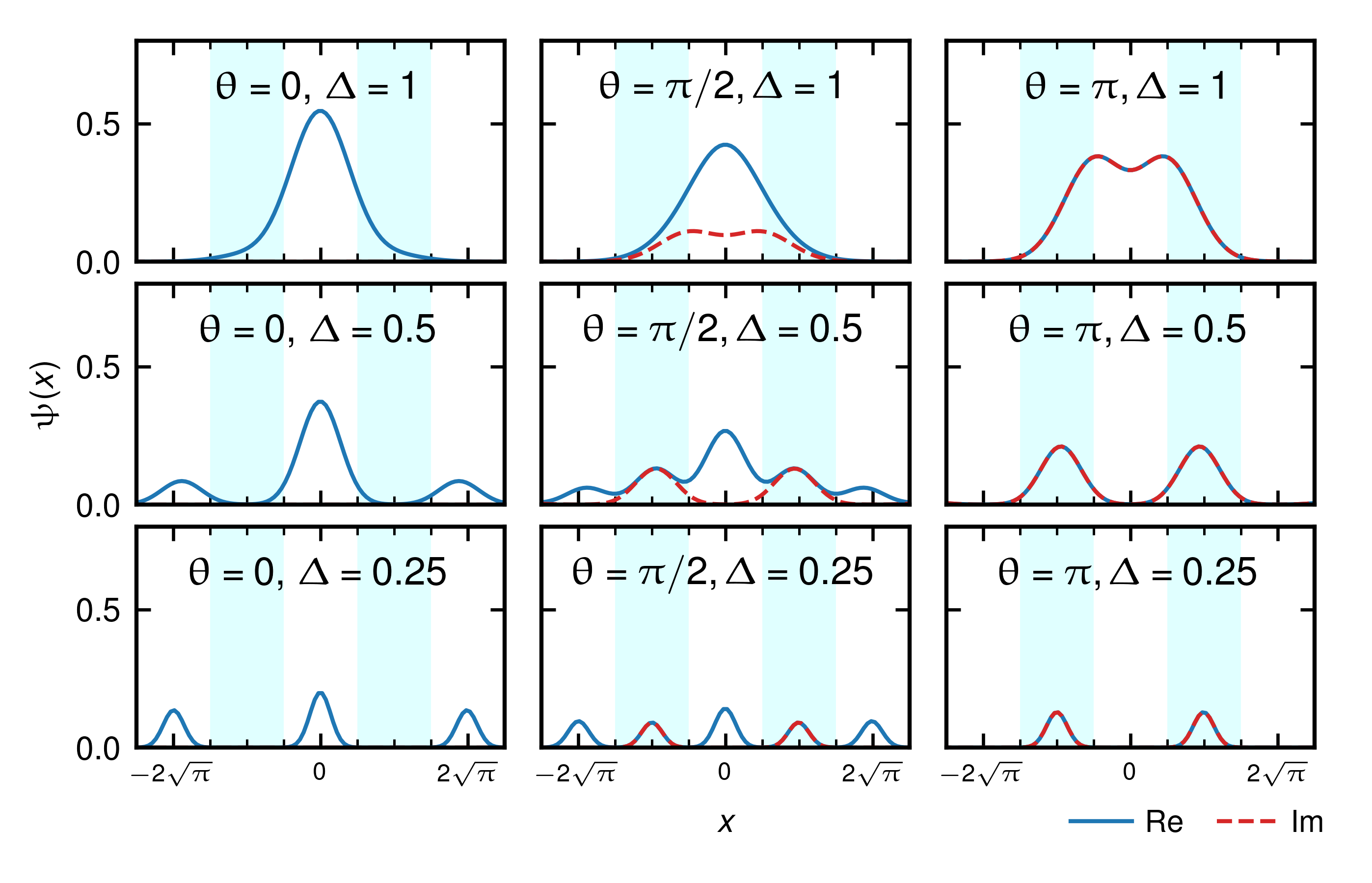}
    \caption{Wavefunctions of a realistic GKP state with various levels of squeezing $\Delta\in \{1/4,1/2, 1\}$, encoding states with $\phi=\pi/4$ and ${\theta\in \{0,\pi/2, \pi\}}$. Note that the wavefunction with $\theta=0$ corresponds to the $0$-logical state, whereas the wavefunction with $\theta=1$ corresponds to the $1$-logical state, up to a global phase. The state with $\theta=\frac \pi 2$ corresponds to the Pauli $\hat Y$ basis state with eigenvalue $1$.}
    \label{fig:gkpWFatPoints}
\end{figure}

Note that there is an asymmetry in the shape of the contour levels of the ROM in Fig.~\ref{fig:roms-gkp} which arises as the level of squeezing is decreased, i.e., $\Delta$ is increased. This asymmetry can be interpreted as arising from the fact that the norm of the unnormalized $0$-logical state is in general larger than the norm of the unnormalized $1$-logical state. Therefore, when the states are combined in superposition and normalized together, the $0$-logical component contributes more than the $1$-logical component. For values of $\Delta\ll 1$, the norm of each state is approximately equal and hence this asymmetry is no longer present~\cite{gottesman2001,pantaleoni2021}.
In Fig.~\ref{fig:gkpWFatPoints} we plot the wavefunction of this GKP state for various levels of squeezing. This provides a visual explanation as to why the ROM of the stabilizer SSD of the encoded GKP state is asymmetric. We observe that for $\Delta=1$ and $\theta=0$, the state approximates the vacuum state, which is a known resource~\cite{baragiola2019,calcluth2023}. However, at $\theta=\pi$ the state retains two peaks. This difference affects the logical content of the decomposed state. Note that the definition of the wavefunction of a GKP state can affect the logical content of the encoded state and hence also the resulting ROM of the encoded state. We use the same definition of the GKP state analyzed in Ref.~\cite{pantaleoni2021} rather than that of Ref.~\cite{shaw2022}. We provide a more detailed discussion in Appendix \ref{sec:appendix-different-gkp-defs}.  

We also note that the stabilizer SSD state has higher values of ROM for all $\Delta$ and $\theta$, as compared to the ROM of the modular SSD.

Furthermore, we stress that the threshold of the ROM, $\mathcal R^*$, is not a necessary condition for achieving quantum advantage. For example, the $0$-logical state with squeezing $\Delta=1$, which approximates the vacuum state, has a value of ROM below the distillation threshold, meaning it cannot be distilled to the $T$ state. Despite this fact, the state can be distilled to the $H$ state and can therefore still be considered a resource for quantum advantage~\cite{baragiola2019}.

We leave further discussion of these results for specific GKP states to Appendix \ref{appendix:gkp-large-delta}, whereby we also provide a comparison of the ROM of the stabilizer SSD of the GKP states with the WLN of the same states.

Finally, we note that the maximal achievable ROM using GKP states is significantly higher than that which is possible using Gaussian states.

\begin{figure}[htb]
    \centering
    \subfloat[ROM of the stabilizer SSD of an even cat state parameterized by $\alpha=re^{i\Phi}.$ The white dashed lines show the regions where the stabilizer SSD ROM is above the threshold for distillability. \label{subfig:cat-stab}]{%
      \includegraphics[width=0.99\linewidth]{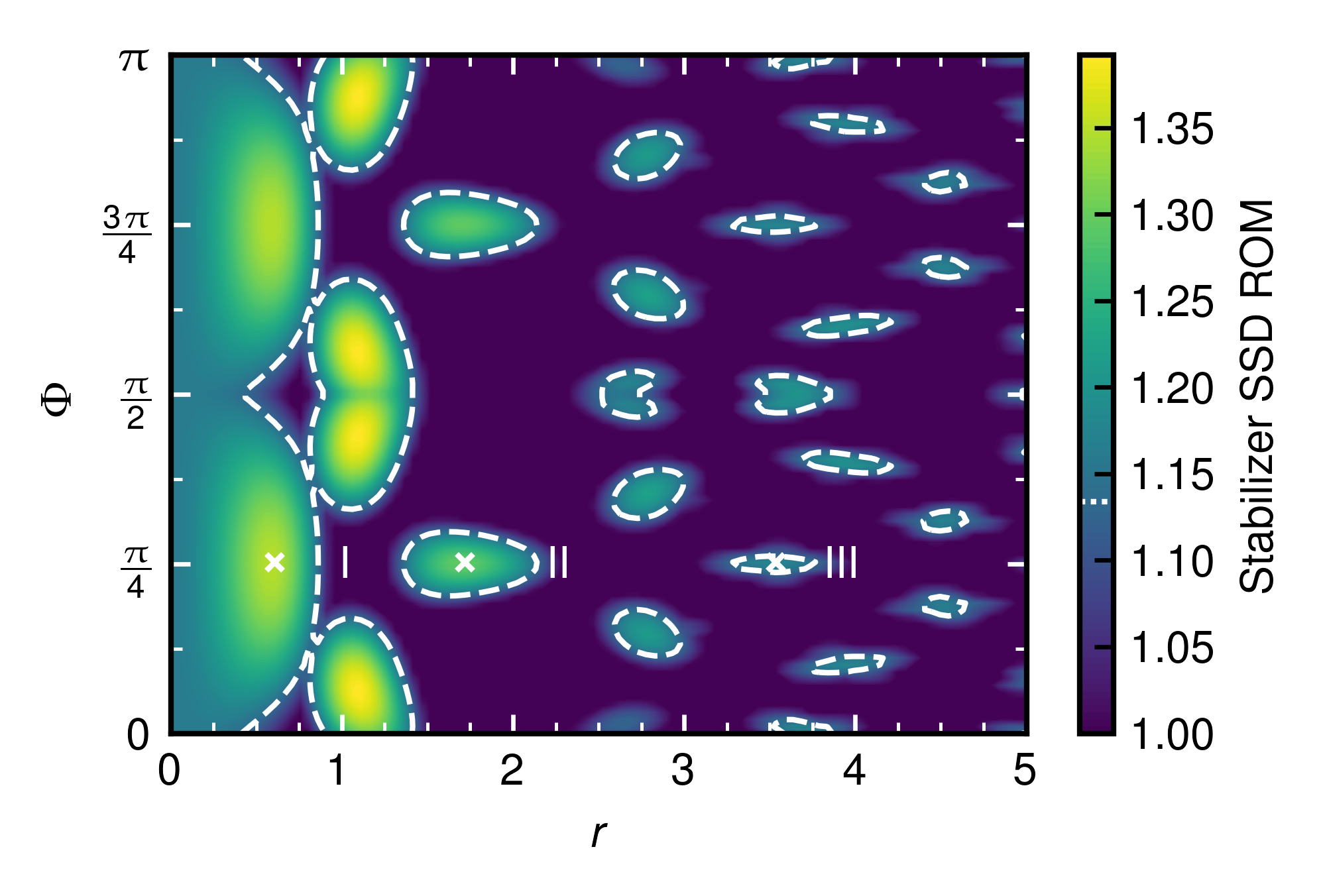}
    }
    \hfill
    \subfloat[ROM of the Gaussian modular SSD (equivalently, the modular SSD) of an even cat state parameterized by $\alpha=re^{i\Phi}.$ \label{subfig:cat-gauss}]{%
      \includegraphics[width=0.99\linewidth]{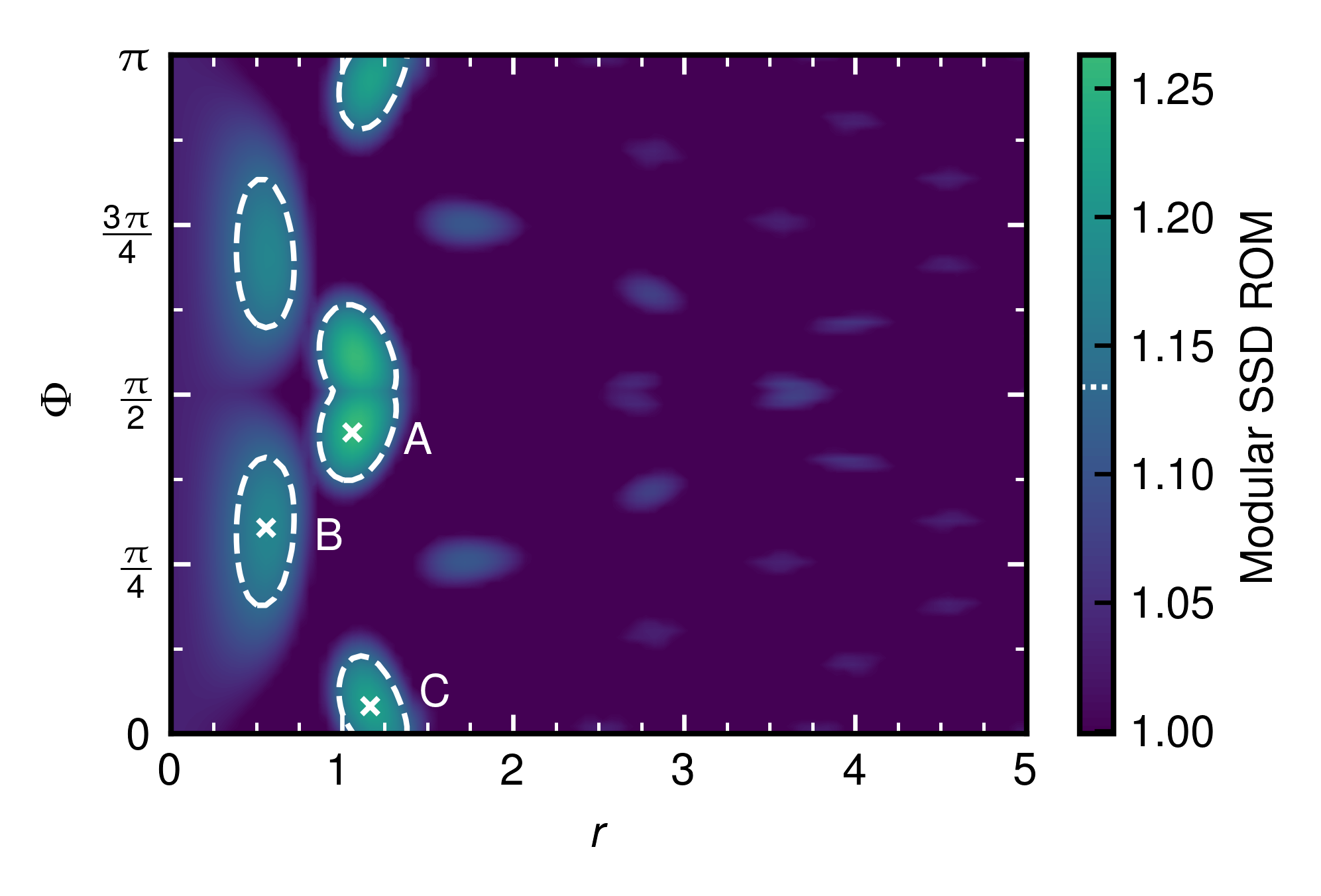}
    }
    \hfill
    
     \caption{Plot a) shows the ROM of the cat states decomposed using stabilizer SSD, while plot b) shows the ROM of the same class of states decomposed with the (Gaussian) modular SSD. The white dashed lines show the regions where the SSD ROM is above the threshold for distillability and hence the states are resourceful for quantum advantage with SGKP circuits. The wavefunctions of the states labeled with a cross are plotted in Fig.~\ref{fig:catWFatPoints}.}
    \label{fig:roms-cat}
\end{figure}

\begin{figure}[htb]
    \centering
    \includegraphics[width=0.99\linewidth]{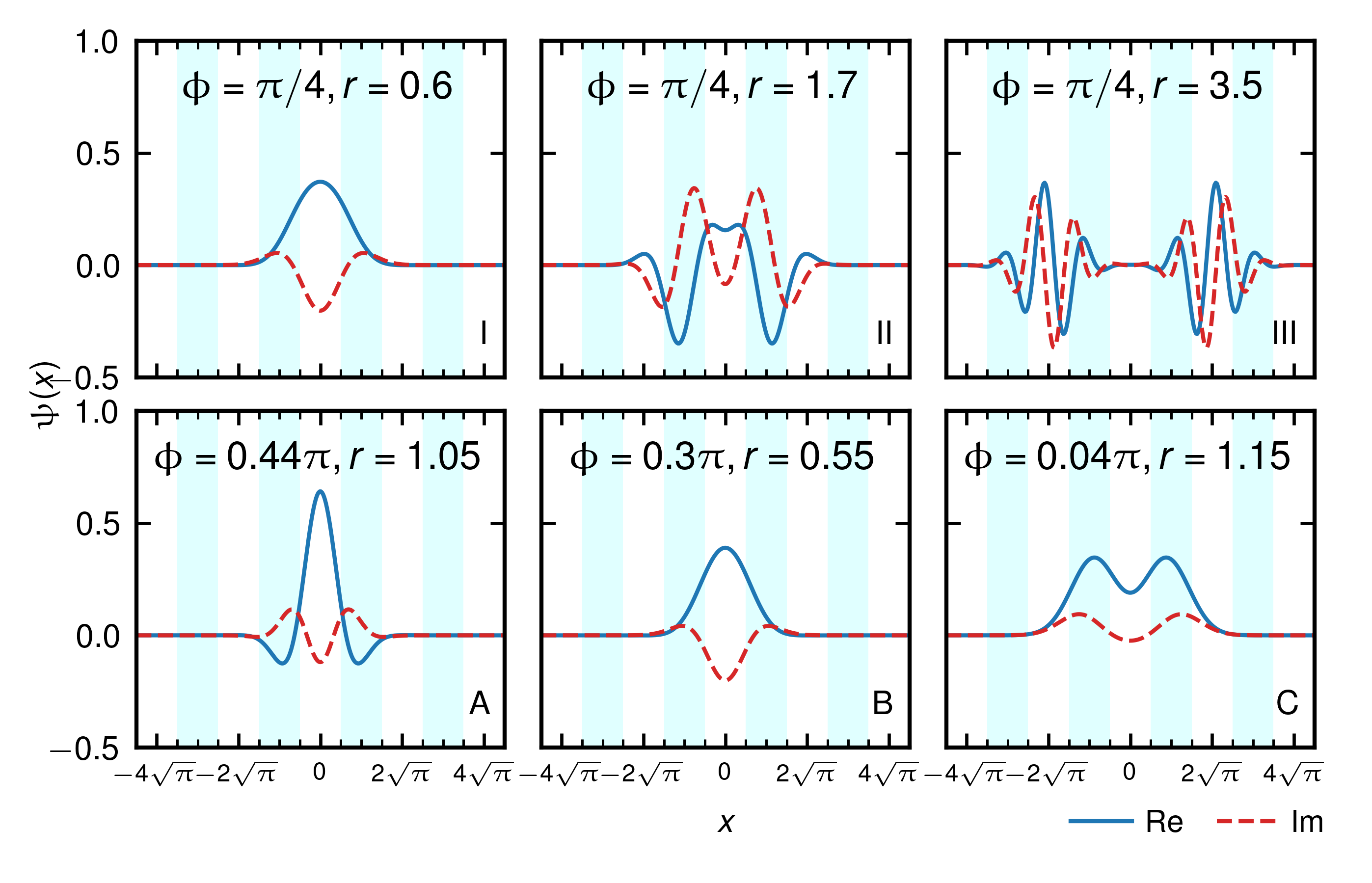}
     \caption{Wavefunctions of the cat state for different choices of $\alpha=re^{i\Phi}$, corresponding to points specified in Fig~\ref{fig:roms-cat}. The regions with a white and blue background represent areas that contribute to the $0$-logical and $1$-logical components of each SSD, respectively.}
    \label{fig:catWFatPoints}
\end{figure}

\subsubsection{Cat states}

After having analysed the two most natural classes of states for our framework --- namely, Gaussian and (realistic) GKP states --- we now move to a class of states with no specific relation to SGKP circuits. In particular, we analyze the even cat state as defined in Eq.~(\ref{eq:even-cat-def}). We parameterize the cat state using the complex number $\alpha$ by separating its magnitude $r$ and  phase $\Phi$, i.e., $\alpha=re^{i\Phi}$. 

Due to the fact that these states have a wavefunction that is symmetric in position, the modular SSD is equivalent to the Gaussian modular SSD. 
The ROM of the stabilizer SSD and the Gaussian modular SSD (equivalently, the modular SSD) of the state, for different values of $r$ and $\Phi$, are plotted in Fig.~\ref{fig:roms-cat}, where the regions above the distillation threshold are enclosed by the dashed white lines.
We find that for most choices of $\alpha$, the ROM of the stabilizer SSD is greater than the ROM of the Gaussian modular SSD.

Note that the value of $\Phi$ corresponds to a rotation in phase space and can be implemented using Gaussian unitary operations, which are included in the set of SGKP circuits. Therefore, we should consider the lower bound of the maximum ROM, as defined in Eq.~(\ref{eq:gkp-rom}), to be the maximum of all angles $\Phi$ for a given $r$. 

We also observe that both the values of the stabilizer SSD ROM and the values of the modular SSD ROM each display symmetry. Specifically, when the state is rotated by $\pi/2$, the values of each respective ROM are equal. This can be seen from the equal values of ROM in each of Figs.~\ref{subfig:cat-stab} and \ref{subfig:cat-gauss} at values of $\Phi$ and $\Phi+\pi/2$. This angle corresponds to a Fourier transform which can equivalently be considered a change of basis of the quadratures,  $\hat q,\hat p$. The stabilizer SSD is known to be symmetric in $\hat q,\hat p$, so this symmetry is to be expected for the stabilizer SSD ROM~\cite{shaw2022}. However, the modular SSD is not symmetric in general. Instead, this symmetry arises from the definition of the cat state.
First note, that the cat state is symmetric under rotations around $\pi$, i.e.,
\begin{align}
    \hat R(\pi)\ket{\bar 0^{\alpha}_{\text{cat}}}=\ket{\bar 0^{\alpha}_{\text{cat}}}.
\end{align}
We also see that the wavefunction of a coherent state with angle $\Phi$ is the complex conjugate of the wavefunction of a state with angle $-\Phi$.
This also implies that the wavefunction of the cat state with angle $\Phi$ is equal to the complex conjugate of the wavefunction with angle $-\Phi$, i.e.
\begin{align}
    \bra{\hat q=x}\ket{\bar 0^{\alpha=re^{i\Phi}}_{\text{cat}}}=\left(\bra{\hat q=x}\ket{\bar 0^{\alpha=re^{-i\Phi}}_{\text{cat}}}\right)^*.
\end{align}
Given that the state is also symmetric under rotations by $\pi$, we see that the wavefunction of the cat state with angle $\Phi$ is equal to the complex conjugate of the cat state with angle $\pi-\Phi$. This also means for a density matrix $\hat \rho$ of a cat state with angle $\Phi$, the corresponding density matrix of a cat state with angle $\pi-\Phi$ can be considered to be $\hat \rho^*$. In terms of the logical density matrix of a cat state $\hat \rho_{L,\Phi}$ with angle $\Phi$, the corresponding density matrix with angle $\pi-\Phi$ is given by $\hat \rho_{L,\pi-\Phi}=\hat \rho_{L,\Phi}^*$, which is equivalent to applying a phase gate to the density matrix $\hat \rho_{L,\Phi}$. The phase gate is Clifford and therefore the ROM of each state is equal.

We also note the decreasing values of ROM for higher values of $r$. This effect can be understood by considering the wavefunction of the state. The value of $r$ corresponds to the distance between the peaks of the wavefunction, and also the width of the individual peaks. As the peaks become further apart, each peak can be binned inside one type of region, rather than across two or more regions. To illustrate this, we have provided plots of the wavefunction of the cat states for some selected values of $r,\Phi$ in Fig.~\ref{fig:catWFatPoints}. In the limit of large $r$, the state consists of two peaks, both contained entirely in either the region corresponding to the $0$-logical state or the region corresponding to the $1$-logical state. This implies that the SSD of the state will be a logical basis state.

Finally, we note that the maximal ROM of the stabilizer SSD of an even cat state is $1.39$, which is higher than the maximal achievable with Gaussian states alone.

\subsubsection{Cubic-phase state}

\begin{figure}[ht]
    \centering
    \subfloat[ROM of the stabilizer SSD of a cubic phase state\label{subfig:cps_stab}]{%
      \includegraphics[width=\linewidth]{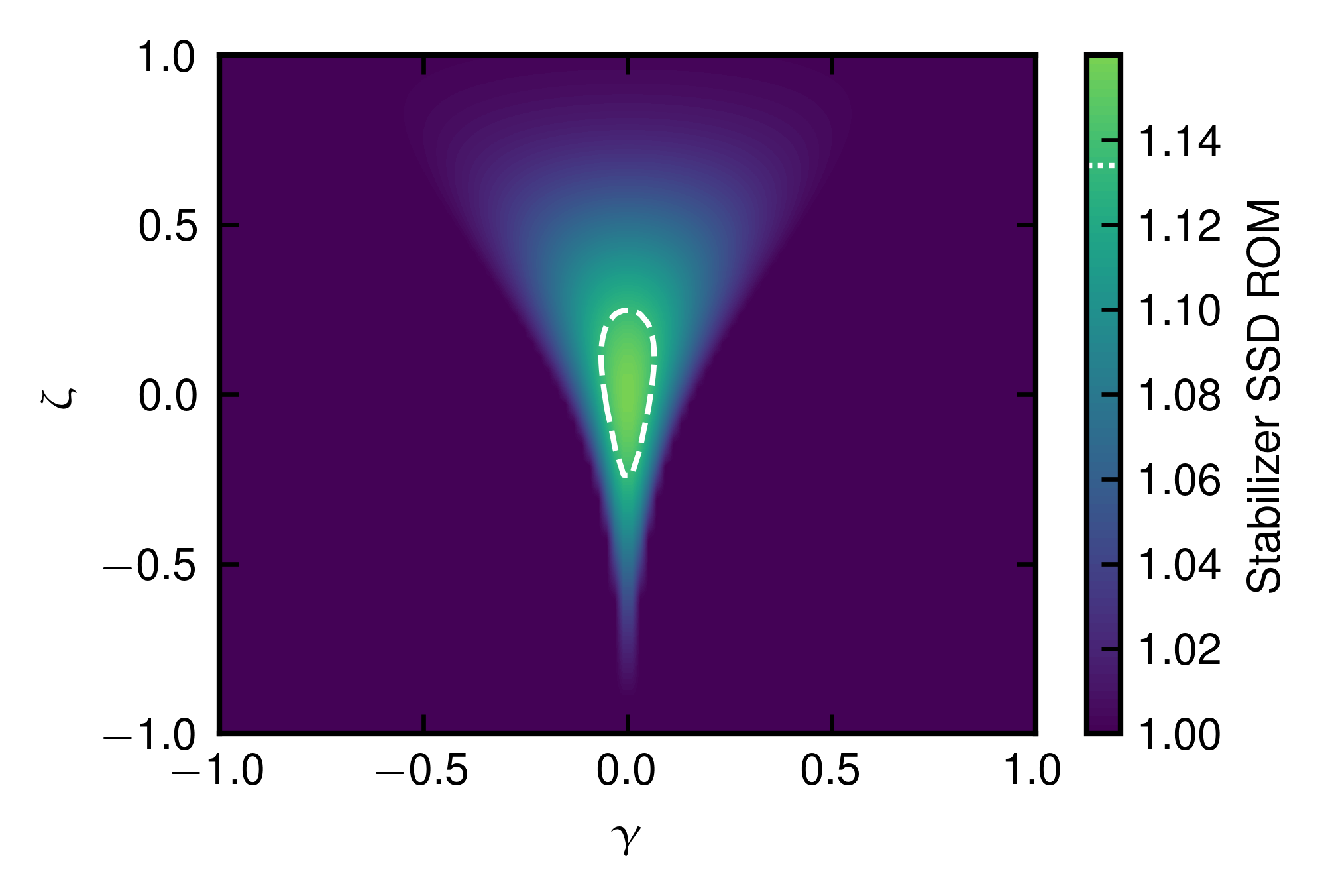}
    }
    \hfill\subfloat[ROM of the Gaussian modular SSD of a cubic phase state\label{subfig:cps_gkpGauss}]{%
      \includegraphics[width=\linewidth]{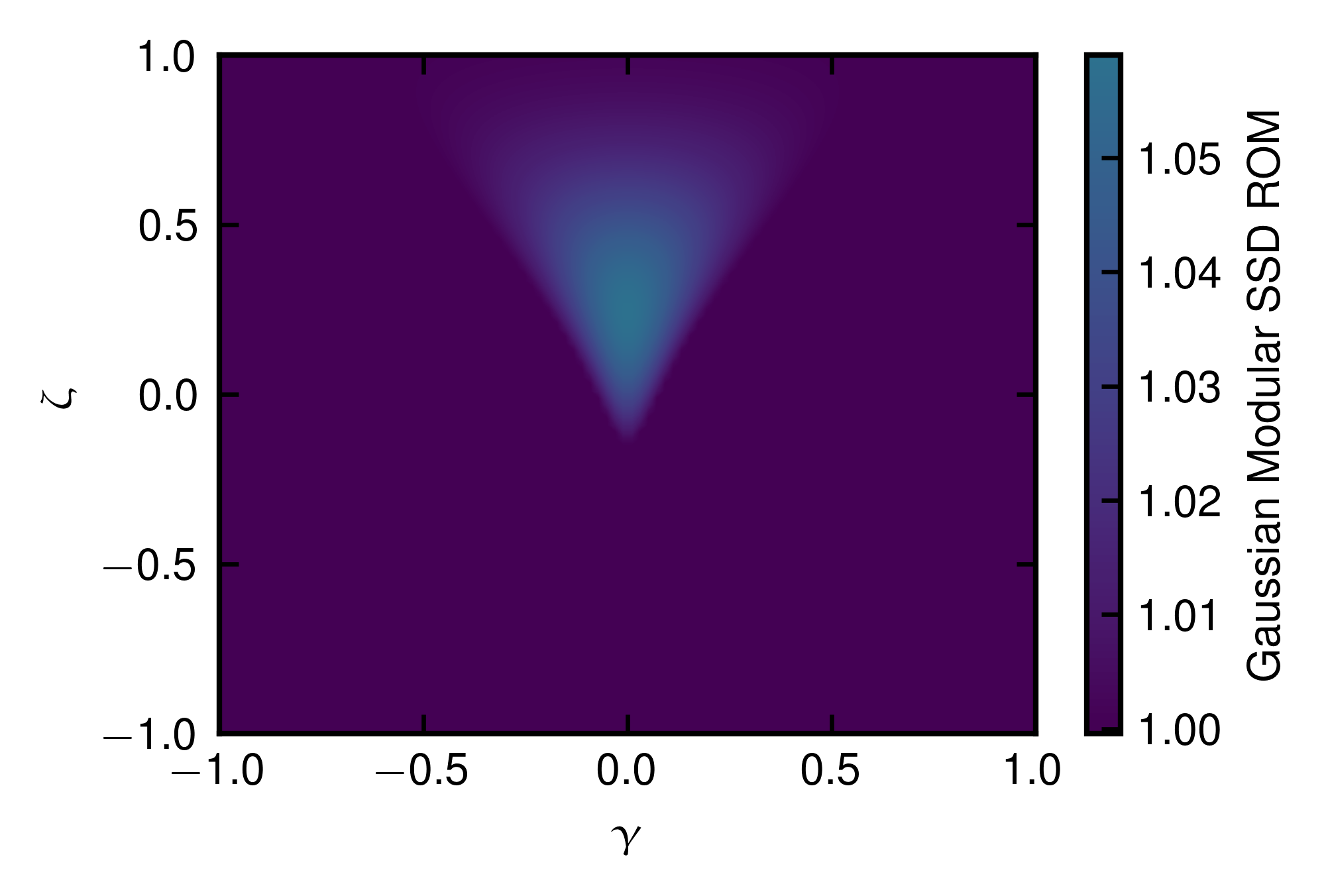}
    }
    \hfill
    
     \caption{The ROM of a decomposed cubic phase state for different values of squeezing $\zeta$ and cubicity $\gamma$. Plot a) shows the ROM of the stabilizer SSD of the state, while plot b) shows the ROM of the Gaussian modular SSD of the state. Note that because the wavefunction of the cubic phase state is not symmetric in position, plot b) is not equivalent to the ROM of the modular SSD of the cubic phase state. The white region in plot a) shows the region of states that are resourceful for SGKP circuits. Note that no states are above the distillation threshold for the Gaussian modular SSD ROM.}
    \label{fig:rom-cps}
\end{figure}

The ROM of the stabilizer SSD and the Gaussian modular SSD of the cubic-phase state is plotted in Fig.~{\ref{fig:rom-cps}}. We find that, counter-intuitively, the ROM of the stabilizer SSD of the state is maximum when both the cubicity and squeezing are zero, i.e., $\gamma=0,\zeta =0$, which corresponds to the vacuum state. Note that this is somewhat surprising since the maximally resourceful state among this family of states is the one for which the state is Gaussian and the WLN is zero. However, given that for SGKP circuits we consider the already highly Wigner negative stabilizer GKP states to be resourceless, we know that negativity is not necessary for the promotion of these circuits to universality. 

Note that unlike the other states considered in this work, the cubic-phase state is not symmetric in the position basis. Therefore, the modular SSD of this state is not equivalent to the Gaussian modular SSD. Hence, evaluating the ROM of the modular SSD does not provide a resource-theoretically meaningful quantifier of the resourcefulness of the cubic-phase state --- as in this case the modular SSD requires non-Gaussian operations, in addition to GKP states, to be implemented. For completeness we provide a plot of the ROM of the modular SSD for the cubic phase state in Appendix \ref{sec:appendix-plot-mod-cps}.

\section{Conclusion}
\label{sec:conclusion}
In quantum computation over DV systems, the fidelity to a target magic state is a well-established criterion for determining whether a state can promote otherwise simulatable Clifford circuits to universality, potentially leading to quantum computational advantage. In CV systems, while the presence of negativities in the Wigner function of a given circuit serves as a necessary condition for universality, it falls short of providing a sufficient criterion. To bridge this gap, we have introduced a resource-theoretically motivated framework, enabling the formulation of a sufficient criterion to assess the resourcefulness of a generic CV state for quantum computation. 

Specifically, we have introduced such a criterion in the framework of SGKP circuits. Our criterion is based on the evaluation of a measure of magic on the encoded logical state associated to a generic CV state $\hat \rho$, upon mapping it to the computational subspace of the GKP code. For resource-theoretically grounded mappings --- such as the stabilizer SSD and the Gaussian modular SSD --- this quantity can be understood as the resourcefulness of the state $\hat \rho$ to promote otherwise simulatable GKP circuits to universality. Applying such a criterion we find that  all pure Gaussian states are equally resourceful for promoting SGKP circuits to universality. Moreover, we found that certain non-Gaussian states, albeit not necessarily all, have a value of ROM higher than the set of Gaussian states. 

Furthermore, our work provides a rigorous and resource-theoretically grounded interpretation of recently introduced methodologies aimed at extracting the binary logical content of generic CV states. In particular, we have elucidated that the mapping established by the stabilizer SSD \cite{shaw2022} can be understood in terms of resourceless operations in the context of SGKP circuits, for any state. This interpretation also holds for the modular SSD \cite{pantaleoni2020}, albeit exclusively for states symmetric in the position representation. Considering the relevant role of SSDs in extracting the logical content of states in the emerging  field of quantum computation over CV systems, we expect that this result will hold independent interest.

We conclude by recalling that the ideal GKP states considered in the SGKP framework are infinitely squeezed. Therefore, in order to provide a conclusive validation of the result presented here from a practical and operational viewpoint --- including the interpretation of SSDs in term of resourceless mapping ---  it would be necessary for our findings to hold also in the presence finite squeezing --- and for the SSDs to be implemented with finite squeezing. We leave this analysis for future work.

\section{Acknowledgments}
We acknowledge discussions with Shahnawaz Ahmed, Timo Hillmann, Laura García-Álvarez, and Oliver Hahn. We also acknowledge the anonymous QIP reviewers for their useful and insightful suggestions to improve the paper. We acknowledge fundings from the HORIZON-EIC-2022-PATHFINDERCHALLENGES-01 programme under Grant Agreement Number 101114899 (Veriqub). Views and opinions expressed are however those of the authors only and do not necessarily reflect those of the European Union. Neither the European Union nor the granting authority can be held responsible for them. G. F. and C. C. acknowledge support from the VR (Swedish Research Council) Grant QuACVA. G. F. and C. C. acknowledge support from the Wallenberg Center for Quantum Technology (WACQT). 

\appendix

\begin{widetext}
\section{Measure of resourcefulness for qubits}
\label{sec:appendix-measure}
In this Appendix, we explicitly derive a threshold for the robustness of magic~\cite{howard2017} of a qubit state for which a supply of such states, to otherwise all-Clifford circuits, are sufficient for universality. While we believe these results are known to the community, we have not seen explicit proof of these relations elsewhere. We use the results of Ref.~\cite{reichardt2009}, which provides a sufficient condition of universality in terms of the Bloch vector of single-qubit states supplied to otherwise all-Clifford circuits. We then use Refs.~\cite{howard2017,seddon2021} to relate this condition to the robustness of magic.
\subsection{Fidelity to T state}
We first begin by explicitly deriving the threshold given in Theorem 1 of Ref.~\cite{reichardt2009}. Two arbitrary DV states, $\hat \rho_1$ and $\hat \rho_2$ can be described in terms of their Bloch vectors $\mathbf a^{(1)}$ and $\mathbf a^{(2)}$. This allows us to evaluate the fidelity between these two states as
\begin{align}
    \label{eq:fidelity-bloch}
    F=\frac{1}{2}(1+\mathbf a^{(1)} \cdot \mathbf a^{(2)}).
\end{align}
We can choose $\mathbf a^{(1)}=\mathbf a_T$ to be the Bloch vector of the $T$ state, i.e.,
\begin{align}
    \mathbf a_T=\frac{1}{\sqrt 3}\mqty(1\\1\\1).
\end{align}
Note also that arbitrary reflections in each axis correspond to different types of $T$ states given by
\begin{align}
    \mathbf a_T^{\pm \pm \pm}=\frac{1}{\sqrt 3}\mqty(\pm 1\\\pm 1\\\pm 1).
\end{align}
The fidelity to the closest $T$ state can be evaluated by choosing different combinations of $\pm\pm\pm$ such that the fidelity is maximized. This fidelity to any choice of $T$ state is given by
\begin{align}
    F_T^{\pm\pm\pm}(\hat \rho_\mathbf a)=\frac{1}{2}+\frac{1}{2\sqrt 3}\mqty(\pm 1\\\pm 1\\\pm 1) \cdot \mathbf a=\frac{1}{2}+\frac{1}{2\sqrt 3}(\pm a_1\pm a_2\pm a_3),
\end{align}
where the values of $a_1,a_2,a_3$ can be positive or negative, but when maximizing the fidelity we pick the one such that each coefficient becomes positive. This means that the fidelity to the closest $T$ state is given by
\begin{align}
    F_T^{\text{max}}(\hat \rho_\mathbf a)=\max_{\pm\pm\pm} F_T^{\pm\pm\pm}(\hat \rho_\mathbf a)=\frac{1}{2}+\frac{1}{2\sqrt 3}( |a_1|+|a_2|+|a_3|),
\end{align}
which can equivalently be written as
\begin{align}
   F_T^{\text{max}}(\hat \rho_\mathbf a)=\frac{1}{2}+\frac{1}{2\sqrt 3} ||\mathbf a||_1.
\end{align}
The condition for the state to be above the threshold is given by
\begin{align}
    F_T^{\text{max}}(\hat \rho_\mathbf a) > \frac{1}{2}\left(1+\frac{\sqrt 3}{\sqrt 7}\right),
\end{align}
i.e.,
\begin{align}
    &\frac{1}{2\sqrt{3}} ||\mathbf a||_1 > \frac{\sqrt 3}{2\sqrt{7}}\\
    \implies &  ||\mathbf a||_1 > \frac{3}{\sqrt{7}}.
\end{align}
Hence, we recover the threshold in terms of the norm of the Bloch vector, as given in Theorem 1 of Ref.~\cite{reichardt2009}.

The ROM for non-stabilizer states is simply defined as~\cite{howard2017}
\begin{align}
    \mathcal R(\hat \rho)=&|\langle X\rangle|+|\langle Y\rangle|+|\langle Z\rangle|\\
    =&|\Tr(\hat \rho X)|+|\Tr(\hat \rho Y)|+|\Tr(\hat \rho Z)|\\
    =&|a_1|+|a_2|+|a_3|\\
    =&||\mathbf a||_1.
\end{align}

Clearly, if $\mathcal R(\hat \rho)>\frac{ 3}{\sqrt 7}$ we satisfy the condition for $T$-type magic state distillation.

\subsection{Fidelity to H state}
For completeness, we also now provide an explicit relation between the robustness of magic and the threshold of the fidelity to the $H$ state for magic state distillation. We begin with an explicit derivation of Theorem 2 of Ref.~\cite{reichardt2009}.
We now consider the fidelity of an arbitrary Bloch vector with the Bloch vector of the $H$ state. I.e., we choose $\mathbf a^{(1)}=\mathbf a_H$ in Eq.~(\ref{eq:fidelity-bloch}), where
\begin{align}
    \mathbf a_H=\frac{1}{\sqrt 2}\mqty(1\\1\\0),
\end{align}
which can also be transformed under single-qubit Clifford operations as
\begin{align}
    \mathbf a_H^{(\pm,\pm,3)}=\frac{1}{\sqrt 2}\mqty(\pm 1\\ \pm 1\\0)\\
    \mathbf a_H^{(\pm,\pm,2)}=\frac{1}{\sqrt 2}\mqty(\pm 1\\ 0\\\pm 1)\\
    \mathbf a_H^{(\pm,\pm,1)}=\frac{1}{\sqrt 2}\mqty(0\\ \pm 1\\ \pm 1).
\end{align}
The fidelity to an arbitrary state $\hat \rho$ with Bloch vector $\mathbf a$ is therefore given by
\begin{align}
    F^{(\pm,\pm,j)}_H=\frac{1}{2}(1+\mathbf a \cdot \mathbf a_H^{(\pm,\pm,j)}),
\end{align}
which can be evaluated as
\begin{align}
    F^{(\pm,\pm,3)}_H(\hat \rho_\mathbf a)=\frac{1}{2}(1+\frac{1}{\sqrt 2}(\pm a_1\pm a_2))\\
    F^{(\pm,\pm,2)}_H(\hat \rho_\mathbf a)=\frac{1}{2}(1+\frac{1}{\sqrt 2}(\pm a_1\pm a_3))\\
    F^{(\pm,\pm,1)}_H(\hat \rho_\mathbf a)=\frac{1}{2}(1+\frac{1}{\sqrt 2}(\pm a_2\pm a_3)).
\end{align}
The maximum value of the fidelity to any $H$ state is thus given by
\begin{align}
     F_H^{\text{max}}(\hat \rho_\mathbf a)=\max_{\pm\pm\pm}F^{\pm\pm\pm}_H( \rho_\mathbf a)=\frac{1}{2}\left(1+\frac{1}{\sqrt 2}\max(|a_1|+|a_2|,|a_2|+|a_3|,|a_1|+|a_3|)\right).
\end{align}
Given that the distillation threshold for the $H$ state is tight~\cite{reichardt2005}, we know that
\begin{align}
    F_H^{\text{max}}(\hat \rho_\mathbf a)>F^*=\frac{1}{2}(1+\frac{1}{\sqrt 2})
\end{align}
meaning the threshold for $H$ state distillation can be expressed as~\cite{reichardt2005}
\begin{align}\max(|a_1|+|a_2|,|a_2|+|a_3|,|a_1|+|a_3|)>1.
\end{align}

Unlike the case of the $T$ state, the robustness of magic is not directly related to this quantity. Instead, we must consider the minimum robustness of magic of an arbitrary state required to satisfy this inequality.

Formally, we need to identify $\mathcal R_H^*$ such that
\begin{align}
|a_1|+|a_2|+|a_3| > \mathcal R_H^* \implies 
\max(|a_1|+|a_2|,|a_2|+|a_3|,|a_1|+|a_3|)>1.
\end{align}
The best possible bound can be found by identifying when $|a_1|+|a_2|=|a_2|+|a_3|=|a_1|+|a_3|=1$, which implies that all $|a_1|=|a_2|=|a_3|=1/2$. Therefore, $\mathcal R_H^*=\frac{3}{2}$. This is significantly higher than the bound found in terms of the fidelity to the nearest $T$ state. Also note, that, unlike the previous bound, this does not identify all qubits states that have a value of fidelity to the closest $H$ state above the distillation threshold.
\section{Stabilizer subsystem decomposition}
\label{sec:appendix-stab}
 In this appendix, we  formalize the definition of the stabilizer SSD given in Eq.~(\ref{eq:def-stab-ssd}). 
 Note that the result of the stabilizer SSD is identical to that of Ref.~\cite{shaw2022} whereby it is defined in terms of the Zak basis. In Ref.~\cite{shaw2022}, it was also stated that the stabilizer SSD is equivalent to GKP error correction. However, here we provide the formal definition of the stabilizer SSD in the context of GKP error correction. We also provide details on calculating the density matrix of a qubit state after this mapping, from the density matrix in the position basis of a CV state. We also provide a circuit diagram illustrating the procedure to implement this mapping.
 
We first formalize the definition of the stabilizer SSD by inspecting Eq.~(\ref{eq:def-stab-ssd}), which maps the CV state $\hat \rho$ to a qubit state $\hat \rho_{\Pi}$.
Note, however, that the right-hand side is a CV state and hence, we should expect a CV state on the left-hand side. Formally, we can resolve this by defining the CV state after the transformation as
\begin{align}
    \label{eq:cv-stab-ssd}
    \hat \rho_\Pi^{\text{CV}} =& \frac{1}{\sqrt\pi} \int^{\sqrt\pi/2}_{-\sqrt\pi/2} \dd t_q\int^{\sqrt\pi/2}_{-\sqrt\pi/2} \dd t_p\hat \Pi \hat V(-\mathbf t)\hat \rho V^\dagger(-\mathbf t)\hat \Pi
\end{align}
which only has support on the GKP basis. We can therefore transform the basis of $\hat \rho_\Pi^{\text{CV}}$ from $\ket{j_{\text{GKP}}}\to\ket{j}$. We can express the qubit density matrix as
\begin{align}
     \hat{\rho}_\Pi= \left(\ket{0}\bra{0_{\text{GKP}}}+\ket{1}\bra{1_{\text{GKP}}}\right)\hat \rho_\Pi^{\text{CV}}\left(\ket{0_{\text{GKP}}}\bra{0}+\ket{1_{\text{GKP}}}\bra{1}\right).
\end{align}
Given this definition of the stabilizer SSD, in the following subsection we identify a general method to calculate the stabilizer SSD of a CV mode, from the density matrix of the mode in the position basis.

\subsection{Position basis representation of the stabilizer SSD}
\label{sec:appendix-stab-pos}
We now express the stabilizer SSD for a single mode in the position basis. We begin by  writing a general expression for each of the four elements of the stabilizer SSD, before simplifying the expression for each term. We see from Eq.~(\ref{eq:cv-stab-ssd}) that the general density matrix element of the qubit resulting from the stabilizer SSD is given by
\begin{align}
    \bra{l}\hat \rho_\Pi\ket{l'} 
     =&\frac{1}{\sqrt \pi} \int^{\sqrt\pi/2}_{-\sqrt\pi/2} \dd t_q\int^{\sqrt\pi/2}_{-\sqrt\pi/2}  \dd t_p  \bra{l_{\text{GKP}}} e^{-it_p\hat q}e^{it_q\hat p}\hat \rho e^{-it_q\hat p}e^{it_p\hat q}\ket{l'_{\text{GKP}}} \nonumber\\
     =& \frac{1}{\sqrt\pi}\int^{\sqrt\pi/2}_{-\sqrt\pi/2} \dd t_q\int^{\sqrt\pi/2}_{-\sqrt\pi/2} \dd t_p  \sum_{n,n'}\bra{\hat q=(2n+l)\sqrt\pi} e^{-it_p(2n+l)\sqrt\pi}e^{it_q\hat p}\hat \rho e^{-it_q\hat p}e^{it_p(2n'+l')\sqrt\pi}\ket{\hat q=(2n'+l')\sqrt\pi} \nonumber\\
     =&\frac{1}{\sqrt\pi} \int^{\sqrt\pi/2}_{-\sqrt\pi/2} \dd t_q\int^{\sqrt\pi/2}_{-\sqrt\pi/2} \dd t_p  \sum_{n,n'}e^{-it_p(2n+l)\sqrt\pi}e^{it_p(2n'+l')\sqrt\pi}\bra{\hat q=(2n+l)\sqrt\pi+t_q} \hat \rho \ket{\hat q=(2n'+l')\sqrt\pi+t_q}.
\end{align}
We can evaluate each term of the qubit density matrix individually. First, we integrate over $t_p$, whereby we use that 
\begin{align}
    \int^{\sqrt\pi/2}_{-\sqrt\pi/2}\dd x e^{-ix s\sqrt\pi}=&\int^{\sqrt\pi/2}_{-\sqrt\pi/2}\dd x \cos(-x s\sqrt\pi)+i\sin(-x s\sqrt\pi)\nonumber\\
    =&\left[\frac{\sin(-s\sqrt\pi)}{-s\sqrt\pi}\right]_{-\sqrt\pi/2}^{\sqrt\pi/2} \nonumber\\
    =&\frac{2\sin(s\pi/2)}{s\sqrt\pi},
    \label{eq:integration-over-period}
\end{align}
which will be zero for any even integer $s$, unless $s\to 0$, at which point it will approach $1/\sqrt\pi$~\cite{Mathematica}. 
Therefore, the integral over $t_p$ can be evaluated as
\begin{align}
    \bra{l}\hat \rho_\Pi\ket{l}=&\frac{1}{\sqrt\pi} \int^{\sqrt\pi/2}_{-\sqrt\pi/2} \dd t_q  \sum_{n,n'}\left(\int^{\sqrt\pi/2}_{-\sqrt\pi/2} \dd t_p e^{-it_p(2n-2n')\sqrt\pi}\right)\bra{\hat q=(2n+l)\sqrt\pi+t_q} \hat \rho \ket{\hat q=(2n'+l')\sqrt\pi+t_q}  \nonumber\\
     =& \int^{\sqrt\pi/2}_{-\sqrt\pi/2} \dd t_q  \sum_{n,n'}\delta_{n,n'}\bra{\hat q=(2n+l)\sqrt\pi+t_q} \hat \rho \ket{\hat q=(2n'+l)\sqrt\pi+t_q} \nonumber\\
     =& \int^{\sqrt\pi/2}_{-\sqrt\pi/2} \dd t_q  \sum_{n}\bra{\hat q=(2n+l)\sqrt\pi+t_q} \hat \rho \ket{\hat q=(2n+l)\sqrt\pi+t_q}.
    \label{eq:stab-ssd-diagonal}
\end{align}
Now we evaluate the two off-diagonal terms. These two terms are equal up to conjugation and, therefore, it suffices to identify the value of a single off-diagonal term. We calculate the term $\bra{0}\hat \rho_\Pi\ket{1}$ by integrating over $t_p$, and using Eq.~(\ref{eq:integration-over-period}) to find
\begin{align}
    \int^{\sqrt\pi/2}_{-\sqrt\pi/2} \dd t_p e^{-it_p(2n-2n'-1)\sqrt\pi}=& \frac{2\sin((2n-2n'-1)\pi/2)}{(2n-2n'-1)\sqrt\pi}= -\frac{2\cos((n-n')\pi)}{(2n-2n'-1)\sqrt\pi}= (-1)^{n-n'}\frac{2}{(1-2n+2n')\sqrt\pi},
\end{align}
such that
\begin{align}
    \bra{0}\hat \rho_\Pi\ket{1} 
     =&\frac{1}{\sqrt\pi} \int^{\sqrt\pi/2}_{-\sqrt\pi/2} \dd t_q \sum_{n,n'}(-1)^{n-n'}\frac{2}{(1-2n+2n')\sqrt\pi}\bra{\hat q=2n\sqrt\pi+t_q} \hat\rho \ket{\hat q=(2n'+1)\sqrt\pi+t_q}.
\end{align}
This can be simplified further by making the substitution $n \to n+n'$,
\begin{align}
    \label{eq:stab-ssd-offdiagonal}
    \bra{0}\hat \rho_\Pi\ket{1}
    =& \frac{1}{\pi}\int^{\sqrt\pi/2}_{-\sqrt\pi/2} \dd t_q  \sum_{n,n'}(-1)^{n}\frac{2}{(1-2n)}\bra{\hat q=2(n+n')\sqrt\pi+t_q} \hat\rho \ket{\hat q=(2n'+1)\sqrt\pi+t_q}.
\end{align}

Together, Eq.~(\ref{eq:stab-ssd-diagonal}) and Eq.~(\ref{eq:stab-ssd-offdiagonal}), provide a method to evaluate the stabilizer SSD directly from the density matrix of any single-mode CV state.

\subsection{Circuit implementation of the stabilizer SSD}
\label{sec:appendix-stab-circuit}
We now demonstrate that the stabilizer SSD~\cite{shaw2022} is equivalent to performing GKP error correction~\cite{gottesman2001}, whereby the measurement results are discarded. Although this result was previously shown in Ref.~\cite{shaw2022}, we here give the exact details of the procedure in terms of the error correction circuit and using only the definition of the stabilizer SSD in the position basis.

Error correction according to the GKP protocol~\cite{gottesman2001} is performed on a mode by measuring the momentum stabilizer $e^{2i\sqrt\pi \hat q}$ and the position stabilizer $e^{2i\sqrt\pi \hat p}$, and then shifting the mode depending on the phase of these measurement outcomes. 
Measurement of a stabilizer in the GKP code consists of a homodyne measurement of a coupled ideal GKP state.

If we measure a value of $t_q$ in the momentum stabilizer measurement and a measurement of $t_p$ in the position stabilizer measurement, we then shift the mode by the measured value modulo $\sqrt\pi$, where the modulus is taken over the interval $(-\sqrt\pi/2,\sqrt\pi/2]$. We can express the measurement results using the same notation for modular variables as provided in Sec.\ref{sec:background-modular-ssd} (and that of Ref.~\cite{pantaleoni2021}) as $t=\cfloor{t}_{\sqrt\pi}+\{t\}_{\sqrt\pi}$, where, as a reminder, $\cfloor{t}_{\sqrt\pi} =\sqrt\pi \lfloor \tfrac{t}{\sqrt\pi}-\tfrac 1 2\rfloor$ is the centered floor function and $\{t\}_{\sqrt\pi}=t_q-\cfloor{t}_{\sqrt\pi}$ is the remainder. Note that the remainder can equivalently be expressed as $\{t\}_{\sqrt\pi}=t \mod \sqrt\pi$, where the modulus is taken over the interval $(-\sqrt\pi/2,\sqrt\pi/2]$. Note that $\cfloor{t}_{\sqrt\pi}$ is an integer multiple of $\sqrt\pi$, while $\{t\}_{\sqrt\pi}$ is a real number on the interval $(-\sqrt\pi/2,\sqrt\pi/2]$. A circuit diagram implementing GKP error correction is provided in Fig.~\ref{fig:gkp-ec-mod}.

\begin{figure}[ht]
        \centering
        $$
        \Qcircuit @C=0.9em @R=.7em {
        \lstick{\hat\rho}              & \qw & \ctrl{1}    & \qw               &  \gate{\hat X(-\{t_q\}_{\sqrt\pi})}      & \qw           & \targMinus & \qw               & \qw                      &  \gate{\hat Z(-\{t_p\}_{\sqrt\pi})}   & \qw  & \qw \\
        \lstick{\ket{0_{\text{GKP}}}}   & \qw & \control\qw & \measureD{\hat p} & \control \cw \cwx   & \lstick{t_q} \\
        \lstick{\ket{0_{\text{GKP}}}}   & \qw & \qw         &\qw                &\qw  & \qw                 & \ctrl{-2}\qw  & \qw   & \measureD{\hat p} & \control \cw \cwx[-2]    & \lstick{t_p} 
        }
        $$
         \caption{GKP error correction as a circuit. We draw the operation $e^{i\hat q_3\hat p_1}$ using the symbol $\ominus$, which can be considered the inverse of the SUM gate~\cite{noh2022}.}
         \label{fig:gkp-ec-mod}
    \end{figure}
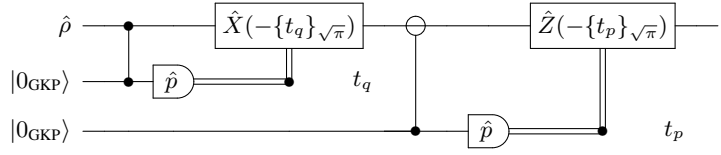
We can analyze the action of this circuit in terms of two Kraus operators: one which implements the position error correction $\hat K^q$ and one which implements the momentum error correction $\hat K^p$. These can be expressed as
\begin{align}
    \label{eq:kraus-ops}
    \hat{K}^p(t_q)=&e^{i\hat p_1 \{t_q\}_{\sqrt\pi}}\bra{\hat p_2=t_q}e^{i\hat q_1\hat q_2}\ket{0_{\text{GKP}}}_2, \nonumber \\
    \hat{K}^q(t_p)
    =&e^{-i\hat q_1 \{t_p\}_{\sqrt\pi}}\bra{\hat p_3=t_p}e^{i\hat q_3\hat p_1}\ket{0_{\text{GKP}}}_3.
\end{align}

We now demonstrate that the combined action of these Kraus operators is to implement the operation $\hat \Pi \hat V(-\mathbf t)$, which is present in the expression for the stabilizer SSD given in Eq.~(\ref{eq:cv-stab-ssd}). We do so by first simplifying the expressions of the two Kraus operators and we then demonstrate that combined, they give the desired action.

We begin by evaluating the first operator by inserting the wavefunction of the $0$-logical GKP state
\begin{align}
    \hat{K}^p(t_q)=&e^{i\hat p_1 \{t_q\}_{\sqrt\pi}}\bra{\hat p_2=t_q}e^{i\hat q_1\hat q_2}\sum_n\ket{\hat q_2=2n\sqrt\pi}\nonumber\\
    =&e^{i\hat p_1 \{t_q\}_{\sqrt\pi}}\bra{\hat p_2=t_q}\sum_ne^{2in\sqrt\pi\hat q_1}\ket{\hat q_2=2n\sqrt\pi}\nonumber\\
    \propto&e^{i\hat p_1 \{t_q\}_{\sqrt\pi}}\sum_ne^{2in\sqrt\pi\hat q_1}e^{-i2n\sqrt\pi t_q}\nonumber\\
    =&e^{i\hat p \{t_q\}_{\sqrt\pi}}\sum_ne^{2in\sqrt\pi(\hat q-t_q)},
\end{align}
where we have dropped the index in the last line as the effect of the Kraus operator only applies to the mode being error corrected.
Note that we can also use the fact that the wavefunction of the $0$-logical state in the momentum basis $\tilde \psi(x)$ can be expressed as~\cite{baragiola2019}
\begin{align}
    \tilde \psi_{0,L}(x) = \sum_n e^{2in\sqrt\pi x}=\sum_n \delta(x-n\sqrt\pi),
\end{align}
to simplify the expression further. We find that
    \begin{align}
    \hat{K}^p(t_q)
    \propto&e^{i\hat p \{t_q\}_{\sqrt\pi}}\tilde \psi_{0,L}(\hat q-t_q)\nonumber\\
    =&e^{i\hat p \{t_q\}_{\sqrt\pi}}e^{-i t_q \hat p}\tilde \psi_{0,L}(\hat q)e^{i t_q \hat p}\nonumber\\
    =&e^{-i\hat p(t_q- \{t_q\}_{\sqrt\pi})}\tilde \psi_{0,L}(\hat q)e^{i(\cfloor{t_q}_{\sqrt\pi}+\{t_q\}_{\sqrt\pi}) \hat p}\nonumber\\
    =&e^{-i\hat p\cfloor{t_q}_{\sqrt\pi}}\tilde \psi_{0,L}(\hat q)e^{i(\cfloor{t_q}_{\sqrt\pi}+\{t_q\}_{\sqrt\pi}) \hat p}\nonumber\\
    =&\tilde \psi_{0,L}(\hat q-\cfloor{t_q}_{\sqrt\pi})e^{i\{t_q\}_{\sqrt\pi} \hat p}\nonumber\\
    =&\tilde \psi_{0,L}(\hat q)e^{i\{t_q\}_{\sqrt\pi} \hat p}
\end{align}
where in the final line we have used the fact that $\cfloor{t_q}_{\sqrt\pi}$ is an integer multiple of $\sqrt\pi$ and the wavefunction in the momentum basis is periodic in $\sqrt\pi$.
The second Kraus operator also simplifies using the same methods,
\begin{align}
    \hat{K}^q(t_p)
    =&e^{-i\hat q_1 \{t_p\}_{\sqrt\pi}}\bra{\hat p_3=t_p}e^{i\hat q_3\hat p_1}\ket{0_{\text{GKP}}}_3\nonumber\\
    =&e^{-i\hat q_1 \{t_p\}_{\sqrt\pi}}\bra{\hat p_3=t_p}\sum_ne^{2in\sqrt\pi \hat p_1}\ket{\hat q_3=2n\sqrt\pi}\nonumber\\
    \propto&e^{-i\hat q \{t_p\}_{\sqrt\pi}}\sum_ne^{2in\sqrt\pi \hat p}e^{-i2nt_p\sqrt\pi}\nonumber\\
    =&e^{-i\hat q \{t_p\}_{\sqrt\pi}}\tilde\psi_{0,L}(\hat p-t_p)\nonumber\\
    =&e^{-i\hat q \{t_p\}_{\sqrt\pi}}\tilde\psi_{0,L}(\hat p-t_p)e^{i\hat q \{t_p\}_{\sqrt\pi}}e^{-i\hat q \{t_p\}_{\sqrt\pi}}\nonumber\\
    =&\tilde\psi_{0,L}(\hat p-t_p+\{t_p\}_{\sqrt\pi})e^{-i\hat q \{t_p\}_{\sqrt\pi}}\nonumber\\
    =&\tilde\psi_{0,L}(\hat p-\cfloor{t_p}_{\sqrt\pi})e^{-i\hat q \{t_p\}_{\sqrt\pi}}\nonumber\\
    =&\tilde\psi_{0,L}(\hat p)e^{-i\hat q \{t_p\}_{\sqrt\pi}}.
\end{align}
Combining these two operators allows us to find an expression for the combined Kraus operator as
\begin{align}
    \hat K(\mathbf t)=&\tilde \psi_{0,L}(\hat p)e^{-i\{t_p\}_{\sqrt\pi}\hat q}\tilde \psi_{0,L}(\hat q)e^{i\{t_q\}_{\sqrt\pi} \hat p}\nonumber\\
    =&\tilde \psi_{0,L}(\hat p)\tilde \psi_{0,L}(\hat q)e^{-i\{t_p\}_{\sqrt\pi}\hat q}e^{i\{t_q\}_{\sqrt\pi} \hat p}\nonumber\\
    =&\hat \Pi \hat V(-\{\mathbf t\}_{\sqrt\pi}),
\end{align}
where we have used that the GKP projector $\hat \Pi$, defined in Eq.~(\ref{eq:gkp-projector}), is equivalent to~\cite{baragiola2019}
\begin{align}
    \tilde \psi_{0,L}(\hat p)\tilde \psi_{0,L}(\hat q)
    =& \sum_{n,n'}\left( \ket{\hat p=n\sqrt\pi}\bra{\hat p=n\sqrt\pi}\right)\left( \ket{\hat q=n'\sqrt\pi}\bra{\hat q=n'\sqrt\pi}\right)\nonumber\\
    \propto& \sum_{n,n'} \ket{\hat p=n\sqrt\pi}e^{-inn'\pi}\bra{\hat q=n'\sqrt\pi}\nonumber\\
    =& \sum_{n,n'} \ket{\hat p=n\sqrt\pi}e^{-2inn'\pi}\bra{\hat q=2n'\sqrt\pi}+ \ket{\hat p=n\sqrt\pi}e^{-in(2n'+1)\pi}\bra{\hat q=(2n'+1)\sqrt\pi}\nonumber\\
    =& \sum_{n,n'} \ket{\hat p=n\sqrt\pi}\bra{\hat q=2n'\sqrt\pi}+ (-1)^n\ket{\hat p=n\sqrt\pi}\bra{\hat q=(2n'+1)\sqrt\pi}\nonumber\\
    =& \ket{0_{\text{GKP}}}\bra{0_{\text{GKP}}}+\ket{1_{\text{GKP}}}\bra{1_{\text{GKP}}}\nonumber\\
    =&\hat \Pi,
\end{align}
and the displacement operator implements a displacement whereby both elements of the vector are taken modulo $\sqrt\pi$ on the interval $(-\sqrt\pi/2,\sqrt\pi/2]$, i.e., $\{\mathbf t\}_{\sqrt\pi}=(\{t_q\}_{\sqrt\pi},\{t_p\}_{\sqrt\pi})$.

The statistical mixture of the output state after a round of GKP error correction, whereby the measured values are ignored, can be evaluated as
\begin{align}
    \hat \rho^{\text{CV}}_{\Pi} \propto \int_{\mathbb R^2} \dd \mathbf t \hat K(\mathbf t)\hat \rho \hat K^\dagger(\mathbf t).
\end{align}
However, due to the fact that the Kraus operator is periodic in both elements of $\mathbf t$ with a period of $\sqrt\pi$, centered around the origin, we can evaluate the output state by integrating over a single period,
\begin{align}
    \hat\rho^{\text{CV}}_{\Pi}\propto \int^{\sqrt\pi/2}_{-\sqrt\pi/2} \dd t_q \int^{\sqrt\pi/2}_{-\sqrt\pi/2} \dd t_p  \hat K^\dagger (\mathbf t)\hat \rho \hat K(\mathbf t).
\end{align}
This also allows us to simplify the expression of the Kraus operator to $\hat K(\mathbf t)=\hat \Pi\hat V(-\mathbf t)$, where $\hat K$ is now only defined over this interval. This is precisely the same expression (up to normalization) as the density matrix we identified at the beginning of this appendix, in Eq.~(\ref{eq:cv-stab-ssd}). 

\section{Modular subsystem decomposition}
\label{sec:appendix-mod}
In this appendix, we demonstrate similar properties for the modular SSD. We begin by writing the full expression for the modular SSD, as given in Eq.~(\ref{eq:definition-mssd}) and show how to calculate the modular SSD of an arbitrary CV state from its density matrix in the position basis. We  then demonstrate that this decomposition can be implemented using a CV circuit. Note, however, that this implementation includes non-Gaussian operations, in addition to the non-Gaussian GKP states. Following this, we also express the modular SSD in terms of the stabilizer SSD, to highlight their connection.

\subsection{Position basis representation of the modular SSD}
\label{sec:appendix-mod-pos}
In this subsection, we provide an expression to evaluate the modular SSD of a CV state in the position basis. To do so, we first recall the definition of the modular SSD. Using the identity defined over modular variables, given in Eq.~(\ref{eq:identity-mod-vars}), we can express any CV state as
\begin{align}
    \mathbbm 1_{\text{CV}}\hat \rho \mathbbm 1_{\text{CV}}=&\sum_{l,l'} \sum_{m_{\mathcal G},m_{\mathcal G}'}\int \dd u_{\mathcal G} \dd u_{\mathcal G}'  \ket{l,m_{\mathcal G},u_{\mathcal G}}\bra{l,m_{\mathcal G},u_{\mathcal G}}\hat \rho   \ket{l',m_{\mathcal G}',u_{\mathcal G}'}\bra{l',m_{\mathcal G}',u_{\mathcal G}'}.
\end{align}
The modular SSD of any CV state can then be calculated, according to Eq.~(\ref{eq:definition-mssd}), by tracing out the Gauge part of the state, i.e.,~\cite{pantaleoni2020}
\begin{align}
    \label{eq:ssd-logical}
    \hat \rho_L=\Tr_{\mathcal G}[\hat \rho]=&\sum_{m_{\mathcal G}\in \mathbb Z}\int^{\alpha/2}_{-\alpha/2} \dd u_{\mathcal G} \prescript{}{\mathcal G}{\bra{m_{\mathcal G},u_{\mathcal G}}}\hat \rho \ket{m_{\mathcal G},u_{\mathcal G}}_{\mathcal G} \nonumber\\
    =&\sum_{l,l'}\sum_{m_{\mathcal G}\in \mathbb Z}\int^{\alpha/2}_{-\alpha/2} \dd u_{\mathcal G} \ket{l}\bra{l,m_{\mathcal G},u_{\mathcal G}}\hat \rho   \ket{l',m_{\mathcal G},u_{\mathcal G}}\bra{l'}.
\end{align}
We can convert this expression to the position basis by using that $\ket{l}_L\ket{m_{\mathcal G},u_{\mathcal G}}_{\mathcal G}=\ket{\hat q=\alpha l+d\alpha m_{\mathcal G}+u_{\mathcal G}}$~\cite{pantaleoni2020}, such that the modular SSD can be expressed as
\begin{align}
    \label{eq:ssd-logical-basis}
    \hat \rho_L=\Tr_{\mathcal G}[\hat \rho]
    =&\sum_{l,l'}\ket{l}\sum_{m_{\mathcal G}\in \mathbb Z}\int^{\alpha/2}_{-\alpha/2} \dd u_{\mathcal G} \bra{\hat q=\alpha l+d\alpha m_{\mathcal G}+u_{\mathcal G}}\hat \rho\ket{\hat q=\alpha l'+d\alpha m_{\mathcal G}+u_{\mathcal G}}\bra{l'}.
\end{align}

If we take the GKP peak separation $\alpha=\sqrt\pi$ and $d=2$ we can evaluate the elements of the density matrix of the resulting logical qubit as
\begin{align}
    \label{eq:ssd-logical-ll-specific}
    \bra{l}\hat \rho_L\ket{l'}
    =&\sum_{m_{\mathcal G}\in \mathbb Z}\int^{\sqrt\pi/2}_{-\sqrt\pi/2} \dd u_{\mathcal G} \bra{\hat q=\sqrt\pi l+2\sqrt\pi m_{\mathcal G}+u_{\mathcal G}}\hat \rho\ket{\hat q=\sqrt\pi l'+2\sqrt\pi m_{\mathcal G}+u_{\mathcal G}}.
\end{align}
By changing the summation and integration variable labels, $u_{\mathcal G}\to t_q$ and $m_{\mathcal G} \to n$, we can equivalently express this as
\begin{align}
    \label{eq:ssd-logical-ll-tq}
    \bra{l}\hat \rho_L\ket{l'}
    =&\sum_{n\in \mathbb Z}\int^{\sqrt\pi/2}_{-\sqrt\pi/2} \dd t_q \bra{\hat q=2\sqrt\pi n+\sqrt\pi l+t_q}\hat \rho\ket{\hat q=2\sqrt\pi n+\sqrt\pi l'+t_q}.
\end{align}

Note that for the diagonal elements $l=l'$ of the resulting modular SSD state, this corresponds with the stabilizer SSD, i.e.,
\begin{align}
    \label{eq:ssd-logical-ll-equal}
    \bra{l}\hat \rho_L\ket{l}
    =&\bra{l}\hat \rho_\Pi\ket{l}.
\end{align}

\subsection{Circuit implementation of the modular SSD}
\label{sec:appendix-mod-circuit}
In this subsection, we demonstrate that the modular SSD can be implemented as a circuit involving GKP stabilizer states and non-Gaussian operations.
By doing so, we also demonstrate that the modular SSD can be interpreted as the average of the GKP error correction map following a logical $\hat Z$ rotation. While the connection has been previously explored in Ref.~\cite{shaw2022} and Ref.~\cite{pantaleoni2022}, we here directly derive the relationship in terms of the position basis representation of the input CV state.

We begin by rewriting the expression for the elements of the density matrix after the modular SSD, given in Eq.~(\ref{eq:ssd-logical-ll-tq}), as
\begin{align}
    \label{eq:ssd-logical-ll-specific-repeated}
    \bra{l}\hat \rho_L\ket{l'}
    =&\sum_{n\in \mathbb Z}\int^{\sqrt\pi/2}_{-\sqrt\pi/2} \dd t_q \bra{\hat q=2\sqrt\pi n+\sqrt\pi l+t_q}\hat \rho\ket{\hat q=2\sqrt\pi n+\sqrt\pi l'+t_q}\\
    =&\sum_{n\in \mathbb Z}\int^{\sqrt\pi/2}_{-\sqrt\pi/2} \dd t_q \bra{\hat q=(2n+l)\sqrt\pi}e^{i\hat p t_q}\hat \rho e^{-i\hat p t_q}\ket{\hat q=(2n+l')\sqrt\pi}\\
    \propto&\sum_{n,n'\in \mathbb Z}\int^{\sqrt\pi/2}_{-\sqrt\pi/2} \dd t_q \delta(2\sqrt\pi n-2\sqrt\pi n') \bra{\hat q=(2n+l)\sqrt\pi}e^{i\hat p t_q}\hat \rho e^{-i\hat p t_q}\ket{\hat q=(2n'+l')\sqrt\pi}\\
    \propto&\sum_{n,n'\in \mathbb Z}\int^{\infty}_{-\infty} \dd s \int^{\sqrt\pi/2}_{-\sqrt\pi/2} \dd t_q e^{-i2\sqrt\pi s (n-n')} \bra{\hat q=(2n+l)\sqrt\pi}e^{i\hat p t_q}\hat \rho e^{-i\hat p t_q}\ket{\hat q=(2n'+l')\sqrt\pi }\\
    \propto&\sum_{n,n'\in \mathbb Z}\int^{\infty}_{-\infty} \dd s \int^{\sqrt\pi/2}_{-\sqrt\pi/2} \dd t_q \bra{\hat q=(2n+l)\sqrt\pi}e^{-i2\sqrt\pi s n} e^{i\hat p t_q}\hat \rho e^{-i\hat p t_q}e^{i2\sqrt\pi s n'} \ket{\hat q=(2n'+l')\sqrt\pi}.
\end{align}
Note that we have dropped normalization constants in this expression, however, this expression will preserve the relative values of $\bra{l}\hat \rho_L\ket{l'}$ and normalization can be restored by normalizing the density matrix at the end of the calculation.
Next, we identify a period in the integrand of this expression, by shifting the value of $s\to s+\sqrt\pi$ to find
\begin{align}
    &\sum_{n,n'\in \mathbb Z}\int^{\infty}_{-\infty} \dd s \int^{\sqrt\pi/2}_{-\sqrt\pi/2} \dd t_q \bra{\hat q=(2n+l)\sqrt\pi}e^{-i2\sqrt\pi (s+\sqrt\pi) n} e^{i\hat p t_q}\hat \rho e^{-i\hat p t_q}e^{i2\sqrt\pi (s+\sqrt\pi) n'} \ket{\hat q=(2n'+l')\sqrt\pi}\\
    =&\sum_{n,n'\in \mathbb Z}\int^{\infty}_{-\infty} \dd s \int^{\sqrt\pi/2}_{-\sqrt\pi/2} \dd t_q \bra{\hat q=(2n+l)\sqrt\pi}e^{-i2\sqrt\pi s n} e^{-i2\pi n} e^{i\hat p t_q}\hat \rho e^{-i\hat p t_q}e^{i2\sqrt\pi s n'} e^{i2\pi  n'} \ket{\hat q=(2n'+l')\sqrt\pi}.
\end{align}
Given that $n,n'$ are both integers, we see that $e^{2\pi i n'}=e^{-2\pi i n}=1$, which means that the integrand is periodic in $s$ with period $\sqrt\pi$. Hence, we can integrate over only this period, which allows us to express the elements of the qubit density matrix as
\begin{align}
    \bra{l}\hat \rho_L\ket{l'}
    \propto&\sum_{n,n'\in \mathbb Z}\int^{\sqrt\pi/2}_{-\sqrt\pi/2} \dd s \int^{\sqrt\pi/2}_{-\sqrt\pi/2} \dd t_q \bra{\hat q=(2n+l)\sqrt\pi}e^{-i2\sqrt\pi s n} e^{i\hat p t_q}\hat \rho e^{-i\hat p t_q}e^{i2\sqrt\pi s n'} \ket{\hat q=(2n'+l')\sqrt\pi}.
\end{align}
We then rewrite the integration variable $s=t_p$ and rewrite the exponents as operators, to find
\begin{align}
    \bra{l}\hat \rho_L\ket{l'}
    \propto&\sum_{n,n'\in \mathbb Z}\int^{\sqrt\pi/2}_{-\sqrt\pi/2} \dd t_p \int^{\sqrt\pi/2}_{-\sqrt\pi/2} \dd t_q \bra{\hat q=(2n+l)\sqrt\pi}e^{-i(\hat q-l\sqrt\pi) t_p} e^{i\hat p t_q}\hat \rho e^{-i\hat p t_q}e^{i(\hat q-l'\sqrt\pi) t_p} \ket{\hat q=(2n'+l')\sqrt\pi}\\
    =&\int^{\sqrt\pi/2}_{-\sqrt\pi/2} \dd t_p \int^{\sqrt\pi/2}_{-\sqrt\pi/2} \dd t_q e^{i(l-l')\sqrt\pi t_p}\bra{l_{\text{GKP}}}e^{-i\hat q t_p} e^{i\hat p t_q}\hat \rho e^{-i\hat p t_q}e^{i\hat q t_p} \ket{l'_{\text{GKP}}}\\
    =&\int^{\sqrt\pi/2}_{-\sqrt\pi/2} \dd t_p \int^{\sqrt\pi/2}_{-\sqrt\pi/2} \dd t_q e^{i(l-l')\sqrt\pi t_p}\bra{l_{\text{GKP}}}\hat V (-\mathbf t)\hat \rho \hat V^\dagger (-\mathbf t)\ket{l'_{\text{GKP}}}.
\end{align}
Note that for $l=l'$ the elements are equal to the elements of the stabilizer SSD as in Eq.~(\ref{eq:ssd-logical-ll-equal}) and therefore, the state must also have the same normalization factor, i.e., 
\begin{align}
    \label{eq:normalized-mod-ssd}
    \bra{l}\hat \rho_L\ket{l'}
    =&\frac{1}{\sqrt\pi}\int^{\sqrt\pi/2}_{-\sqrt\pi/2} \dd t_p \int^{\sqrt\pi/2}_{-\sqrt\pi/2} \dd t_q e^{i(l-l')\sqrt\pi t_p}\bra{l_{\text{GKP}}}\hat \Pi \hat V (-\mathbf t)\hat \rho \hat V^\dagger (-\mathbf t)\hat \Pi \ket{l'_{\text{GKP}}}.
\end{align}

We now demonstrate how to convert an arbitrary CV state to the modular SSD of the state, encoded in the GKP basis.
The remaining exponent term in Eq.~(\ref{eq:normalized-mod-ssd}) can be considered to be a logical $\hat Z$ rotation~\cite{pantaleoni2022} whereby
\begin{align}
    \hat R_Z (\theta)=\ket{0_{\text{GKP}}}\bra{0_{\text{GKP}}}+e^{i\theta}\ket{1_{\text{GKP}}}\bra{1_{\text{GKP}}}.
\end{align}
We can therefore express a transformation of a CV state to a GKP-encoded qubit state following the modular SSD as

\begin{align}
    \label{eq:ssd-logical-projector}
    \hat \rho_L^{\text{CV}}
    =&\frac{1}{\sqrt\pi}\int^{\sqrt\pi/2}_{-\sqrt\pi/2} \dd t_p \int^{\sqrt\pi/2}_{-\sqrt\pi/2} \dd t_q  \hat R_Z (t_p\sqrt\pi)\hat \Pi \hat V(-\mathbf t) \hat \rho \hat V^\dagger(-\mathbf t)\hat \Pi \hat R_Z^\dagger (t_p\sqrt\pi)\nonumber\\
    =&\frac{1}{\sqrt\pi}\int^{\sqrt\pi}_{-\sqrt\pi} \dd t_p \int^{\sqrt\pi/2}_{-\sqrt\pi/2} \dd t_q  \hat R_Z (t_p\sqrt\pi) \hat \rho_\Pi(\mathbf t)\hat R_Z^\dagger (t_p\sqrt\pi),
\end{align}
where we have used Eq.~(\ref{eq:def-stab-ssd-t}) to express the state $\hat \rho$ after GKP error-correction as $\hat \rho_\Pi(\mathbf t)$.
This means that the modular SSD can be understood as performing GKP error correction followed by a logical $\hat Z$ rotation and an integration over all possible outcomes. We provide a circuit diagram of this circuit in Figure \ref{fig:ssd-circuit1}. Note that, as in the case of the stabilizer SSD, it is possible to convert the CV state, given in Eq.~(\ref{eq:ssd-logical-projector}), to a normalized qubit state by transforming the encoded basis to the qubit basis.

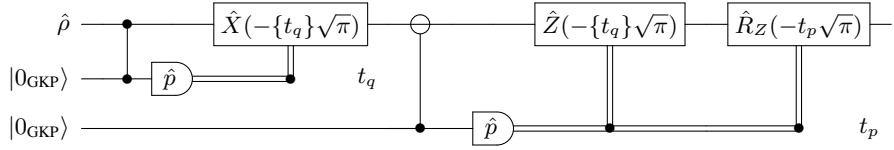
\begin{figure}[ht]
        \centering
        $$
        \Qcircuit @C=0.9em @R=.7em {
        \lstick{\hat \rho}              & \qw & \ctrl{1}    & \qw               &  \gate{\hat X(-\{t_q\}{\sqrt\pi} )}      & \qw           & \targMinus & \qw               & \qw                      &  \gate{\hat Z(-\{t_q\}{\sqrt\pi})}   & \qw & \gate{\hat R_Z(-t_p\sqrt\pi)} & \qw \\
        \lstick{\ket{0_{\text{GKP}}}}   & \qw & \control\qw & \measureD{\hat p} & \control \cw \cwx   & \lstick{t_q} \\
        \lstick{\ket{0_{\text{GKP}}}}   & \qw & \qw         &\qw                &\qw  & \qw                 & \ctrl{-2}\qw  & \qw   & \measureD{\hat p} & \control \cw \cwx[-2]& \cw& \control \cw \cwx[-2]    & \lstick{t_p} 
        }
        $$
         \caption{Modular subsystem decomposition as a circuit. We use the same notation for the inverse SUM gate as presented in Fig.~\ref{fig:gkp-ec-mod}.}
         \label{fig:ssd-circuit1}
    \end{figure}

\subsection{Modular SSD in terms of a Gaussian and non-Gaussian part}
\label{sec:appendix-mod-g-and-ng}
We  now demonstrate that the modular SSD can be expressed in terms of a summation of a part involving only Gaussian Kraus operators and a part involving non-Gaussian Kraus operators. Note that the state $\hat\rho$ may be non-Gaussian, and the GKP error correction routine also requires access to non-Gaussian GKP states.
 
Given that $R_z(\theta)=\cos(\theta/2)\mathbbm 1-i\sin(\theta/2)\hat Z$ we can write
\begin{align}
    \label{eq:ssd-logical-projector-gandng}
    \hat \rho_L^{\text{CV}}
    \propto&\int^{\sqrt\pi/2}_{-\sqrt\pi/2} \dd t_p \int^{\sqrt\pi/2}_{-\sqrt\pi/2} \dd t_q  \hat R_Z (t_p\sqrt\pi) \hat \rho_\Pi(\mathbf t)\hat R_Z^\dagger (t_p\sqrt\pi)\\
    =&\int^{\sqrt\pi/2}_{-\sqrt\pi/2} \dd t_p \int^{\sqrt\pi/2}_{-\sqrt\pi/2} \dd t_q  \left(\cos(t_p\sqrt\pi/2)\mathbbm 1-i\sin(t_p\sqrt\pi/2)\hat Z\right)\hat \rho_\Pi(\mathbf t)\left(\cos(t_p\sqrt\pi/2)\mathbbm 1+i\sin(t_p\sqrt\pi/2)\hat Z^\dagger\right)\\
    \propto&\hat \rho_L^{\text{CV},\text{G}}+\hat \rho_L^{\text{CV},\text{NG}}
\end{align}
whereby the Gaussian part --- i.e., consisting of only Gaussian Kraus operators --- is given by
\begin{align}
    \label{eq:gaussian-part}
    \hat \rho_L^{\text{CV},\text{G}}\propto&\int^{\sqrt\pi/2}_{-\sqrt\pi/2} \dd t_p \int^{\sqrt\pi/2}_{-\sqrt\pi/2} \dd t_q  \cos^2(t_p\sqrt\pi/2)\hat \rho_\Pi(\mathbf t)+\sin^2(t_p\sqrt\pi/2)\hat Z\hat \rho_\Pi(\mathbf t)\hat Z^\dagger,
\end{align}
while the non-Gaussian part is given by
\begin{align}
    \hat \rho_L^{\text{CV},\text{NG}}\propto&\int^{\sqrt\pi/2}_{-\sqrt\pi/2} \dd t_p \int^{\sqrt\pi/2}_{-\sqrt\pi/2} \dd t_q \frac{i}{2} \sin(t_p\sqrt\pi) \left(\hat \rho_\Pi(\mathbf t)\hat Z^\dagger-\hat Z\hat \rho_\Pi(\mathbf t)\right).
\end{align}

\section{Gaussian modular subsystem decomposition}
\label{sec:appendix-gauss}
We propose an alternative map that, for symmetric states, is equivalent to the modular SSD. The motivation for introducing such a map is that this map can be implemented in terms of  SGKP circuits, in contrast to modular SSD. The map is given in Eq.~(\ref{eq:gaussian-part}) as the Gaussian part of the modular SSD. The logical qubit density matrix of the state after Gaussian modular SSD can be expressed as
\begin{align}
    \label{eq:unnormalized-gaus-mod}
    \bra{l}\hat \rho_L^{\text{G}}\ket{l'}=&\frac{1}{\mathcal N_{\text{G}}}\int^{\sqrt\pi/2}_{-\sqrt\pi/2} \dd t_p \int^{\sqrt\pi/2}_{-\sqrt\pi/2} \dd t_q \bra{l'_{\text{GKP}}}\left( \cos^2(t_p\sqrt\pi/2)\hat \rho_\Pi(\mathbf t)+\sin^2(t_p\sqrt\pi/2)\hat Z\hat \rho_\Pi(\mathbf t)\hat Z^\dagger\right)\ket{l_{\text{GKP}}}.
\end{align}
where $\mathcal N_{\text{G}}$ is a normalization constant, which, as we later show, is equal to $\sqrt\pi$.

\subsection{Position basis representation of the Gaussian modular SSD}
\label{sec:appendix-gauss-pos}
We now demonstrate how to calculate the density matrix $\hat \rho_L^{\text{G}}$ of the Gaussian modular SSD from the position basis representation of a general CV state. We  do so by first calculating the diagonal components, followed by calculating the off-diagonal components.

We begin with evaluating the diagonal components $\bra{l}\hat \rho_L^{\text{G}}\ket{l}$ as
\begin{align}
\bra{l}\hat \rho_L^{\text{G}}\ket{l}= &\frac{1}{\mathcal N_{\text{G}}}\int^{\sqrt\pi/2}_{-\sqrt\pi/2} \dd t_p \int^{\sqrt\pi/2}_{-\sqrt\pi/2} \dd t_q  \cos^2(t_p\sqrt\pi/2) \bra{l_{\text{GKP}}}\hat \rho_\Pi(\mathbf t)\ket{l_{\text{GKP}}}+\sin^2(t_p\sqrt\pi/2) \bra{l_{\text{GKP}}}\hat Z\hat \rho_\Pi(\mathbf t)\hat Z^\dagger\ket{l_{\text{GKP}}}\nonumber\\
=&\frac{1}{\mathcal N_{\text{G}}}\int^{\sqrt\pi/2}_{-\sqrt\pi/2} \dd t_p \int^{\sqrt\pi/2}_{-\sqrt\pi/2} \dd t_q  \cos^2(t_p\sqrt\pi/2) \bra{l_{\text{GKP}}}\hat \rho_\Pi(\mathbf t)\ket{l_{\text{GKP}}}+\sin^2(t_p\sqrt\pi/2) \bra{l_{\text{GKP}}}(-1)^l\hat \rho_\Pi(\mathbf t)(-1)^l\ket{l_{\text{GKP}}}\nonumber\\
=&\frac{1}{\mathcal N_{\text{G}}}\int^{\sqrt\pi/2}_{-\sqrt\pi/2} \dd t_p \int^{\sqrt\pi/2}_{-\sqrt\pi/2} \dd t_q  \left(\cos^2(t_p\sqrt\pi/2) +\sin^2(t_p\sqrt\pi/2)\right) \bra{l_{\text{GKP}}}\hat \rho_\Pi(\mathbf t)\ket{l_{\text{GKP}}}\nonumber\\
=&\frac{1}{\mathcal N_{\text{G}}}\int^{\sqrt\pi/2}_{-\sqrt\pi/2} \dd t_p \int^{\sqrt\pi/2}_{-\sqrt\pi/2} \dd t_q   \bra{l_{\text{GKP}}}\hat \rho_\Pi(\mathbf t)\ket{l_{\text{GKP}}}
\end{align}
which coincides with the diagonal elements of the stabilizer SSD and hence also the modular SSD, i.e.
\begin{align}
    \label{eq:diagonal-equal}
    \bra{l}\hat \rho_L^{\text{G}}\ket{l}=\bra{l}\hat \rho_\Pi\ket{l}=\bra{l}\hat \rho_L \ket{l}.
\end{align}
Taking the trace of both the Gaussian modular SSD and the modular SSD implies that the normalization constant $\mathcal N_{\text{G}}=\sqrt\pi$.

Meanwhile, for the off-diagonal element $\bra{0}\hat \rho_L^{\text{G}}\ket{1}$, we can simplify the expression as 
\begin{align}
\bra{0}\hat \rho_L^{\text{G}}\ket{1}=&\frac{1}{\sqrt\pi}\int^{\sqrt\pi/2}_{-\sqrt\pi/2} \dd t_p \int^{\sqrt\pi/2}_{-\sqrt\pi/2} \dd t_q  \cos^2(t_p\sqrt\pi/2) \bra{0_{\text{GKP}}}\hat \rho_\Pi(\mathbf t)\ket{1_{\text{GKP}}}+\sin^2(t_p\sqrt\pi/2) \bra{0_{\text{GKP}}}\hat Z\hat \rho_\Pi(\mathbf t)\hat Z^\dagger\ket{1_{\text{GKP}}}\nonumber \\
=&\frac{1}{\sqrt\pi}\int^{\sqrt\pi/2}_{-\sqrt\pi/2} \dd t_p \int^{\sqrt\pi/2}_{-\sqrt\pi/2} \dd t_q  \cos^2(t_p\sqrt\pi/2) \bra{0_{\text{GKP}}}\hat \rho_\Pi(\mathbf t)\ket{1_{\text{GKP}}}-\sin^2(t_p\sqrt\pi/2) \bra{0_{\text{GKP}}}\hat \rho_\Pi(\mathbf t)\ket{1_{\text{GKP}}}\nonumber \\
=&\frac{1}{\sqrt\pi}\int^{\sqrt\pi/2}_{-\sqrt\pi/2} \dd t_p \int^{\sqrt\pi/2}_{-\sqrt\pi/2} \dd t_q  \left(\cos^2(t_p\sqrt\pi/2) -\sin^2(t_p\sqrt\pi/2)\right) \bra{0_{\text{GKP}}}\hat \rho_\Pi(\mathbf t)\ket{1_{\text{GKP}}}\nonumber \\
=&\frac{1}{\sqrt\pi}\int^{\sqrt\pi/2}_{-\sqrt\pi/2} \dd t_p \int^{\sqrt\pi/2}_{-\sqrt\pi/2} \dd t_q  \cos(t_p\sqrt\pi)\bra{0_{\text{GKP}}}\hat \rho_\Pi(\mathbf t)\ket{1_{\text{GKP}}}.
\end{align}

We can then insert Eq.~(\ref{eq:def-stab-ssd-t}) into this expression to find
\begin{align}
\bra{0}\hat \rho_L^{\text{G}}\ket{1}
     =&\frac{1}{\sqrt\pi}\int^{\sqrt\pi/2}_{-\sqrt\pi/2} \dd t_p \int^{\sqrt\pi/2}_{-\sqrt\pi/2} \dd t_q  \cos(t_p\sqrt\pi)\bra{0_{\text{GKP}}} \hat V(-\mathbf t) \hat \rho \hat V^\dagger(-\mathbf t)\ket{1_{\text{GKP}}}\nonumber \\
    =&\frac{1}{\sqrt\pi}\int^{\sqrt\pi/2}_{-\sqrt\pi/2} \dd t_p \int^{\sqrt\pi/2}_{-\sqrt\pi/2} \dd t_q  \cos(t_p\sqrt\pi)\bra{0_{\text{GKP}}}e^{-it_p\hat q}e^{it_q\hat p } \hat \rho e^{-it_q\hat p }e^{it_p\hat q}\ket{1_{\text{GKP}}}\nonumber \\
    =&\frac{1}{\sqrt\pi}\int^{\sqrt\pi/2}_{-\sqrt\pi/2} \dd t_p \int^{\sqrt\pi/2}_{-\sqrt\pi/2} \dd t_q  \cos(t_p\sqrt\pi)\sum_{m,m'}\bra{\hat q=2m\sqrt\pi}e^{-it_p\hat q}e^{it_q\hat p } \hat \rho e^{-it_q\hat p }e^{it_p\hat q}\ket{\hat q=(2m'+1)\sqrt\pi}\nonumber \\
    =&\frac{1}{\sqrt\pi}\int^{\sqrt\pi/2}_{-\sqrt\pi/2} \dd t_p \int^{\sqrt\pi/2}_{-\sqrt\pi/2} \dd t_q  \cos(t_p\sqrt\pi)\sum_{m,m'}\bra{\hat q=2m\sqrt\pi}e^{-it_p2m\sqrt\pi}e^{it_q\hat p } \hat \rho e^{-it_q\hat p }e^{it_p(2m'+1)\sqrt\pi}\ket{\hat q=(2m'+1)\sqrt\pi}\nonumber \\
    =&\frac{1}{\sqrt\pi}\int^{\sqrt\pi/2}_{-\sqrt\pi/2} \dd t_p \int^{\sqrt\pi/2}_{-\sqrt\pi/2} \dd t_q  \cos(t_p\sqrt\pi)\sum_{m,m'}e^{-it_p(2m)\sqrt\pi} e^{it_p(2m'+1)\sqrt\pi}\bra{\hat q=(2m)\sqrt\pi+t_q}\hat \rho \ket{\hat q=(2m'+1)\sqrt\pi+t_q}.
    \label{eq:rhoG01}
\end{align}
We can integrate the relevant terms over $t_p$ by using that
\begin{align}
    \int^{\sqrt\pi/2}_{-\sqrt\pi/2} \dd x\cos(x\sqrt\pi)e^{-ixs\sqrt\pi} =& \int^{\sqrt\pi/2}_{-\sqrt\pi/2} \dd x\cos(x\sqrt\pi)(\cos(-xs\sqrt\pi)+i\sin(-xs\sqrt\pi))\nonumber \\
    =& \int^{\sqrt\pi/2}_{-\sqrt\pi/2} \dd x\cos(x\sqrt\pi)\cos(xs\sqrt\pi)\nonumber \\
    =& \frac{2 s\sin(s\pi)}{\sqrt\pi (1-s^2)}.
\end{align}
Note that the denominator will be zero when $s=\pm 1$. The numerator will be zero for any integer $s$. In the limit of $s\to 1$ or $s\to -1$ we find~\cite{Mathematica},
\begin{align}
    \lim_{s\to \pm 1}\frac{2 s\sin(s\pi)}{\sqrt\pi (1-s^2)}=\sqrt\pi.
\end{align}
Therefore, for any $m,m' \in \mathbb Z$ we have
\begin{align}
    \int^{\sqrt\pi/2}_{-\sqrt\pi/2} \dd t_p\cos(t_p\sqrt\pi)e^{-it_p(2m)\sqrt\pi} e^{it_p(2m'+1)\sqrt\pi}=\frac{\sqrt\pi}{2}\delta_{m,m'}+\frac{\sqrt\pi}{2}\delta_{m-1,m'}.
\end{align}
Using this result to integrate Eq.~(\ref{eq:rhoG01}), we find
\begin{align}
\label{eq:gssd-offdiag-ext}
\bra{0}\hat \rho_L^{\text{G}}\ket{1}=&\frac{1}{2}\int^{\sqrt\pi/2}_{-\sqrt\pi/2}  \dd t_q  \sum_{m,m'}(\delta_{m,m'}+\delta_{m-1,m'})\bra{\hat q=2m\sqrt\pi+t_q}\hat \rho \ket{\hat q=(2m'+1)\sqrt\pi+t_q} \nonumber\\
=&\frac{1}{2}\int^{\sqrt\pi/2}_{-\sqrt\pi/2}  \dd t_q  \sum_{m}\bra{\hat q=2m\sqrt\pi+t_q}\hat \rho \ket{(2m+1)\sqrt\pi+t_q}+\bra{2m\sqrt\pi+t_q}\hat \rho \ket{\hat q=(2m-1)\sqrt\pi+t_q}.
\end{align}

The other off-diagonal term is equal to the Hermitian conjugate of this term, i.e.,
\begin{align}
    \label{eq:gssd-offidag}
    \bra{1}\hat \rho_{L}^{\text{G}}\ket{0}
    =&\left(\bra{0}\hat \rho_{L}^{\text{G}}\ket{1}\right)^*.
\end{align}

\subsection{Circuit implementation of the Gaussian modular SSD}
\label{sec:appendix-gauss-circuit}
Inspecting Eq.~(\ref{eq:gaussian-part}), we see that we can consider the action of the Gaussian modular SSD as first implementing the GKP error-correction routine, which transforms the state 
\begin{align}
    \hat \rho \to \hat\rho_\Pi(\mathbf t)
\end{align}
with some measurement values $\mathbf t$, followed by a Gaussian channel defined as
\begin{align}
    \label{eq:gauss-channel}
    \varepsilon_{t_p}(\hat\rho)=\cos^2(t_p\sqrt\pi/2)\hat \rho+\sin^2(t_p\sqrt\pi/2)\hat Z\hat \rho\hat Z^\dagger,
\end{align}
whereby $\hat Z=e^{i\sqrt\pi \hat q}$ is the logical $\hat Z$ operation in the GKP encoding. The channel can be interpreted as implementing a $\hat Z$ flip on the encoded qubit state with probability $\sin^2(t_p\sqrt\pi/2)$.

Therefore, we can prepare the Gaussian modular SSD state, encoded in the GKP basis, by performing GKP error correction followed by applying the channel given in Eq.~(\ref{eq:gauss-channel}). This is equivalent to performing GKP error correction, followed by applying a logical $Z$ operation with a probability $p_Z(t_p)$ and then discarding $\mathbf t$ such that the resulting state is a mixed state over the possible measurement values of $\mathbf t$. The circuit diagram to prepare $\hat \rho_L^{\text{G}}$ from $\hat \rho$ is given in Fig.~(\ref{fig:gauss-ssd-prep}).

\begin{figure}[ht]
        \centering
        $$
        \Qcircuit @C=0.9em @R=.7em {
        \lstick{\hat \rho}              & \qw & \ctrl{1}    & \qw               &  \gate{\hat X(-\{t_q\}_{\sqrt\pi})}      & \qw           & \targMinus & \qw               & \qw                      &  \gate{\hat Z(-\{t_p\}_{\sqrt\pi})}   & \qw & \gate{\mathcal{E}_{t_p}} & \qw \\
        \lstick{\ket{0_{\text{GKP}}}}   & \qw & \control\qw & \measureD{\hat p} & \control \cw \cwx   & \lstick{t_q} \\
        \lstick{\ket{0_{\text{GKP}}}}   & \qw & \qw         &\qw                &\qw  & \qw                 & \ctrl{-2}\qw  & \qw   & \measureD{\hat p} & \control \cw \cwx[-2]   &\cw & \control \cw \cwx[-2]    & \lstick{t_p} 
        }
        $$
         \caption{Gaussian modular subsystem decomposition as a circuit. We use the same notation for the inverse SUM gate as presented in Fig.~\ref{fig:gkp-ec-mod}. The Gaussian channel $\varepsilon_{t_p}$ is defined in Eq.~(\ref{eq:gauss-channel}).}
         \label{fig:gauss-ssd-prep}
    \end{figure}
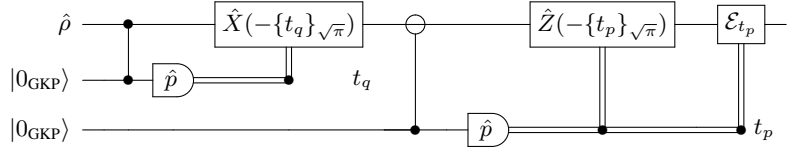

\section{Equivalence of modular SSD and Gaussian modular SSD for symmetric density matrices}
\label{sec:appendix-equivalence-symmetric}
In this appendix, we demonstrate that the modular SSD and the Gaussian modular SSD are equivalent when the density matrix of the CV state is symmetric, i.e., when
\begin{align}
    \label{eq:symmetric-meaning}
    \bra{\hat q=x}\hat \rho\ket{\hat q=x'}=\bra{\hat q=-x}\hat \rho\ket{\hat q=-x'}.
\end{align}
Note that, for a pure state this would mean that the wavefunction is symmetric in position, i.e., $\psi(x)=\psi(-x)$, or that the wavefunction is antisymmetric, i.e., $\psi(-x)=-\psi(x)$.

First, we recall Eq.~(\ref{eq:diagonal-equal}), which informs us that for all input states, the diagonal elements of the Gaussian modular SSD are equal to the corresponding elements of the modular SSD. Therefore, to demonstrate that the two decompositions are equivalent, for the case of a symmetric CV state, it suffices to show that the off-diagonal elements of the two decompositions are equal. As a reminder, the off-diagonal element of the two states after each decomposition can be expressed as
\begin{align}
\label{eq:gauss-mod-offdiag-repeated}
\bra{0}\hat \rho_L^{\text{G}}\ket{1}=&\frac{1}{2}\int^{\sqrt\pi/2}_{-\sqrt\pi/2}  \dd t_q  \sum_{n\in\mathbbm Z}\bra{\hat q=2n\sqrt\pi+t_q}\hat \rho \ket{\hat q=(2n+1)\sqrt\pi+t_q}+\bra{\hat q=2n\sqrt\pi+t_q}\hat \rho \ket{\hat q=(2n-1)\sqrt\pi+t_q}
\end{align}
and
\begin{align}
    \bra{0}\hat \rho_L\ket{1}
    =&\sum_{n\in \mathbb Z}\int^{\sqrt\pi/2}_{-\sqrt\pi/2} \dd t_q \bra{\hat q=2\sqrt\pi n+t_q}\hat \rho\ket{\hat q=2\sqrt\pi n+\sqrt\pi+t_q}.
\end{align}

Inspecting $\bra{0}\hat\rho_L^{\text{G}}\ket{1}$, given in Eq.~(\ref{eq:gauss-mod-offdiag-repeated}), we see that we can split the expression into the sum of two terms, for each $\pm 1$, expressed together as
\begin{align}
\label{eq:gauss-mod-offdiag-twoterms}
\frac{1}{2}\int^{\sqrt\pi/2}_{-\sqrt\pi/2}  \dd t_q  \sum_{n\in\mathbbm Z}\bra{\hat q=2n\sqrt\pi+t_q}\hat \rho \ket{\hat q=(2n\pm 1)\sqrt\pi+t_q}.
\end{align}
We can use the symmetric condition, given by Eq.~(\ref{eq:symmetric-meaning}), to rewrite the second term as
\begin{align}
   &\frac{1}{2}\int^{\sqrt\pi/2}_{-\sqrt\pi/2}  \dd t_q  \sum_{n\in\mathbbm Z}\bra{\hat q=2n\sqrt\pi+t_q}\hat \rho \ket{\hat q=(2n- 1)\sqrt\pi+t_q}\\
   =&\frac{1}{2}\int^{\sqrt\pi/2}_{-\sqrt\pi/2}  \dd t_q  \sum_{n\in\mathbbm Z}\bra{\hat q=-2n\sqrt\pi-t_q}\hat \rho \ket{\hat q=(-2n+1)\sqrt\pi-t_q}.
\end{align}
By substituting $n\to-n$ and $t_q\to -t_q$ in this term, we find that it is equal to the first term, i.e.,
\begin{align}
   &\frac{1}{2}\int^{\sqrt\pi/2}_{-\sqrt\pi/2}  \dd t_q  \sum_{n\in\mathbbm Z}\bra{\hat q=2n\sqrt\pi+t_q}\hat \rho \ket{\hat q=(2n- 1)\sqrt\pi+t_q}\\
   =&\frac{1}{2}\int^{\sqrt\pi/2}_{-\sqrt\pi/2}  \dd t_q  \sum_{n\in\mathbbm Z}\bra{\hat q=2n\sqrt\pi+t_q}\hat \rho \ket{\hat q=(2n+ 1)\sqrt\pi+t_q}
\end{align}
and thus we can conclude that the sum of these two terms equals $\bra{0}\hat \rho_L\ket{1}$. Therefore, for symmetric states, we find that the modular SSD is indeed equivalent to the Gaussian modular SSD, i.e., $\hat \rho^{\text{G}}_L=\hat \rho_L$.

\section{Effect of the choice of mapping on the ROM of the logical state}
\label{sec:appendix-mapping-choice}
The logical resourcefulness of different CV states depends on the choice of mapping. We also find that different mappings can result in different hierarchies of states. We illustrate this with an example, given in Fig. \ref{fig:comparison-rom-stab-mod}, where we plot the ROM of the logical state resulting from mapping a specific GKP state, Eq.~(\ref{eq:realistic-gkp-theta}), with Bloch angles $\phi=\pi/4$ and $\theta=\pi/25$, with varying values of $\Delta$. The ROM of the logical states reached via each decomposition are also plotted in Fig.~\ref{fig:roms-gkp}, whereby the ROM of the stabilizer SSD and modular SSD of each state in Fig.~\ref{fig:comparison-rom-stab-mod} can be considered a vertical cross-section through Fig.~\ref{subfig:gkp_stab} and Fig.~\ref{subfig:gkp_gkpGauss}, respectively. We see that the ROM can be below the threshold for distillation when using one mapping, while it is above the distillation threshold for the other.

Furthermore, we plot the gradient of the ROM for each mapping in the inset figure. We find that there are some values of $\Delta$, e.g., $\Delta\approx 0.7$, whereby the gradient of the stabilizer SSD is positive, while the gradient of the modular SSD is negative. This difference means that the two different decompositions result in logical states whereby increasing $\Delta$ increases ROM for the logical state resulting from the stabilizer SSD, meanwhile, the same increase in $\Delta$ decreases ROM for the logical state resulting from the modular SSD. 

\begin{figure}[h]
    \centering
    \includegraphics[width=0.5\linewidth]{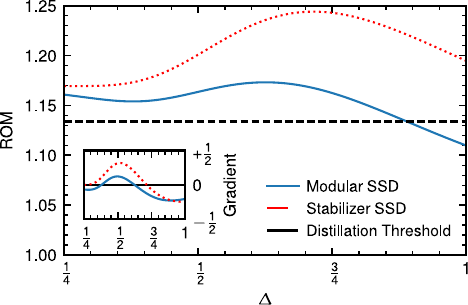}
     \caption{Comparison of ROM of the resulting logical state found by the modular SSD (blue, solid lines)   and the stabilizer SSD (red, dashed lines) of a GKP state, Eq.~(\ref{eq:realistic-gkp-theta}), with Bloch angles $\phi=\pi/4$ and $\theta=\pi/25$ and varying squeezing $\Delta$. The inset plot shows the gradient of the ROM of each decomposition, for each value of $\Delta$.} 
    \label{fig:comparison-rom-stab-mod}
\end{figure}

\section{Alternative definitions of a realistic GKP state}
\label{sec:appendix-different-gkp-defs}

We summarize two alternative definitions of the realistic GKP state. For a detailed analysis of the different definitions of realistic GKP states and their equivalence, refer to Ref.~\cite{matsuura2020}. We also give an explanation for why the logical state found by the stabilizer SSD of the realistic GKP state given in Eq.~(\ref{eq:realistic-gkp-def}) differs from that found in Ref.~\cite{shaw2022}. 

Throughout our work, we use the definition of the realistic GKP state given in Eq.~(\ref{eq:realistic-gkp-def}), which is the same definition used in Ref.~\cite{pantaleoni2021} whereby the modular SSD of the GKP state was first analyzed.

In Ref.~\cite{shaw2022}, whereby the stabilizer SSD is defined, the realistic GKP state is defined in terms of the finite energy operator $e^{-\Delta^{\prime 2} \hat a^\dagger \hat a}$ applied to the ideal GKP state, given in Eq.~(\ref{eq:ideal-gkp-def}). For high squeezing, i.e., $\Delta=\Delta' \ll 1$, the fidelity of the two corresponding wavefunctions is very high. However, for low squeezing, the two states diverge. 
A key difference is that for the definition in terms of the finite energy operator, any GKP state will approach the vacuum in the limit that $\Delta '\to \infty$. Meanwhile, for the definition used throughout this work, the equivalent low-squeezing limit is reached when $\Delta\to 1$, whereby the $0$-logical state has high fidelity to the vacuum state. However, the $1$-logical state does not approach the vacuum state.

\section{Analysis of GKP state with large values of the squeezing parameter}
\label{appendix:gkp-large-delta}
In this appendix we provide a more detailed analysis of the ROM of the stabilizer SSD of two specific logical GKP states, following the discussion in Sec.~\ref{sec:analysis-gkp}. 

We now focus on two specific logical states shown in Fig.~\ref{subfig:gkp_stab}, with different levels of squeezing. Specifically, we plot the ROM of the stabilizer SSD of the $0$-logical and $T$-logical states for different levels of squeezing $\Delta$ in Fig.~\ref{fig:rom-vs-wig-lines}. These lines display the same values as shown in Fig.~\ref{subfig:gkp_stab} for $\theta=0$ and $\theta=\arccos(1/\sqrt 3)$. We also plot the WLN of these states for each $\Delta$.

\begin{figure}
    \centering
    \includegraphics[width=0.6\linewidth]{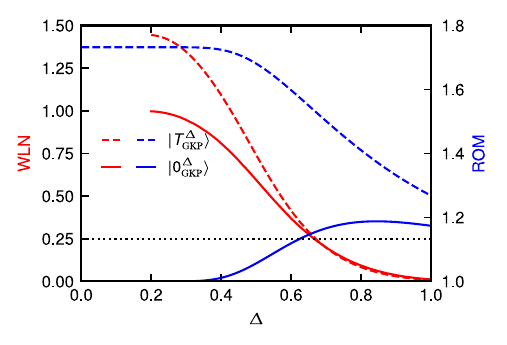}
     \caption{Comparison of WLN (red) and ROM of the resulting state given by the stabilizer SSD (blue) for the $0$-logical (solid lines) and $T$-logical (dashed lines) GKP state defined in terms of its squeezing parameter $\Delta$. Note that the value of the ROM of the $0$-logical state begins to decrease after $\Delta>0.78$.}
    \label{fig:rom-vs-wig-lines}
\end{figure}

First, note that in the limit of infinite squeezing, $\Delta\to 0$, the WLN of each state converges to the value of the logarithmic negativity of a unit cell of the respective Wigner function~\cite{yamasaki2020}. For the $0$-logical state the logarithmic negativity converges to $\log_2(2)= 1$, meanwhile for the $T$-logical state it converges to $\log_2(1+\sqrt 3)\approx 1.45$. The reason for this is that each cell contributes a finite value of negativity~\cite{garcia-alvarez2019,hahn2022} which is normalized over the full Wigner function. Approximating the Wigner function of a highly squeezed GKP state as consisting of a large number of cells, each contributing an equal cell WLN that is normalized over the number of cells, then the total negativity will be independent of the number of cells.

As seen in Figure \ref{fig:rom-vs-wig-lines}, for the $T$-logical state both WLN and ROM increase as expected at increasing squeezing level (i.e., decreasing $\Delta$).  

Instead, for the $0$-logical state, the ROM is decreasing at increasing squeezing level while the WLN is increasing, for $\Delta<0.78$.  In the context of all Gaussian circuits, including the supply of Gaussian states such as the vacuum, highly squeezed GKP states can be considered a sufficient resource for universality~\cite{baragiola2019}. This is supported by the high Wigner negativity of near-ideal GKP states. Meanwhile, in the context of SGKP circuits, noise introduced by finite squeezing can be considered the resource~\cite{calcluth2023}. Intuitively, this is because the finite squeezing increases the overlap of a $0$-logical GKP state with a magic $T$-logical GKP state. This is supported by the increasing ROM when the GKP states are increasingly noisy, i.e., for increasing $\Delta$. 

We also observe that the ROM of the stabilizer SSD of the $0$-logical state begins to decrease for values of $\Delta>0.78$. 
To understand why this occurs, we must consider the state obtained by the stabilizer SSD.
Following the expression for the stabilizer SSD given in Appendix \ref{sec:appendix-stab}, we find that for a pure state $\psi(x)$, the unnormalized coefficients of the density matrix $\hat \rho_\Pi$ of the resulting logical state can be expressed as
\begin{align}
    &\bar{\rho}_\Pi^{00}=\sum_n\int^{\sqrt\pi/2}_{-\sqrt\pi/2} \dd t_q |\psi(t_q+2\sqrt\pi n)|^2\\
    &\bar{\rho}_\Pi^{11}=\sum_n\int^{\sqrt\pi/2}_{-\sqrt\pi/2}\dd t_q  |\psi\left((2n+1)\sqrt\pi+t_q\right)|^2\\
    &\bar{\rho}_\Pi^{01}\nonumber\\
    &=\sum_{n,n'}\int^{\sqrt\pi/2}_{-\sqrt\pi/2} \dd t_q k_{n,n'}\rho(t_q+2n\sqrt\pi,t_q+(2n'+1)\sqrt\pi),
\end{align}
where
\begin{align}
    k_{n,n'}=(-1)^{n-n'}\frac{2}{1-2 n+2n'},
\end{align}
and $\rho(x,x')=\psi(x)\psi^*(x')$.
When the pure CV state has the property that $\psi(x)\approx0$ for $|x|>\sqrt\pi$, we find that the summation over $n$ only contributes for $n=0$. Meanwhile, the summation over $n'$ only contributes when $n'=0$ or $n'=-1$. Hence, the off-diagonal coefficient of the density matrix of the state after stabilizer SSD can be expressed as
\begin{align}
    \bar\rho_\Pi^{01}&\approx2\int^{\sqrt\pi/2}_{-\sqrt\pi/2}\psi(t_q)\psi^*(t_q+\sqrt\pi)+\psi(t_q)\psi^*(t_q-\sqrt\pi).
\end{align}
Furthermore, we can also make the change of variable ${t_q\to -t_q}$ in the second term, which implies that, for a symmetric state, the integral over each term must be equal and, hence,
\begin{align}
    \label{eq:rho01_gkp}
    \bar\rho_\Pi^{01}&\approx4\int^{\sqrt\pi/2}_{-\sqrt\pi/2}\psi(t_q)\psi^*(t_q+\sqrt\pi).
\end{align}

\begin{figure}
    \centering
    \includegraphics[width=0.5\linewidth]{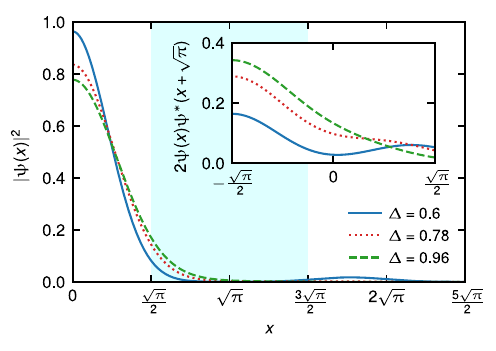}
     \caption{The modulus square of the wavefunction of the $0$-logical GKP state for large values of $\Delta$. We omit the values for $x<0$ because the state is symmetric in position. The values in the white regions of the plot contribute to the relative value of $ \rho_\Pi^{00}$, while the blue region contributes to the relative value of $ \rho_\Pi^{11}$.
    The inset figure shows the product of the wavefunction and the wavefunction offset by $\sqrt\pi$. The integral of the curve in this region contributes to the relative value of $ \rho_\Pi^{01}= \rho_\Pi^{10}$.}
    \label{fig:rom-vs-wig-explained}
\end{figure}

The modulus square of the wavefunction $|\psi(x)|^2$ and the product of the wavefunction with the conjugate of its displaced product, multiplied by a factor of $2$, $2\psi(x)\psi(x+\sqrt\pi)$, are both plotted in Fig.~\ref{fig:rom-vs-wig-explained}. The area under the curve of $|\psi(x)|^2$, in the white regions of the main plot, equals the approximate value of $\bar \rho_\Pi^{00}/2$, where the factor of $1/2$ arises due to the fact that the plot is in the range $x\ge 0$. Meanwhile, the area under the curve of the blue regions of the main plot equals the approximate value of $\bar \rho_\Pi^{11}/2$. Furthermore, by plotting twice the value of the integrated in Eq.~(\ref{eq:rho01_gkp}), the area under the curve of the inset plot equals the approximate value of $\bar \rho_\Pi^{01}/2$. Hence, each corresponding area is proportional to the relevant density matrix element.

Given that the ROM of the state can be expressed as in Eq.~(\ref{eq:rom-single-coefs}) and using the fact that the $0$-logical GKP state has a real wavefunction for any $\Delta$ and, therefore, also real $\bar \rho_\Pi^{01}$, we find that the ROM is proportional to
\begin{align}
    \mathcal R^{(1)}(\hat \rho_\Pi)\propto 2|\bar \rho_\Pi^{01}|+|\bar \rho_\Pi^{00}-\bar \rho_\Pi^{11}|.
\end{align}
Therefore, the stabilizer SSD of a $0$-logical GKP state will have maximum ROM when the sum of the difference of the areas under the curve in each region of the main plot and the area under the curve of the inset plot is maximized. We observe that although the value of $\rho_{\Pi}^{00}$ increases as $\Delta$ increases, the value of $\rho_{\Pi}^{01}$ also decreases. The value $\Delta=0.78$ is the optimal value of this summation and is therefore also the optimal value of $\Delta$, for this type of state, with maximum ROM.

\section{Modular SSD of the cubic phase state}
\label{sec:appendix-plot-mod-cps}
We include here for completeness a plot of the ROM of the modular SSD for the cubic phase state in Fig.~\ref{fig:modular-cps}. However, we stress that in general, the ROM of the decomposed state has no operational meaning. For the cubic phase state, we cannot make the connection to the Gaussian modular SSD due to the fact that the state does not have a density matrix that is symmetric in position. Note that the plot includes squeezed vacuum states along the axis $\gamma=0$. These states are symmetric in position and therefore their modular SSD decomposition is equivalent to the Gaussian modular SSD in Fig.~\ref{subfig:cps_gkpGauss} along the same axis. This axis, in turn, also includes the vacuum state at $\gamma=0,\zeta=0$. As is the case for the Gaussian modular SSD ROM, the state prepared using the modular SSD from the vacuum state is not above the distillation threshold. However, two new distillation regions appear, characterized by low squeezing and moderate cubicity. 

\begin{figure}[ht]
    \centering
    \includegraphics[width=0.5\linewidth]{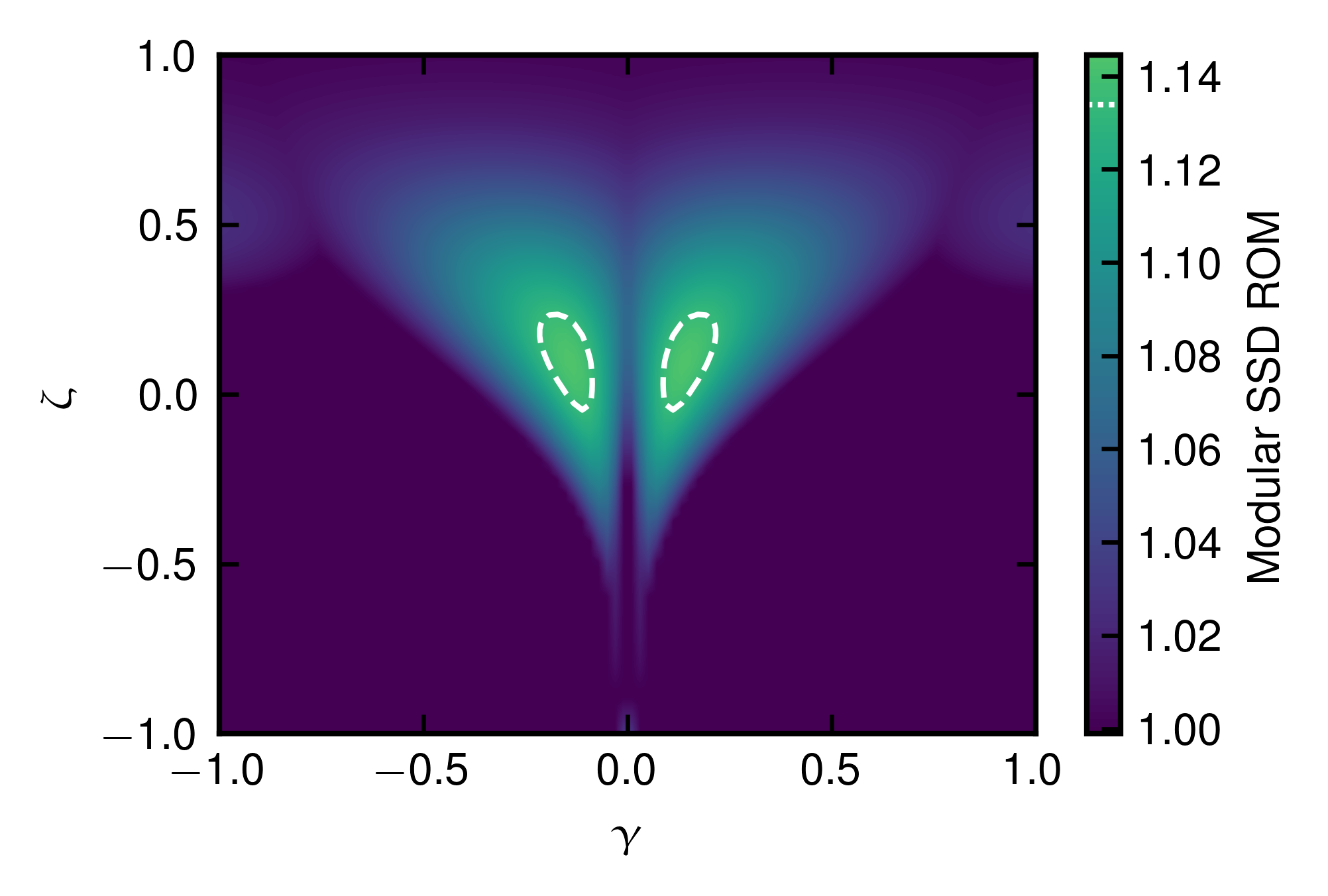}
     \caption{ROM of the modular SSD of the cubic phase state. The white dashed lines show the distillation threshold of the qubit state prepared from the modular SSD. However, this threshold is not a criterion for universality.}
    \label{fig:modular-cps}
\end{figure}

\end{widetext}
\bibliographystyle{apsrev4-2}
\bibliography{./main.bib}

\begin{thebibliography}{90}%
\makeatletter
\providecommand \@ifxundefined [1]{%
 \@ifx{#1\undefined}
}%
\providecommand \@ifnum [1]{%
 \ifnum #1\expandafter \@firstoftwo
 \else \expandafter \@secondoftwo
 \fi
}%
\providecommand \@ifx [1]{%
 \ifx #1\expandafter \@firstoftwo
 \else \expandafter \@secondoftwo
 \fi
}%
\providecommand \natexlab [1]{#1}%
\providecommand \enquote  [1]{``#1''}%
\providecommand \bibnamefont  [1]{#1}%
\providecommand \bibfnamefont [1]{#1}%
\providecommand \citenamefont [1]{#1}%
\providecommand \href@noop [0]{\@secondoftwo}%
\providecommand \href [0]{\begingroup \@sanitize@url \@href}%
\providecommand \@href[1]{\@@startlink{#1}\@@href}%
\providecommand \@@href[1]{\endgroup#1\@@endlink}%
\providecommand \@sanitize@url [0]{\catcode `\\12\catcode `\$12\catcode `\&12\catcode `\#12\catcode `\^12\catcode `\_12\catcode `\%12\relax}%
\providecommand \@@startlink[1]{}%
\providecommand \@@endlink[0]{}%
\providecommand \url  [0]{\begingroup\@sanitize@url \@url }%
\providecommand \@url [1]{\endgroup\@href {#1}{\urlprefix }}%
\providecommand \urlprefix  [0]{URL }%
\providecommand \Eprint [0]{\href }%
\providecommand \doibase [0]{https://doi.org/}%
\providecommand \selectlanguage [0]{\@gobble}%
\providecommand \bibinfo  [0]{\@secondoftwo}%
\providecommand \bibfield  [0]{\@secondoftwo}%
\providecommand \translation [1]{[#1]}%
\providecommand \BibitemOpen [0]{}%
\providecommand \bibitemStop [0]{}%
\providecommand \bibitemNoStop [0]{.\EOS\space}%
\providecommand \EOS [0]{\spacefactor3000\relax}%
\providecommand \BibitemShut  [1]{\csname bibitem#1\endcsname}%
\let\auto@bib@innerbib\@empty
\bibitem [{\citenamefont {Horodecki}\ \emph {et~al.}(2003)\citenamefont {Horodecki}, \citenamefont {Horodecki},\ and\ \citenamefont {Oppenheim}}]{horodecki2003}%
  \BibitemOpen
  \bibfield  {author} {\bibinfo {author} {\bibfnamefont {M.}~\bibnamefont {Horodecki}}, \bibinfo {author} {\bibfnamefont {P.}~\bibnamefont {Horodecki}},\ and\ \bibinfo {author} {\bibfnamefont {J.}~\bibnamefont {Oppenheim}},\ }\href {https://doi.org/10.1103/PhysRevA.67.062104} {\bibfield  {journal} {\bibinfo  {journal} {Phys. Rev. A}\ }\textbf {\bibinfo {volume} {67}},\ \bibinfo {pages} {062104} (\bibinfo {year} {2003})}\BibitemShut {NoStop}%
\bibitem [{\citenamefont {Brunner}\ \emph {et~al.}(2014)\citenamefont {Brunner}, \citenamefont {Cavalcanti}, \citenamefont {Pironio}, \citenamefont {Scarani},\ and\ \citenamefont {Wehner}}]{brunner2014bell}%
  \BibitemOpen
  \bibfield  {author} {\bibinfo {author} {\bibfnamefont {N.}~\bibnamefont {Brunner}}, \bibinfo {author} {\bibfnamefont {D.}~\bibnamefont {Cavalcanti}}, \bibinfo {author} {\bibfnamefont {S.}~\bibnamefont {Pironio}}, \bibinfo {author} {\bibfnamefont {V.}~\bibnamefont {Scarani}},\ and\ \bibinfo {author} {\bibfnamefont {S.}~\bibnamefont {Wehner}},\ }\href {https://doi.org/10.1103/RevModPhys.86.419} {\bibfield  {journal} {\bibinfo  {journal} {Reviews of modern physics}\ }\textbf {\bibinfo {volume} {86}},\ \bibinfo {pages} {419} (\bibinfo {year} {2014})}\BibitemShut {NoStop}%
\bibitem [{\citenamefont {Streltsov}\ \emph {et~al.}(2017)\citenamefont {Streltsov}, \citenamefont {Adesso},\ and\ \citenamefont {Plenio}}]{streltsov2017}%
  \BibitemOpen
  \bibfield  {author} {\bibinfo {author} {\bibfnamefont {A.}~\bibnamefont {Streltsov}}, \bibinfo {author} {\bibfnamefont {G.}~\bibnamefont {Adesso}},\ and\ \bibinfo {author} {\bibfnamefont {M.~B.}\ \bibnamefont {Plenio}},\ }\href {https://doi.org/10.1103/RevModPhys.89.041003} {\bibfield  {journal} {\bibinfo  {journal} {Rev. Mod. Phys.}\ }\textbf {\bibinfo {volume} {89}},\ \bibinfo {pages} {041003} (\bibinfo {year} {2017})}\BibitemShut {NoStop}%
\bibitem [{\citenamefont {Chitambar}\ and\ \citenamefont {Gour}(2019)}]{chitambar2019}%
  \BibitemOpen
  \bibfield  {author} {\bibinfo {author} {\bibfnamefont {E.}~\bibnamefont {Chitambar}}\ and\ \bibinfo {author} {\bibfnamefont {G.}~\bibnamefont {Gour}},\ }\href {https://doi.org/10.1103/RevModPhys.91.025001} {\bibfield  {journal} {\bibinfo  {journal} {Rev. Mod. Phys.}\ }\textbf {\bibinfo {volume} {91}},\ \bibinfo {pages} {025001} (\bibinfo {year} {2019})}\BibitemShut {NoStop}%
\bibitem [{\citenamefont {Budroni}\ \emph {et~al.}(2022)\citenamefont {Budroni}, \citenamefont {Cabello}, \citenamefont {G{\"u}hne}, \citenamefont {Kleinmann},\ and\ \citenamefont {Larsson}}]{budroni2022kochen}%
  \BibitemOpen
  \bibfield  {author} {\bibinfo {author} {\bibfnamefont {C.}~\bibnamefont {Budroni}}, \bibinfo {author} {\bibfnamefont {A.}~\bibnamefont {Cabello}}, \bibinfo {author} {\bibfnamefont {O.}~\bibnamefont {G{\"u}hne}}, \bibinfo {author} {\bibfnamefont {M.}~\bibnamefont {Kleinmann}},\ and\ \bibinfo {author} {\bibfnamefont {J.-{\AA}.}\ \bibnamefont {Larsson}},\ }\href {https://doi.org/10.1103/RevModPhys.94.045007} {\bibfield  {journal} {\bibinfo  {journal} {Reviews of Modern Physics}\ }\textbf {\bibinfo {volume} {94}},\ \bibinfo {pages} {045007} (\bibinfo {year} {2022})}\BibitemShut {NoStop}%
\bibitem [{\citenamefont {Bravyi}\ and\ \citenamefont {Kitaev}(2005)}]{bravyi2005}%
  \BibitemOpen
  \bibfield  {author} {\bibinfo {author} {\bibfnamefont {S.}~\bibnamefont {Bravyi}}\ and\ \bibinfo {author} {\bibfnamefont {A.}~\bibnamefont {Kitaev}},\ }\href {https://doi.org/10.1103/PhysRevA.71.022316} {\bibfield  {journal} {\bibinfo  {journal} {Phys. Rev. A}\ }\textbf {\bibinfo {volume} {71}},\ \bibinfo {pages} {022316} (\bibinfo {year} {2005})}\BibitemShut {NoStop}%
\bibitem [{\citenamefont {Harrow}\ and\ \citenamefont {Montanaro}(2017)}]{harrow2017}%
  \BibitemOpen
  \bibfield  {author} {\bibinfo {author} {\bibfnamefont {A.~W.}\ \bibnamefont {Harrow}}\ and\ \bibinfo {author} {\bibfnamefont {A.}~\bibnamefont {Montanaro}},\ }\href {https://doi.org/10.1038/nature23458} {\bibfield  {journal} {\bibinfo  {journal} {Nature}\ }\textbf {\bibinfo {volume} {549}},\ \bibinfo {pages} {203} (\bibinfo {year} {2017})}\BibitemShut {NoStop}%
\bibitem [{\citenamefont {Gottesman}(1997)}]{gottesman1997}%
  \BibitemOpen
  \bibfield  {author} {\bibinfo {author} {\bibfnamefont {D.}~\bibnamefont {Gottesman}},\ }\emph {\bibinfo {title} {Stabilizer Codes and Quantum Error Correction}},\ \href@noop {} {Ph.D. thesis},\ \bibinfo  {school} {California Institute of Technology} (\bibinfo {year} {1997}),\ \bibinfo {note} {\href{https://arxiv.org/abs/quant-ph/9705052}{arXiv:quant-ph/9705052}}\BibitemShut {NoStop}%
\bibitem [{\citenamefont {Nielsen}\ and\ \citenamefont {Chuang}(2000)}]{nielsen2000}%
  \BibitemOpen
  \bibfield  {author} {\bibinfo {author} {\bibfnamefont {M.~A.}\ \bibnamefont {Nielsen}}\ and\ \bibinfo {author} {\bibfnamefont {I.~L.}\ \bibnamefont {Chuang}},\ }\href@noop {} {\emph {\bibinfo {title} {Quantum Computation and Quantum Information}}}\ (\bibinfo  {publisher} {{Cambridge University Press}},\ \bibinfo {year} {2000})\BibitemShut {NoStop}%
\bibitem [{\citenamefont {Bravyi}\ and\ \citenamefont {Haah}(2012)}]{bravyi2012}%
  \BibitemOpen
  \bibfield  {author} {\bibinfo {author} {\bibfnamefont {S.}~\bibnamefont {Bravyi}}\ and\ \bibinfo {author} {\bibfnamefont {J.}~\bibnamefont {Haah}},\ }\href {https://doi.org/10.1103/PhysRevA.86.052329} {\bibfield  {journal} {\bibinfo  {journal} {Phys. Rev. A}\ }\textbf {\bibinfo {volume} {86}},\ \bibinfo {pages} {052329} (\bibinfo {year} {2012})}\BibitemShut {NoStop}%
\bibitem [{\citenamefont {Litinski}(2019)}]{litinski2019}%
  \BibitemOpen
  \bibfield  {author} {\bibinfo {author} {\bibfnamefont {D.}~\bibnamefont {Litinski}},\ }\href {https://doi.org/10.22331/q-2019-12-02-205} {\bibfield  {journal} {\bibinfo  {journal} {Quantum}\ }\textbf {\bibinfo {volume} {3}},\ \bibinfo {pages} {205} (\bibinfo {year} {2019})}\BibitemShut {NoStop}%
\bibitem [{\citenamefont {Reichardt}(2009)}]{reichardt2009}%
  \BibitemOpen
  \bibfield  {author} {\bibinfo {author} {\bibfnamefont {B.}~\bibnamefont {Reichardt}},\ }\href {https://doi.org/10.26421/QIC9.11-12-7} {\bibfield  {journal} {\bibinfo  {journal} {Quantum Inf. Comput.}\ }\textbf {\bibinfo {volume} {9}},\ \bibinfo {pages} {1030} (\bibinfo {year} {2009})}\BibitemShut {NoStop}%
\bibitem [{\citenamefont {Campbell}\ and\ \citenamefont {Howard}(2017)}]{campbell2017}%
  \BibitemOpen
  \bibfield  {author} {\bibinfo {author} {\bibfnamefont {E.~T.}\ \bibnamefont {Campbell}}\ and\ \bibinfo {author} {\bibfnamefont {M.}~\bibnamefont {Howard}},\ }\href {https://doi.org/10.1103/PhysRevA.95.022316} {\bibfield  {journal} {\bibinfo  {journal} {Phys. Rev. A}\ }\textbf {\bibinfo {volume} {95}},\ \bibinfo {pages} {022316} (\bibinfo {year} {2017})}\BibitemShut {NoStop}%
\bibitem [{\citenamefont {Ferraro}\ \emph {et~al.}(2005)\citenamefont {Ferraro}, \citenamefont {Olivares},\ and\ \citenamefont {Paris}}]{ferraro2005}%
  \BibitemOpen
  \bibfield  {author} {\bibinfo {author} {\bibfnamefont {A.}~\bibnamefont {Ferraro}}, \bibinfo {author} {\bibfnamefont {S.}~\bibnamefont {Olivares}},\ and\ \bibinfo {author} {\bibfnamefont {M.~G.~A.}\ \bibnamefont {Paris}},\ }\href@noop {} {\emph {\bibinfo {title} {Gaussian States in Quantum Information}}}\ (\bibinfo  {publisher} {{Bibliopolis}},\ \bibinfo {address} {{Napoli}},\ \bibinfo {year} {2005})\BibitemShut {NoStop}%
\bibitem [{\citenamefont {Menicucci}\ \emph {et~al.}(2006)\citenamefont {Menicucci}, \citenamefont {{van Loock}}, \citenamefont {Gu}, \citenamefont {Weedbrook}, \citenamefont {Ralph},\ and\ \citenamefont {Nielsen}}]{menicucci2006}%
  \BibitemOpen
  \bibfield  {author} {\bibinfo {author} {\bibfnamefont {N.~C.}\ \bibnamefont {Menicucci}}, \bibinfo {author} {\bibfnamefont {P.}~\bibnamefont {{van Loock}}}, \bibinfo {author} {\bibfnamefont {M.}~\bibnamefont {Gu}}, \bibinfo {author} {\bibfnamefont {C.}~\bibnamefont {Weedbrook}}, \bibinfo {author} {\bibfnamefont {T.~C.}\ \bibnamefont {Ralph}},\ and\ \bibinfo {author} {\bibfnamefont {M.~A.}\ \bibnamefont {Nielsen}},\ }\href {https://doi.org/10.1103/PhysRevLett.97.110501} {\bibfield  {journal} {\bibinfo  {journal} {Phys. Rev. Lett.}\ }\textbf {\bibinfo {volume} {97}},\ \bibinfo {pages} {110501} (\bibinfo {year} {2006})}\BibitemShut {NoStop}%
\bibitem [{\citenamefont {Weedbrook}\ \emph {et~al.}(2012)\citenamefont {Weedbrook}, \citenamefont {Pirandola}, \citenamefont {{Garc{\'\i}a-Patr{\'o}n}}, \citenamefont {Cerf}, \citenamefont {Ralph}, \citenamefont {Shapiro},\ and\ \citenamefont {Lloyd}}]{weedbrook2012}%
  \BibitemOpen
  \bibfield  {author} {\bibinfo {author} {\bibfnamefont {C.}~\bibnamefont {Weedbrook}}, \bibinfo {author} {\bibfnamefont {S.}~\bibnamefont {Pirandola}}, \bibinfo {author} {\bibfnamefont {R.}~\bibnamefont {{Garc{\'\i}a-Patr{\'o}n}}}, \bibinfo {author} {\bibfnamefont {N.~J.}\ \bibnamefont {Cerf}}, \bibinfo {author} {\bibfnamefont {T.~C.}\ \bibnamefont {Ralph}}, \bibinfo {author} {\bibfnamefont {J.~H.}\ \bibnamefont {Shapiro}},\ and\ \bibinfo {author} {\bibfnamefont {S.}~\bibnamefont {Lloyd}},\ }\href {https://doi.org/10.1103/RevModPhys.84.621} {\bibfield  {journal} {\bibinfo  {journal} {Rev. Mod. Phys.}\ }\textbf {\bibinfo {volume} {84}},\ \bibinfo {pages} {621} (\bibinfo {year} {2012})}\BibitemShut {NoStop}%
\bibitem [{\citenamefont {Adesso}\ \emph {et~al.}(2014)\citenamefont {Adesso}, \citenamefont {Ragy},\ and\ \citenamefont {Lee}}]{adesso2014continuous}%
  \BibitemOpen
  \bibfield  {author} {\bibinfo {author} {\bibfnamefont {G.}~\bibnamefont {Adesso}}, \bibinfo {author} {\bibfnamefont {S.}~\bibnamefont {Ragy}},\ and\ \bibinfo {author} {\bibfnamefont {A.~R.}\ \bibnamefont {Lee}},\ }\href {https://doi.org/10.1142/S1230161214400010} {\bibfield  {journal} {\bibinfo  {journal} {Open Systems \& Information Dynamics}\ }\textbf {\bibinfo {volume} {21}},\ \bibinfo {pages} {1440001} (\bibinfo {year} {2014})}\BibitemShut {NoStop}%
\bibitem [{\citenamefont {Bartlett}\ \emph {et~al.}(2002)\citenamefont {Bartlett}, \citenamefont {Sanders}, \citenamefont {Braunstein},\ and\ \citenamefont {Nemoto}}]{bartlett2002}%
  \BibitemOpen
  \bibfield  {author} {\bibinfo {author} {\bibfnamefont {S.~D.}\ \bibnamefont {Bartlett}}, \bibinfo {author} {\bibfnamefont {B.~C.}\ \bibnamefont {Sanders}}, \bibinfo {author} {\bibfnamefont {S.~L.}\ \bibnamefont {Braunstein}},\ and\ \bibinfo {author} {\bibfnamefont {K.}~\bibnamefont {Nemoto}},\ }\href {https://doi.org/10.1103/PhysRevLett.88.097904} {\bibfield  {journal} {\bibinfo  {journal} {Phys. Rev. Lett.}\ }\textbf {\bibinfo {volume} {88}},\ \bibinfo {pages} {097904} (\bibinfo {year} {2002})}\BibitemShut {NoStop}%
\bibitem [{\citenamefont {Gottesman}\ \emph {et~al.}(2001)\citenamefont {Gottesman}, \citenamefont {Kitaev},\ and\ \citenamefont {Preskill}}]{gottesman2001}%
  \BibitemOpen
  \bibfield  {author} {\bibinfo {author} {\bibfnamefont {D.}~\bibnamefont {Gottesman}}, \bibinfo {author} {\bibfnamefont {A.}~\bibnamefont {Kitaev}},\ and\ \bibinfo {author} {\bibfnamefont {J.}~\bibnamefont {Preskill}},\ }\href {https://doi.org/10.1103/PhysRevA.64.012310} {\bibfield  {journal} {\bibinfo  {journal} {Phys. Rev. A}\ }\textbf {\bibinfo {volume} {64}},\ \bibinfo {pages} {012310} (\bibinfo {year} {2001})}\BibitemShut {NoStop}%
\bibitem [{\citenamefont {Gu}\ \emph {et~al.}(2009)\citenamefont {Gu}, \citenamefont {Weedbrook}, \citenamefont {Menicucci}, \citenamefont {Ralph},\ and\ \citenamefont {{van Loock}}}]{gu2009}%
  \BibitemOpen
  \bibfield  {author} {\bibinfo {author} {\bibfnamefont {M.}~\bibnamefont {Gu}}, \bibinfo {author} {\bibfnamefont {C.}~\bibnamefont {Weedbrook}}, \bibinfo {author} {\bibfnamefont {N.~C.}\ \bibnamefont {Menicucci}}, \bibinfo {author} {\bibfnamefont {T.~C.}\ \bibnamefont {Ralph}},\ and\ \bibinfo {author} {\bibfnamefont {P.}~\bibnamefont {{van Loock}}},\ }\href {https://doi.org/10.1103/PhysRevA.79.062318} {\bibfield  {journal} {\bibinfo  {journal} {Phys. Rev. A}\ }\textbf {\bibinfo {volume} {79}},\ \bibinfo {pages} {062318} (\bibinfo {year} {2009})}\BibitemShut {NoStop}%
\bibitem [{\citenamefont {Ghose}\ and\ \citenamefont {Sanders}(2007)}]{ghose2007}%
  \BibitemOpen
  \bibfield  {author} {\bibinfo {author} {\bibfnamefont {S.}~\bibnamefont {Ghose}}\ and\ \bibinfo {author} {\bibfnamefont {B.~C.}\ \bibnamefont {Sanders}},\ }\href {https://doi.org/10.1080/09500340601101575} {\bibfield  {journal} {\bibinfo  {journal} {J. Mod. Opt.}\ }\textbf {\bibinfo {volume} {54}},\ \bibinfo {pages} {855} (\bibinfo {year} {2007})}\BibitemShut {NoStop}%
\bibitem [{\citenamefont {Miyata}\ \emph {et~al.}(2016)\citenamefont {Miyata}, \citenamefont {Ogawa}, \citenamefont {Marek}, \citenamefont {Filip}, \citenamefont {Yonezawa}, \citenamefont {Yoshikawa},\ and\ \citenamefont {Furusawa}}]{miyata2016}%
  \BibitemOpen
  \bibfield  {author} {\bibinfo {author} {\bibfnamefont {K.}~\bibnamefont {Miyata}}, \bibinfo {author} {\bibfnamefont {H.}~\bibnamefont {Ogawa}}, \bibinfo {author} {\bibfnamefont {P.}~\bibnamefont {Marek}}, \bibinfo {author} {\bibfnamefont {R.}~\bibnamefont {Filip}}, \bibinfo {author} {\bibfnamefont {H.}~\bibnamefont {Yonezawa}}, \bibinfo {author} {\bibfnamefont {J.-i.}\ \bibnamefont {Yoshikawa}},\ and\ \bibinfo {author} {\bibfnamefont {A.}~\bibnamefont {Furusawa}},\ }\href {https://doi.org/10.1103/PhysRevA.93.022301} {\bibfield  {journal} {\bibinfo  {journal} {Phys. Rev. A}\ }\textbf {\bibinfo {volume} {93}},\ \bibinfo {pages} {022301} (\bibinfo {year} {2016})}\BibitemShut {NoStop}%
\bibitem [{\citenamefont {Baragiola}\ \emph {et~al.}(2019)\citenamefont {Baragiola}, \citenamefont {Pantaleoni}, \citenamefont {Alexander}, \citenamefont {Karanjai},\ and\ \citenamefont {Menicucci}}]{baragiola2019}%
  \BibitemOpen
  \bibfield  {author} {\bibinfo {author} {\bibfnamefont {B.~Q.}\ \bibnamefont {Baragiola}}, \bibinfo {author} {\bibfnamefont {G.}~\bibnamefont {Pantaleoni}}, \bibinfo {author} {\bibfnamefont {R.~N.}\ \bibnamefont {Alexander}}, \bibinfo {author} {\bibfnamefont {A.}~\bibnamefont {Karanjai}},\ and\ \bibinfo {author} {\bibfnamefont {N.~C.}\ \bibnamefont {Menicucci}},\ }\href {https://doi.org/10.1103/PhysRevLett.123.200502} {\bibfield  {journal} {\bibinfo  {journal} {Phys. Rev. Lett.}\ }\textbf {\bibinfo {volume} {123}},\ \bibinfo {pages} {200502} (\bibinfo {year} {2019})}\BibitemShut {NoStop}%
\bibitem [{\citenamefont {Vasconcelos}\ \emph {et~al.}(2010)\citenamefont {Vasconcelos}, \citenamefont {Sanz},\ and\ \citenamefont {Glancy}}]{vasconcelos2010}%
  \BibitemOpen
  \bibfield  {author} {\bibinfo {author} {\bibfnamefont {H.~M.}\ \bibnamefont {Vasconcelos}}, \bibinfo {author} {\bibfnamefont {L.}~\bibnamefont {Sanz}},\ and\ \bibinfo {author} {\bibfnamefont {S.}~\bibnamefont {Glancy}},\ }\href {https://doi.org/10.1364/OL.35.003261} {\bibfield  {journal} {\bibinfo  {journal} {Optics Letters}\ }\textbf {\bibinfo {volume} {35}},\ \bibinfo {pages} {3261} (\bibinfo {year} {2010})}\BibitemShut {NoStop}%
\bibitem [{\citenamefont {Weigand}\ and\ \citenamefont {Terhal}(2018)}]{Weigand2018}%
  \BibitemOpen
  \bibfield  {author} {\bibinfo {author} {\bibfnamefont {D.~J.}\ \bibnamefont {Weigand}}\ and\ \bibinfo {author} {\bibfnamefont {B.~M.}\ \bibnamefont {Terhal}},\ }\href {https://doi.org/10.1103/PhysRevA.97.022341} {\bibfield  {journal} {\bibinfo  {journal} {Phys. Rev. A}\ }\textbf {\bibinfo {volume} {97}},\ \bibinfo {pages} {022341} (\bibinfo {year} {2018})}\BibitemShut {NoStop}%
\bibitem [{\citenamefont {Mari}\ and\ \citenamefont {Eisert}(2012)}]{mari2012}%
  \BibitemOpen
  \bibfield  {author} {\bibinfo {author} {\bibfnamefont {A.}~\bibnamefont {Mari}}\ and\ \bibinfo {author} {\bibfnamefont {J.}~\bibnamefont {Eisert}},\ }\href {https://doi.org/10.1103/PhysRevLett.109.230503} {\bibfield  {journal} {\bibinfo  {journal} {Phys. Rev. Lett.}\ }\textbf {\bibinfo {volume} {109}},\ \bibinfo {pages} {230503} (\bibinfo {year} {2012})}\BibitemShut {NoStop}%
\bibitem [{\citenamefont {Veitch}\ \emph {et~al.}(2013)\citenamefont {Veitch}, \citenamefont {Wiebe}, \citenamefont {Ferrie},\ and\ \citenamefont {Emerson}}]{veitch2013}%
  \BibitemOpen
  \bibfield  {author} {\bibinfo {author} {\bibfnamefont {V.}~\bibnamefont {Veitch}}, \bibinfo {author} {\bibfnamefont {N.}~\bibnamefont {Wiebe}}, \bibinfo {author} {\bibfnamefont {C.}~\bibnamefont {Ferrie}},\ and\ \bibinfo {author} {\bibfnamefont {J.}~\bibnamefont {Emerson}},\ }\href {https://doi.org/10.1088/1367-2630/15/1/013037} {\bibfield  {journal} {\bibinfo  {journal} {New J. Phys.}\ }\textbf {\bibinfo {volume} {15}},\ \bibinfo {pages} {013037} (\bibinfo {year} {2013})}\BibitemShut {NoStop}%
\bibitem [{\citenamefont {{Rahimi-Keshari}}\ \emph {et~al.}(2015)\citenamefont {{Rahimi-Keshari}}, \citenamefont {Lund},\ and\ \citenamefont {Ralph}}]{rahimi-keshari2015}%
  \BibitemOpen
  \bibfield  {author} {\bibinfo {author} {\bibfnamefont {S.}~\bibnamefont {{Rahimi-Keshari}}}, \bibinfo {author} {\bibfnamefont {A.~P.}\ \bibnamefont {Lund}},\ and\ \bibinfo {author} {\bibfnamefont {T.~C.}\ \bibnamefont {Ralph}},\ }\href {https://doi.org/10.1103/PhysRevLett.114.060501} {\bibfield  {journal} {\bibinfo  {journal} {Phys. Rev. Lett.}\ }\textbf {\bibinfo {volume} {114}},\ \bibinfo {pages} {060501} (\bibinfo {year} {2015})}\BibitemShut {NoStop}%
\bibitem [{\citenamefont {Albarelli}\ \emph {et~al.}(2018)\citenamefont {Albarelli}, \citenamefont {Genoni}, \citenamefont {Paris},\ and\ \citenamefont {Ferraro}}]{albarelli2018}%
  \BibitemOpen
  \bibfield  {author} {\bibinfo {author} {\bibfnamefont {F.}~\bibnamefont {Albarelli}}, \bibinfo {author} {\bibfnamefont {M.~G.}\ \bibnamefont {Genoni}}, \bibinfo {author} {\bibfnamefont {M.~G.~A.}\ \bibnamefont {Paris}},\ and\ \bibinfo {author} {\bibfnamefont {A.}~\bibnamefont {Ferraro}},\ }\href {https://doi.org/10.1103/PhysRevA.98.052350} {\bibfield  {journal} {\bibinfo  {journal} {Phys. Rev. A}\ }\textbf {\bibinfo {volume} {98}},\ \bibinfo {pages} {052350} (\bibinfo {year} {2018})}\BibitemShut {NoStop}%
\bibitem [{\citenamefont {Takagi}\ and\ \citenamefont {Zhuang}(2018)}]{takagi2018}%
  \BibitemOpen
  \bibfield  {author} {\bibinfo {author} {\bibfnamefont {R.}~\bibnamefont {Takagi}}\ and\ \bibinfo {author} {\bibfnamefont {Q.}~\bibnamefont {Zhuang}},\ }\href {https://doi.org/10.1103/PhysRevA.97.062337} {\bibfield  {journal} {\bibinfo  {journal} {Phys. Rev. A}\ }\textbf {\bibinfo {volume} {97}},\ \bibinfo {pages} {062337} (\bibinfo {year} {2018})}\BibitemShut {NoStop}%
\bibitem [{\citenamefont {Chabaud}\ \emph {et~al.}(2020)\citenamefont {Chabaud}, \citenamefont {Markham},\ and\ \citenamefont {Grosshans}}]{chabaud2020}%
  \BibitemOpen
  \bibfield  {author} {\bibinfo {author} {\bibfnamefont {U.}~\bibnamefont {Chabaud}}, \bibinfo {author} {\bibfnamefont {D.}~\bibnamefont {Markham}},\ and\ \bibinfo {author} {\bibfnamefont {F.}~\bibnamefont {Grosshans}},\ }\href {https://doi.org/10.1103/PhysRevLett.124.063605} {\bibfield  {journal} {\bibinfo  {journal} {Phys. Rev. Lett.}\ }\textbf {\bibinfo {volume} {124}},\ \bibinfo {pages} {063605} (\bibinfo {year} {2020})}\BibitemShut {NoStop}%
\bibitem [{Note1()}]{Note1}%
  \BibitemOpen
  \bibinfo {note} {We restrict to Gaussian operations parameterized by rational symplectic matrices $\protect \text {Sp}(2n,\protect \mathbb Q)$ and all real phase space displacements $\protect \text {HW}(n)$, which were shown to be simulatable in Ref.~\cite {calcluth2023}.}\BibitemShut {Stop}%
\bibitem [{\citenamefont {{Bermejo-Vega}}\ \emph {et~al.}(2016)\citenamefont {{Bermejo-Vega}}, \citenamefont {Lin},\ and\ \citenamefont {Van Den~Nest}}]{bermejo-vega2016}%
  \BibitemOpen
  \bibfield  {author} {\bibinfo {author} {\bibfnamefont {J.}~\bibnamefont {{Bermejo-Vega}}}, \bibinfo {author} {\bibfnamefont {C.~Y.-Y.}\ \bibnamefont {Lin}},\ and\ \bibinfo {author} {\bibfnamefont {M.}~\bibnamefont {Van Den~Nest}},\ }\href {https://doi.org/10.26421/QIC16.5-6-1} {\bibfield  {journal} {\bibinfo  {journal} {Quantum Inf. Comput.}\ }\textbf {\bibinfo {volume} {16}},\ \bibinfo {pages} {361} (\bibinfo {year} {2016})}\BibitemShut {NoStop}%
\bibitem [{\citenamefont {{Bermejo-Vega}}(2016)}]{Juani-thesis}%
  \BibitemOpen
  \bibfield  {author} {\bibinfo {author} {\bibfnamefont {J.}~\bibnamefont {{Bermejo-Vega}}},\ }\href@noop {} {\bibfield  {journal} {\bibinfo  {journal} {PhD Thesis, Technische Universit\"at M\"unchen Max-Planck-Institut f\"ur Quantenoptik}\ } (\bibinfo {year} {2016})},\ \bibinfo {note} {\href{https://arxiv.org/abs/1611.09274}{arXiv:1611.09274}}\BibitemShut {NoStop}%
\bibitem [{\citenamefont {Calcluth}\ \emph {et~al.}(2022)\citenamefont {Calcluth}, \citenamefont {Ferraro},\ and\ \citenamefont {Ferrini}}]{calcluth2022}%
  \BibitemOpen
  \bibfield  {author} {\bibinfo {author} {\bibfnamefont {C.}~\bibnamefont {Calcluth}}, \bibinfo {author} {\bibfnamefont {A.}~\bibnamefont {Ferraro}},\ and\ \bibinfo {author} {\bibfnamefont {G.}~\bibnamefont {Ferrini}},\ }\href {https://doi.org/10.22331/q-2022-12-01-867} {\bibfield  {journal} {\bibinfo  {journal} {Quantum}\ }\textbf {\bibinfo {volume} {6}},\ \bibinfo {pages} {867} (\bibinfo {year} {2022})}\BibitemShut {NoStop}%
\bibitem [{\citenamefont {Calcluth}\ \emph {et~al.}(2023)\citenamefont {Calcluth}, \citenamefont {Ferraro},\ and\ \citenamefont {Ferrini}}]{calcluth2023}%
  \BibitemOpen
  \bibfield  {author} {\bibinfo {author} {\bibfnamefont {C.}~\bibnamefont {Calcluth}}, \bibinfo {author} {\bibfnamefont {A.}~\bibnamefont {Ferraro}},\ and\ \bibinfo {author} {\bibfnamefont {G.}~\bibnamefont {Ferrini}},\ }\href {https://doi.org/10.1103/PhysRevA.107.062414} {\bibfield  {journal} {\bibinfo  {journal} {Phys. Rev. A}\ }\textbf {\bibinfo {volume} {107}},\ \bibinfo {pages} {062414} (\bibinfo {year} {2023})}\BibitemShut {NoStop}%
\bibitem [{\citenamefont {Pantaleoni}\ \emph {et~al.}(2020)\citenamefont {Pantaleoni}, \citenamefont {Baragiola},\ and\ \citenamefont {Menicucci}}]{pantaleoni2020}%
  \BibitemOpen
  \bibfield  {author} {\bibinfo {author} {\bibfnamefont {G.}~\bibnamefont {Pantaleoni}}, \bibinfo {author} {\bibfnamefont {B.~Q.}\ \bibnamefont {Baragiola}},\ and\ \bibinfo {author} {\bibfnamefont {N.~C.}\ \bibnamefont {Menicucci}},\ }\href {https://doi.org/10.1103/PhysRevLett.125.040501} {\bibfield  {journal} {\bibinfo  {journal} {Phys. Rev. Lett.}\ }\textbf {\bibinfo {volume} {125}},\ \bibinfo {pages} {040501} (\bibinfo {year} {2020})}\BibitemShut {NoStop}%
\bibitem [{\citenamefont {Pantaleoni}\ \emph {et~al.}(2021)\citenamefont {Pantaleoni}, \citenamefont {Baragiola},\ and\ \citenamefont {Menicucci}}]{pantaleoni2021}%
  \BibitemOpen
  \bibfield  {author} {\bibinfo {author} {\bibfnamefont {G.}~\bibnamefont {Pantaleoni}}, \bibinfo {author} {\bibfnamefont {B.~Q.}\ \bibnamefont {Baragiola}},\ and\ \bibinfo {author} {\bibfnamefont {N.~C.}\ \bibnamefont {Menicucci}},\ }\href {https://doi.org/10.1103/PhysRevA.104.012430} {\bibfield  {journal} {\bibinfo  {journal} {Phys. Rev. A}\ }\textbf {\bibinfo {volume} {104}},\ \bibinfo {pages} {012430} (\bibinfo {year} {2021})}\BibitemShut {NoStop}%
\bibitem [{\citenamefont {Shaw}\ \emph {et~al.}(2022)\citenamefont {Shaw}, \citenamefont {Doherty},\ and\ \citenamefont {Grimsmo}}]{shaw2022}%
  \BibitemOpen
  \bibfield  {author} {\bibinfo {author} {\bibfnamefont {M.~H.}\ \bibnamefont {Shaw}}, \bibinfo {author} {\bibfnamefont {A.~C.}\ \bibnamefont {Doherty}},\ and\ \bibinfo {author} {\bibfnamefont {A.~L.}\ \bibnamefont {Grimsmo}},\ }\Eprint {https://arxiv.org/abs/2210.14919} {arXiv:2210.14919}  (\bibinfo {year} {2022})\BibitemShut {NoStop}%
\bibitem [{\citenamefont {Tzitrin}\ \emph {et~al.}(2020)\citenamefont {Tzitrin}, \citenamefont {Bourassa}, \citenamefont {Menicucci},\ and\ \citenamefont {Sabapathy}}]{tzitrin2020}%
  \BibitemOpen
  \bibfield  {author} {\bibinfo {author} {\bibfnamefont {I.}~\bibnamefont {Tzitrin}}, \bibinfo {author} {\bibfnamefont {J.~E.}\ \bibnamefont {Bourassa}}, \bibinfo {author} {\bibfnamefont {N.~C.}\ \bibnamefont {Menicucci}},\ and\ \bibinfo {author} {\bibfnamefont {K.~K.}\ \bibnamefont {Sabapathy}},\ }\href {https://doi.org/10.1103/PhysRevA.101.032315} {\bibfield  {journal} {\bibinfo  {journal} {Phys. Rev. A}\ }\textbf {\bibinfo {volume} {101}},\ \bibinfo {pages} {032315} (\bibinfo {year} {2020})}\BibitemShut {NoStop}%
\bibitem [{\citenamefont {Konno}\ \emph {et~al.}(2021)\citenamefont {Konno}, \citenamefont {Asavanant}, \citenamefont {Fukui}, \citenamefont {Sakaguchi}, \citenamefont {Hanamura}, \citenamefont {Marek}, \citenamefont {Filip}, \citenamefont {Yoshikawa},\ and\ \citenamefont {Furusawa}}]{konno2021}%
  \BibitemOpen
  \bibfield  {author} {\bibinfo {author} {\bibfnamefont {S.}~\bibnamefont {Konno}}, \bibinfo {author} {\bibfnamefont {W.}~\bibnamefont {Asavanant}}, \bibinfo {author} {\bibfnamefont {K.}~\bibnamefont {Fukui}}, \bibinfo {author} {\bibfnamefont {A.}~\bibnamefont {Sakaguchi}}, \bibinfo {author} {\bibfnamefont {F.}~\bibnamefont {Hanamura}}, \bibinfo {author} {\bibfnamefont {P.}~\bibnamefont {Marek}}, \bibinfo {author} {\bibfnamefont {R.}~\bibnamefont {Filip}}, \bibinfo {author} {\bibfnamefont {J.-i.}\ \bibnamefont {Yoshikawa}},\ and\ \bibinfo {author} {\bibfnamefont {A.}~\bibnamefont {Furusawa}},\ }\href {https://doi.org/10.1103/PhysRevResearch.3.043026} {\bibfield  {journal} {\bibinfo  {journal} {Phys. Rev. Res.}\ }\textbf {\bibinfo {volume} {3}},\ \bibinfo {pages} {043026} (\bibinfo {year} {2021})}\BibitemShut {NoStop}%
\bibitem [{\citenamefont {Fukui}\ and\ \citenamefont {Takeda}(2022)}]{fukui2022}%
  \BibitemOpen
  \bibfield  {author} {\bibinfo {author} {\bibfnamefont {K.}~\bibnamefont {Fukui}}\ and\ \bibinfo {author} {\bibfnamefont {S.}~\bibnamefont {Takeda}},\ }\href {https://doi.org/10.1088/1361-6455/ac489c} {\bibfield  {journal} {\bibinfo  {journal} {Journal of Physics B: Atomic, Molecular and Optical Physics}\ }\textbf {\bibinfo {volume} {55}},\ \bibinfo {pages} {012001} (\bibinfo {year} {2022})}\BibitemShut {NoStop}%
\bibitem [{\citenamefont {Matsos}\ \emph {et~al.}(2023)\citenamefont {Matsos}, \citenamefont {Valahu}, \citenamefont {Navickas}, \citenamefont {Rao}, \citenamefont {Millican}, \citenamefont {Biercuk},\ and\ \citenamefont {Tan}}]{matsos2023robust}%
  \BibitemOpen
  \bibfield  {author} {\bibinfo {author} {\bibfnamefont {V.}~\bibnamefont {Matsos}}, \bibinfo {author} {\bibfnamefont {C.}~\bibnamefont {Valahu}}, \bibinfo {author} {\bibfnamefont {T.}~\bibnamefont {Navickas}}, \bibinfo {author} {\bibfnamefont {A.}~\bibnamefont {Rao}}, \bibinfo {author} {\bibfnamefont {M.}~\bibnamefont {Millican}}, \bibinfo {author} {\bibfnamefont {M.}~\bibnamefont {Biercuk}},\ and\ \bibinfo {author} {\bibfnamefont {T.}~\bibnamefont {Tan}},\ }\href@noop {} {\bibfield  {journal} {\bibinfo  {journal} {arXiv preprint arXiv:2310.15546}\ } (\bibinfo {year} {2023})}\BibitemShut {NoStop}%
\bibitem [{\citenamefont {Noh}\ \emph {et~al.}(2022{\natexlab{a}})\citenamefont {Noh}, \citenamefont {Chamberland},\ and\ \citenamefont {Brand{\~a}o}}]{noh2022low}%
  \BibitemOpen
  \bibfield  {author} {\bibinfo {author} {\bibfnamefont {K.}~\bibnamefont {Noh}}, \bibinfo {author} {\bibfnamefont {C.}~\bibnamefont {Chamberland}},\ and\ \bibinfo {author} {\bibfnamefont {F.~G.}\ \bibnamefont {Brand{\~a}o}},\ }\href {https://doi.org/10.1103/PRXQuantum.3.010315} {\bibfield  {journal} {\bibinfo  {journal} {PRX Quantum}\ }\textbf {\bibinfo {volume} {3}},\ \bibinfo {pages} {010315} (\bibinfo {year} {2022}{\natexlab{a}})}\BibitemShut {NoStop}%
\bibitem [{\citenamefont {Howard}\ and\ \citenamefont {Campbell}(2017)}]{howard2017}%
  \BibitemOpen
  \bibfield  {author} {\bibinfo {author} {\bibfnamefont {M.}~\bibnamefont {Howard}}\ and\ \bibinfo {author} {\bibfnamefont {E.}~\bibnamefont {Campbell}},\ }\href {https://doi.org/10.1103/PhysRevLett.118.090501} {\bibfield  {journal} {\bibinfo  {journal} {Phys. Rev. Lett.}\ }\textbf {\bibinfo {volume} {118}},\ \bibinfo {pages} {090501} (\bibinfo {year} {2017})}\BibitemShut {NoStop}%
\bibitem [{\citenamefont {Veitch}\ \emph {et~al.}(2014)\citenamefont {Veitch}, \citenamefont {Mousavian}, \citenamefont {Gottesman},\ and\ \citenamefont {Emerson}}]{veitch2014}%
  \BibitemOpen
  \bibfield  {author} {\bibinfo {author} {\bibfnamefont {V.}~\bibnamefont {Veitch}}, \bibinfo {author} {\bibfnamefont {S.~A.~H.}\ \bibnamefont {Mousavian}}, \bibinfo {author} {\bibfnamefont {D.}~\bibnamefont {Gottesman}},\ and\ \bibinfo {author} {\bibfnamefont {J.}~\bibnamefont {Emerson}},\ }\href {https://doi.org/10.1088/1367-2630/16/1/013009} {\bibfield  {journal} {\bibinfo  {journal} {New J. Phys.}\ }\textbf {\bibinfo {volume} {16}},\ \bibinfo {pages} {013009} (\bibinfo {year} {2014})}\BibitemShut {NoStop}%
\bibitem [{\citenamefont {Hahn}\ \emph {et~al.}(2022{\natexlab{a}})\citenamefont {Hahn}, \citenamefont {Ferraro}, \citenamefont {Hultquist}, \citenamefont {Ferrini},\ and\ \citenamefont {{Garc{\'\i}a-{\'A}lvarez}}}]{hahn2022}%
  \BibitemOpen
  \bibfield  {author} {\bibinfo {author} {\bibfnamefont {O.}~\bibnamefont {Hahn}}, \bibinfo {author} {\bibfnamefont {A.}~\bibnamefont {Ferraro}}, \bibinfo {author} {\bibfnamefont {L.}~\bibnamefont {Hultquist}}, \bibinfo {author} {\bibfnamefont {G.}~\bibnamefont {Ferrini}},\ and\ \bibinfo {author} {\bibfnamefont {L.}~\bibnamefont {{Garc{\'\i}a-{\'A}lvarez}}},\ }\href {https://doi.org/10.1103/PhysRevLett.128.210502} {\bibfield  {journal} {\bibinfo  {journal} {Phys. Rev. Lett.}\ }\textbf {\bibinfo {volume} {128}},\ \bibinfo {pages} {210502} (\bibinfo {year} {2022}{\natexlab{a}})}\BibitemShut {NoStop}%
\bibitem [{\citenamefont {Leone}\ \emph {et~al.}(2022)\citenamefont {Leone}, \citenamefont {Oliviero},\ and\ \citenamefont {Hamma}}]{leone2022}%
  \BibitemOpen
  \bibfield  {author} {\bibinfo {author} {\bibfnamefont {L.}~\bibnamefont {Leone}}, \bibinfo {author} {\bibfnamefont {S.~F.}\ \bibnamefont {Oliviero}},\ and\ \bibinfo {author} {\bibfnamefont {A.}~\bibnamefont {Hamma}},\ }\href {https://doi.org/10.1103/PhysRevLett.128.050402} {\bibfield  {journal} {\bibinfo  {journal} {Phys. Rev. Lett.}\ }\textbf {\bibinfo {volume} {128}},\ \bibinfo {pages} {050402} (\bibinfo {year} {2022})}\BibitemShut {NoStop}%
\bibitem [{\citenamefont {Gottesman}(1999)}]{gottesman1999}%
  \BibitemOpen
  \bibfield  {author} {\bibinfo {author} {\bibfnamefont {D.}~\bibnamefont {Gottesman}},\ }\href@noop {} {\emph {\bibinfo {title} {The {{Heisenberg}} Representation of Quantum Computers}}},\ edited by\ \bibinfo {editor} {\bibfnamefont {S.~P.}\ \bibnamefont {Corney}}, \bibinfo {editor} {\bibfnamefont {R.}~\bibnamefont {Delbourgo}},\ and\ \bibinfo {editor} {\bibfnamefont {P.~D.}\ \bibnamefont {Jarvis}},\ Group22: {{Proceedings}} of the {{XXII}} International Colloquium on Group Theoretical Methods in Physics\ (\bibinfo  {publisher} {{Cambridge, MA, International Press}},\ \bibinfo {year} {1999})\ pp.\ \bibinfo {pages} {32--43}\BibitemShut {NoStop}%
\bibitem [{\citenamefont {Kitaev}(1997)}]{kitaev1997}%
  \BibitemOpen
  \bibfield  {author} {\bibinfo {author} {\bibfnamefont {A.~Y.}\ \bibnamefont {Kitaev}},\ }\href {https://doi.org/10.1070/RM1997v052n06ABEH002155} {\bibfield  {journal} {\bibinfo  {journal} {Russian Mathematical Surveys}\ }\textbf {\bibinfo {volume} {52}},\ \bibinfo {pages} {1191} (\bibinfo {year} {1997})}\BibitemShut {NoStop}%
\bibitem [{\citenamefont {Seddon}\ \emph {et~al.}(2021)\citenamefont {Seddon}, \citenamefont {Regula}, \citenamefont {Pashayan}, \citenamefont {Ouyang},\ and\ \citenamefont {Campbell}}]{seddon2021}%
  \BibitemOpen
  \bibfield  {author} {\bibinfo {author} {\bibfnamefont {J.~R.}\ \bibnamefont {Seddon}}, \bibinfo {author} {\bibfnamefont {B.}~\bibnamefont {Regula}}, \bibinfo {author} {\bibfnamefont {H.}~\bibnamefont {Pashayan}}, \bibinfo {author} {\bibfnamefont {Y.}~\bibnamefont {Ouyang}},\ and\ \bibinfo {author} {\bibfnamefont {E.~T.}\ \bibnamefont {Campbell}},\ }\href {https://doi.org/10.1103/PRXQuantum.2.010345} {\bibfield  {journal} {\bibinfo  {journal} {PRX Quantum}\ }\textbf {\bibinfo {volume} {2}},\ \bibinfo {pages} {010345} (\bibinfo {year} {2021})}\BibitemShut {NoStop}%
\bibitem [{\citenamefont {{Garc{\'\i}a-{\'A}lvarez}}\ \emph {et~al.}(2020)\citenamefont {{Garc{\'\i}a-{\'A}lvarez}}, \citenamefont {Calcluth}, \citenamefont {Ferraro},\ and\ \citenamefont {Ferrini}}]{garcia-alvarez2020}%
  \BibitemOpen
  \bibfield  {author} {\bibinfo {author} {\bibfnamefont {L.}~\bibnamefont {{Garc{\'\i}a-{\'A}lvarez}}}, \bibinfo {author} {\bibfnamefont {C.}~\bibnamefont {Calcluth}}, \bibinfo {author} {\bibfnamefont {A.}~\bibnamefont {Ferraro}},\ and\ \bibinfo {author} {\bibfnamefont {G.}~\bibnamefont {Ferrini}},\ }\href {https://doi.org/10.1103/PhysRevResearch.2.043322} {\bibfield  {journal} {\bibinfo  {journal} {Phys. Rev. Research}\ }\textbf {\bibinfo {volume} {2}},\ \bibinfo {pages} {043322} (\bibinfo {year} {2020})}\BibitemShut {NoStop}%
\bibitem [{\citenamefont {Grimsmo}\ \emph {et~al.}(2020)\citenamefont {Grimsmo}, \citenamefont {Combes},\ and\ \citenamefont {Baragiola}}]{grimsmo2020}%
  \BibitemOpen
  \bibfield  {author} {\bibinfo {author} {\bibfnamefont {A.~L.}\ \bibnamefont {Grimsmo}}, \bibinfo {author} {\bibfnamefont {J.}~\bibnamefont {Combes}},\ and\ \bibinfo {author} {\bibfnamefont {B.~Q.}\ \bibnamefont {Baragiola}},\ }\href {https://doi.org/10.1103/PhysRevX.10.011058} {\bibfield  {journal} {\bibinfo  {journal} {Phys. Rev. X}\ }\textbf {\bibinfo {volume} {10}},\ \bibinfo {pages} {011058} (\bibinfo {year} {2020})}\BibitemShut {NoStop}%
\bibitem [{\citenamefont {Cochrane}\ \emph {et~al.}(1999)\citenamefont {Cochrane}, \citenamefont {Milburn},\ and\ \citenamefont {Munro}}]{cochrane1999}%
  \BibitemOpen
  \bibfield  {author} {\bibinfo {author} {\bibfnamefont {P.~T.}\ \bibnamefont {Cochrane}}, \bibinfo {author} {\bibfnamefont {G.~J.}\ \bibnamefont {Milburn}},\ and\ \bibinfo {author} {\bibfnamefont {W.~J.}\ \bibnamefont {Munro}},\ }\href {https://doi.org/10.1103/PhysRevA.59.2631} {\bibfield  {journal} {\bibinfo  {journal} {Phys. Rev. A}\ }\textbf {\bibinfo {volume} {59}},\ \bibinfo {pages} {2631} (\bibinfo {year} {1999})}\BibitemShut {NoStop}%
\bibitem [{\citenamefont {Lloyd}\ and\ \citenamefont {Braunstein}(1999)}]{lloyd1999}%
  \BibitemOpen
  \bibfield  {author} {\bibinfo {author} {\bibfnamefont {S.}~\bibnamefont {Lloyd}}\ and\ \bibinfo {author} {\bibfnamefont {S.~L.}\ \bibnamefont {Braunstein}},\ }\href {https://doi.org/10.1103/PhysRevLett.82.1784} {\bibfield  {journal} {\bibinfo  {journal} {Phys. Rev. Lett.}\ }\textbf {\bibinfo {volume} {82}},\ \bibinfo {pages} {1784} (\bibinfo {year} {1999})}\BibitemShut {NoStop}%
\bibitem [{\citenamefont {Budinger}\ \emph {et~al.}(2022)\citenamefont {Budinger}, \citenamefont {Furusawa},\ and\ \citenamefont {{van Loock}}}]{budinger2022}%
  \BibitemOpen
  \bibfield  {author} {\bibinfo {author} {\bibfnamefont {N.}~\bibnamefont {Budinger}}, \bibinfo {author} {\bibfnamefont {A.}~\bibnamefont {Furusawa}},\ and\ \bibinfo {author} {\bibfnamefont {P.}~\bibnamefont {{van Loock}}},\ }\Eprint {https://arxiv.org/abs/2211.09060} {arXiv:2211.09060}  (\bibinfo {year} {2022})\BibitemShut {NoStop}%
\bibitem [{\citenamefont {{Arvind}}\ \emph {et~al.}(1995)\citenamefont {{Arvind}}, \citenamefont {Dutta}, \citenamefont {Mukunda},\ and\ \citenamefont {Simon}}]{arvind1995}%
  \BibitemOpen
  \bibfield  {author} {\bibinfo {author} {\bibnamefont {{Arvind}}}, \bibinfo {author} {\bibfnamefont {B.}~\bibnamefont {Dutta}}, \bibinfo {author} {\bibfnamefont {N.}~\bibnamefont {Mukunda}},\ and\ \bibinfo {author} {\bibfnamefont {R.}~\bibnamefont {Simon}},\ }\href@noop {} {\bibfield  {journal} {\bibinfo  {journal} {Pramana J. Phys.}\ }\textbf {\bibinfo {volume} {45}},\ \bibinfo {pages} {441} (\bibinfo {year} {1995})}\BibitemShut {NoStop}%
\bibitem [{Note2()}]{Note2}%
  \BibitemOpen
  \bibinfo {note} {Note that the basis states are given in their unnormalized form for two reasons. First, except for the finite squeezing limit of $\Delta \to 0$, the normalization constants do not have a closed analytic form~\cite {gottesman2001}. Second, the encoded logical state is defined in terms of the unnormalized basis states and then the encoded state is normalized.}\BibitemShut {Stop}%
\bibitem [{\citenamefont {Matsuura}\ \emph {et~al.}(2020)\citenamefont {Matsuura}, \citenamefont {Yamasaki},\ and\ \citenamefont {Koashi}}]{matsuura2020}%
  \BibitemOpen
  \bibfield  {author} {\bibinfo {author} {\bibfnamefont {T.}~\bibnamefont {Matsuura}}, \bibinfo {author} {\bibfnamefont {H.}~\bibnamefont {Yamasaki}},\ and\ \bibinfo {author} {\bibfnamefont {M.}~\bibnamefont {Koashi}},\ }\href {https://doi.org/10.1103/PhysRevA.102.032408} {\bibfield  {journal} {\bibinfo  {journal} {Phys. Rev. A}\ }\textbf {\bibinfo {volume} {102}},\ \bibinfo {pages} {032408} (\bibinfo {year} {2020})}\BibitemShut {NoStop}%
\bibitem [{\citenamefont {Fl{\"u}hmann}\ \emph {et~al.}(2019)\citenamefont {Fl{\"u}hmann}, \citenamefont {Nguyen}, \citenamefont {Marinelli}, \citenamefont {Negnevitsky}, \citenamefont {Mehta},\ and\ \citenamefont {Home}}]{fluhmann2019}%
  \BibitemOpen
  \bibfield  {author} {\bibinfo {author} {\bibfnamefont {C.}~\bibnamefont {Fl{\"u}hmann}}, \bibinfo {author} {\bibfnamefont {T.~L.}\ \bibnamefont {Nguyen}}, \bibinfo {author} {\bibfnamefont {M.}~\bibnamefont {Marinelli}}, \bibinfo {author} {\bibfnamefont {V.}~\bibnamefont {Negnevitsky}}, \bibinfo {author} {\bibfnamefont {K.}~\bibnamefont {Mehta}},\ and\ \bibinfo {author} {\bibfnamefont {{\relax JP}.}~\bibnamefont {Home}},\ }\href {https://doi.org/10.1038/s41586-019-0960-6} {\bibfield  {journal} {\bibinfo  {journal} {Nature}\ }\textbf {\bibinfo {volume} {566}},\ \bibinfo {pages} {513} (\bibinfo {year} {2019})}\BibitemShut {NoStop}%
\bibitem [{\citenamefont {{Campagne-Ibarcq}}\ \emph {et~al.}(2020)\citenamefont {{Campagne-Ibarcq}}, \citenamefont {Eickbusch}, \citenamefont {Touzard}, \citenamefont {{Zalys-Geller}}, \citenamefont {Frattini}, \citenamefont {Sivak}, \citenamefont {Reinhold}, \citenamefont {Puri}, \citenamefont {Shankar}, \citenamefont {Schoelkopf}, \citenamefont {Frunzio}, \citenamefont {Mirrahimi},\ and\ \citenamefont {Devoret}}]{campagne-ibarcq2020}%
  \BibitemOpen
  \bibfield  {author} {\bibinfo {author} {\bibfnamefont {P.}~\bibnamefont {{Campagne-Ibarcq}}}, \bibinfo {author} {\bibfnamefont {A.}~\bibnamefont {Eickbusch}}, \bibinfo {author} {\bibfnamefont {S.}~\bibnamefont {Touzard}}, \bibinfo {author} {\bibfnamefont {E.}~\bibnamefont {{Zalys-Geller}}}, \bibinfo {author} {\bibfnamefont {{\relax NE}.}~\bibnamefont {Frattini}}, \bibinfo {author} {\bibfnamefont {{\relax VV}.}~\bibnamefont {Sivak}}, \bibinfo {author} {\bibfnamefont {P.}~\bibnamefont {Reinhold}}, \bibinfo {author} {\bibfnamefont {S.}~\bibnamefont {Puri}}, \bibinfo {author} {\bibfnamefont {S.}~\bibnamefont {Shankar}}, \bibinfo {author} {\bibfnamefont {{\relax RJ}.}~\bibnamefont {Schoelkopf}}, \bibinfo {author} {\bibfnamefont {L.}~\bibnamefont {Frunzio}}, \bibinfo {author} {\bibfnamefont {M.}~\bibnamefont {Mirrahimi}},\ and\ \bibinfo {author} {\bibfnamefont {M.~H.}\ \bibnamefont {Devoret}},\ }\href {https://doi.org/10.1038/s41586-020-2603-3} {\bibfield  {journal} {\bibinfo  {journal} {Nature}\ }\textbf
  {\bibinfo {volume} {584}},\ \bibinfo {pages} {368} (\bibinfo {year} {2020})}\BibitemShut {NoStop}%
\bibitem [{\citenamefont {Kudra}\ \emph {et~al.}(2022)\citenamefont {Kudra}, \citenamefont {Kervinen}, \citenamefont {Strandberg}, \citenamefont {Ahmed}, \citenamefont {Scigliuzzo}, \citenamefont {Osman}, \citenamefont {Lozano}, \citenamefont {Thol{\'e}n}, \citenamefont {Borgani}, \citenamefont {Haviland} \emph {et~al.}}]{kudra2022}%
  \BibitemOpen
  \bibfield  {author} {\bibinfo {author} {\bibfnamefont {M.}~\bibnamefont {Kudra}}, \bibinfo {author} {\bibfnamefont {M.}~\bibnamefont {Kervinen}}, \bibinfo {author} {\bibfnamefont {I.}~\bibnamefont {Strandberg}}, \bibinfo {author} {\bibfnamefont {S.}~\bibnamefont {Ahmed}}, \bibinfo {author} {\bibfnamefont {M.}~\bibnamefont {Scigliuzzo}}, \bibinfo {author} {\bibfnamefont {A.}~\bibnamefont {Osman}}, \bibinfo {author} {\bibfnamefont {D.~P.}\ \bibnamefont {Lozano}}, \bibinfo {author} {\bibfnamefont {M.~O.}\ \bibnamefont {Thol{\'e}n}}, \bibinfo {author} {\bibfnamefont {R.}~\bibnamefont {Borgani}}, \bibinfo {author} {\bibfnamefont {D.~B.}\ \bibnamefont {Haviland}}, \emph {et~al.},\ }\href {https://doi.org/10.1103/PRXQuantum.3.030301} {\bibfield  {journal} {\bibinfo  {journal} {PRX Quantum}\ }\textbf {\bibinfo {volume} {3}},\ \bibinfo {pages} {030301} (\bibinfo {year} {2022})}\BibitemShut {NoStop}%
\bibitem [{\citenamefont {Konno}\ \emph {et~al.}(2023)\citenamefont {Konno}, \citenamefont {Asavanant}, \citenamefont {Hanamura}, \citenamefont {Nagayoshi}, \citenamefont {Fukui}, \citenamefont {Sakaguchi}, \citenamefont {Ide}, \citenamefont {China}, \citenamefont {Yabuno}, \citenamefont {Miki}, \citenamefont {Terai}, \citenamefont {Takase}, \citenamefont {Endo}, \citenamefont {Marek}, \citenamefont {Filip}, \citenamefont {{van Loock}},\ and\ \citenamefont {Furusawa}}]{konno2023}%
  \BibitemOpen
  \bibfield  {author} {\bibinfo {author} {\bibfnamefont {S.}~\bibnamefont {Konno}}, \bibinfo {author} {\bibfnamefont {W.}~\bibnamefont {Asavanant}}, \bibinfo {author} {\bibfnamefont {F.}~\bibnamefont {Hanamura}}, \bibinfo {author} {\bibfnamefont {H.}~\bibnamefont {Nagayoshi}}, \bibinfo {author} {\bibfnamefont {K.}~\bibnamefont {Fukui}}, \bibinfo {author} {\bibfnamefont {A.}~\bibnamefont {Sakaguchi}}, \bibinfo {author} {\bibfnamefont {R.}~\bibnamefont {Ide}}, \bibinfo {author} {\bibfnamefont {F.}~\bibnamefont {China}}, \bibinfo {author} {\bibfnamefont {M.}~\bibnamefont {Yabuno}}, \bibinfo {author} {\bibfnamefont {S.}~\bibnamefont {Miki}}, \bibinfo {author} {\bibfnamefont {H.}~\bibnamefont {Terai}}, \bibinfo {author} {\bibfnamefont {K.}~\bibnamefont {Takase}}, \bibinfo {author} {\bibfnamefont {M.}~\bibnamefont {Endo}}, \bibinfo {author} {\bibfnamefont {P.}~\bibnamefont {Marek}}, \bibinfo {author} {\bibfnamefont {R.}~\bibnamefont {Filip}}, \bibinfo {author} {\bibfnamefont {P.}~\bibnamefont {{van Loock}}},\ and\
  \bibinfo {author} {\bibfnamefont {A.}~\bibnamefont {Furusawa}},\ }\href@noop {} {} (\bibinfo {year} {2023}),\ \Eprint {https://arxiv.org/abs/2309.02306} {arxiv:2309.02306 [quant-ph]} \BibitemShut {NoStop}%
\bibitem [{\citenamefont {Mirrahimi}\ \emph {et~al.}(2014)\citenamefont {Mirrahimi}, \citenamefont {Leghtas}, \citenamefont {Albert}, \citenamefont {Touzard}, \citenamefont {Schoelkopf}, \citenamefont {Jiang},\ and\ \citenamefont {Devoret}}]{mirrahimi2014}%
  \BibitemOpen
  \bibfield  {author} {\bibinfo {author} {\bibfnamefont {M.}~\bibnamefont {Mirrahimi}}, \bibinfo {author} {\bibfnamefont {Z.}~\bibnamefont {Leghtas}}, \bibinfo {author} {\bibfnamefont {V.~V.}\ \bibnamefont {Albert}}, \bibinfo {author} {\bibfnamefont {S.}~\bibnamefont {Touzard}}, \bibinfo {author} {\bibfnamefont {R.~J.}\ \bibnamefont {Schoelkopf}}, \bibinfo {author} {\bibfnamefont {L.}~\bibnamefont {Jiang}},\ and\ \bibinfo {author} {\bibfnamefont {M.~H.}\ \bibnamefont {Devoret}},\ }\href {https://doi.org/10.1088/1367-2630/16/4/045014} {\bibfield  {journal} {\bibinfo  {journal} {New J. Phys.}\ }\textbf {\bibinfo {volume} {16}},\ \bibinfo {pages} {045014} (\bibinfo {year} {2014})}\BibitemShut {NoStop}%
\bibitem [{Note3()}]{Note3}%
  \BibitemOpen
  \bibinfo {note} {As is the case for GKP states, we provide the unnormalized basis states because the encoded qubit state is defined in terms of the two unnormalized basis states.}\BibitemShut {Stop}%
\bibitem [{\citenamefont {Lund}\ \emph {et~al.}(2008)\citenamefont {Lund}, \citenamefont {Ralph},\ and\ \citenamefont {Haselgrove}}]{lund2008}%
  \BibitemOpen
  \bibfield  {author} {\bibinfo {author} {\bibfnamefont {A.~P.}\ \bibnamefont {Lund}}, \bibinfo {author} {\bibfnamefont {T.~C.}\ \bibnamefont {Ralph}},\ and\ \bibinfo {author} {\bibfnamefont {H.~L.}\ \bibnamefont {Haselgrove}},\ }\href {https://doi.org/10.1103/PhysRevLett.100.030503} {\bibfield  {journal} {\bibinfo  {journal} {Phys. Rev. Lett.}\ }\textbf {\bibinfo {volume} {100}},\ \bibinfo {pages} {030503} (\bibinfo {year} {2008})}\BibitemShut {NoStop}%
\bibitem [{\citenamefont {Ralph}\ \emph {et~al.}(2003)\citenamefont {Ralph}, \citenamefont {Gilchrist}, \citenamefont {Milburn}, \citenamefont {Munro},\ and\ \citenamefont {Glancy}}]{ralph2003}%
  \BibitemOpen
  \bibfield  {author} {\bibinfo {author} {\bibfnamefont {T.~C.}\ \bibnamefont {Ralph}}, \bibinfo {author} {\bibfnamefont {A.}~\bibnamefont {Gilchrist}}, \bibinfo {author} {\bibfnamefont {G.~J.}\ \bibnamefont {Milburn}}, \bibinfo {author} {\bibfnamefont {W.~J.}\ \bibnamefont {Munro}},\ and\ \bibinfo {author} {\bibfnamefont {S.}~\bibnamefont {Glancy}},\ }\href {https://doi.org/10.1103/PhysRevA.68.042319} {\bibfield  {journal} {\bibinfo  {journal} {Phys. Rev. A}\ }\textbf {\bibinfo {volume} {68}},\ \bibinfo {pages} {042319} (\bibinfo {year} {2003})}\BibitemShut {NoStop}%
\bibitem [{\citenamefont {Touzard}\ \emph {et~al.}(2018)\citenamefont {Touzard}, \citenamefont {Grimm}, \citenamefont {Leghtas}, \citenamefont {Mundhada}, \citenamefont {Reinhold}, \citenamefont {Axline}, \citenamefont {Reagor}, \citenamefont {Chou}, \citenamefont {Blumoff}, \citenamefont {Sliwa}, \citenamefont {Shankar}, \citenamefont {Frunzio}, \citenamefont {Schoelkopf}, \citenamefont {Mirrahimi},\ and\ \citenamefont {Devoret}}]{touzard2018}%
  \BibitemOpen
  \bibfield  {author} {\bibinfo {author} {\bibfnamefont {S.}~\bibnamefont {Touzard}}, \bibinfo {author} {\bibfnamefont {A.}~\bibnamefont {Grimm}}, \bibinfo {author} {\bibfnamefont {Z.}~\bibnamefont {Leghtas}}, \bibinfo {author} {\bibfnamefont {S.~O.}\ \bibnamefont {Mundhada}}, \bibinfo {author} {\bibfnamefont {P.}~\bibnamefont {Reinhold}}, \bibinfo {author} {\bibfnamefont {C.}~\bibnamefont {Axline}}, \bibinfo {author} {\bibfnamefont {M.}~\bibnamefont {Reagor}}, \bibinfo {author} {\bibfnamefont {K.}~\bibnamefont {Chou}}, \bibinfo {author} {\bibfnamefont {J.}~\bibnamefont {Blumoff}}, \bibinfo {author} {\bibfnamefont {K.~M.}\ \bibnamefont {Sliwa}}, \bibinfo {author} {\bibfnamefont {S.}~\bibnamefont {Shankar}}, \bibinfo {author} {\bibfnamefont {L.}~\bibnamefont {Frunzio}}, \bibinfo {author} {\bibfnamefont {R.~J.}\ \bibnamefont {Schoelkopf}}, \bibinfo {author} {\bibfnamefont {M.}~\bibnamefont {Mirrahimi}},\ and\ \bibinfo {author} {\bibfnamefont {M.~H.}\ \bibnamefont {Devoret}},\ }\href
  {https://doi.org/10.1103/PhysRevX.8.021005} {\bibfield  {journal} {\bibinfo  {journal} {Phys. Rev. X}\ }\textbf {\bibinfo {volume} {8}},\ \bibinfo {pages} {021005} (\bibinfo {year} {2018})}\BibitemShut {NoStop}%
\bibitem [{\citenamefont {Lewenstein}\ \emph {et~al.}(2021)\citenamefont {Lewenstein}, \citenamefont {Ciappina}, \citenamefont {Pisanty}, \citenamefont {{Rivera-Dean}}, \citenamefont {Stammer}, \citenamefont {Lamprou},\ and\ \citenamefont {Tzallas}}]{lewenstein2021}%
  \BibitemOpen
  \bibfield  {author} {\bibinfo {author} {\bibfnamefont {M.}~\bibnamefont {Lewenstein}}, \bibinfo {author} {\bibfnamefont {{\relax MF}.}~\bibnamefont {Ciappina}}, \bibinfo {author} {\bibfnamefont {E.}~\bibnamefont {Pisanty}}, \bibinfo {author} {\bibfnamefont {J.}~\bibnamefont {{Rivera-Dean}}}, \bibinfo {author} {\bibfnamefont {P.}~\bibnamefont {Stammer}}, \bibinfo {author} {\bibfnamefont {T.}~\bibnamefont {Lamprou}},\ and\ \bibinfo {author} {\bibfnamefont {P.}~\bibnamefont {Tzallas}},\ }\href {https://doi.org/10.1038/s41567-021-01317-w} {\bibfield  {journal} {\bibinfo  {journal} {Nature Physics}\ }\textbf {\bibinfo {volume} {17}},\ \bibinfo {pages} {1104} (\bibinfo {year} {2021})}\BibitemShut {NoStop}%
\bibitem [{\citenamefont {Grimm}\ \emph {et~al.}(2020)\citenamefont {Grimm}, \citenamefont {Frattini}, \citenamefont {Puri}, \citenamefont {Mundhada}, \citenamefont {Touzard}, \citenamefont {Mirrahimi}, \citenamefont {Girvin}, \citenamefont {Shankar},\ and\ \citenamefont {Devoret}}]{grimm2020}%
  \BibitemOpen
  \bibfield  {author} {\bibinfo {author} {\bibfnamefont {A.}~\bibnamefont {Grimm}}, \bibinfo {author} {\bibfnamefont {N.~E.}\ \bibnamefont {Frattini}}, \bibinfo {author} {\bibfnamefont {S.}~\bibnamefont {Puri}}, \bibinfo {author} {\bibfnamefont {S.~O.}\ \bibnamefont {Mundhada}}, \bibinfo {author} {\bibfnamefont {S.}~\bibnamefont {Touzard}}, \bibinfo {author} {\bibfnamefont {M.}~\bibnamefont {Mirrahimi}}, \bibinfo {author} {\bibfnamefont {S.~M.}\ \bibnamefont {Girvin}}, \bibinfo {author} {\bibfnamefont {S.}~\bibnamefont {Shankar}},\ and\ \bibinfo {author} {\bibfnamefont {M.~H.}\ \bibnamefont {Devoret}},\ }\href {https://doi.org/10.1038/s41586-020-2587-z} {\bibfield  {journal} {\bibinfo  {journal} {Nature}\ }\textbf {\bibinfo {volume} {584}},\ \bibinfo {pages} {205} (\bibinfo {year} {2020})}\BibitemShut {NoStop}%
\bibitem [{\citenamefont {Leghtas}\ \emph {et~al.}(2015)\citenamefont {Leghtas}, \citenamefont {Touzard}, \citenamefont {Pop}, \citenamefont {Kou}, \citenamefont {Vlastakis}, \citenamefont {Petrenko}, \citenamefont {Sliwa}, \citenamefont {Narla}, \citenamefont {Shankar}, \citenamefont {Hatridge}, \citenamefont {Reagor}, \citenamefont {Frunzio}, \citenamefont {Schoelkopf}, \citenamefont {Mirrahimi},\ and\ \citenamefont {{M. H. Devoret}}}]{leghtas2015}%
  \BibitemOpen
  \bibfield  {author} {\bibinfo {author} {\bibfnamefont {Z.}~\bibnamefont {Leghtas}}, \bibinfo {author} {\bibfnamefont {S.}~\bibnamefont {Touzard}}, \bibinfo {author} {\bibfnamefont {I.~M.}\ \bibnamefont {Pop}}, \bibinfo {author} {\bibfnamefont {A.}~\bibnamefont {Kou}}, \bibinfo {author} {\bibfnamefont {B.}~\bibnamefont {Vlastakis}}, \bibinfo {author} {\bibfnamefont {A.}~\bibnamefont {Petrenko}}, \bibinfo {author} {\bibfnamefont {K.~M.}\ \bibnamefont {Sliwa}}, \bibinfo {author} {\bibfnamefont {A.}~\bibnamefont {Narla}}, \bibinfo {author} {\bibfnamefont {S.}~\bibnamefont {Shankar}}, \bibinfo {author} {\bibfnamefont {M.~J.}\ \bibnamefont {Hatridge}}, \bibinfo {author} {\bibfnamefont {M.}~\bibnamefont {Reagor}}, \bibinfo {author} {\bibfnamefont {L.}~\bibnamefont {Frunzio}}, \bibinfo {author} {\bibfnamefont {R.~J.}\ \bibnamefont {Schoelkopf}}, \bibinfo {author} {\bibfnamefont {M.}~\bibnamefont {Mirrahimi}},\ and\ \bibinfo {author} {\bibnamefont {{M. H. Devoret}}},\ }\href {https://doi.org/10.1126/science.aaa2085}
  {\bibfield  {journal} {\bibinfo  {journal} {Science}\ }\textbf {\bibinfo {volume} {347}},\ \bibinfo {pages} {853} (\bibinfo {year} {2015})}\BibitemShut {NoStop}%
\bibitem [{\citenamefont {Ofek}\ \emph {et~al.}(2016)\citenamefont {Ofek}, \citenamefont {Petrenko}, \citenamefont {Heeres}, \citenamefont {Reinhold}, \citenamefont {Leghtas}, \citenamefont {Vlastakis}, \citenamefont {Liu}, \citenamefont {Frunzio}, \citenamefont {Girvin}, \citenamefont {Jiang}, \citenamefont {Mirrahimi}, \citenamefont {Devoret},\ and\ \citenamefont {Schoelkopf}}]{ofek2016}%
  \BibitemOpen
  \bibfield  {author} {\bibinfo {author} {\bibfnamefont {N.}~\bibnamefont {Ofek}}, \bibinfo {author} {\bibfnamefont {A.}~\bibnamefont {Petrenko}}, \bibinfo {author} {\bibfnamefont {R.}~\bibnamefont {Heeres}}, \bibinfo {author} {\bibfnamefont {P.}~\bibnamefont {Reinhold}}, \bibinfo {author} {\bibfnamefont {Z.}~\bibnamefont {Leghtas}}, \bibinfo {author} {\bibfnamefont {B.}~\bibnamefont {Vlastakis}}, \bibinfo {author} {\bibfnamefont {Y.}~\bibnamefont {Liu}}, \bibinfo {author} {\bibfnamefont {L.}~\bibnamefont {Frunzio}}, \bibinfo {author} {\bibfnamefont {S.~M.}\ \bibnamefont {Girvin}}, \bibinfo {author} {\bibfnamefont {L.}~\bibnamefont {Jiang}}, \bibinfo {author} {\bibfnamefont {M.}~\bibnamefont {Mirrahimi}}, \bibinfo {author} {\bibfnamefont {M.~H.}\ \bibnamefont {Devoret}},\ and\ \bibinfo {author} {\bibfnamefont {R.~J.}\ \bibnamefont {Schoelkopf}},\ }\href {https://doi.org/10.1038/nature18949} {\bibfield  {journal} {\bibinfo  {journal} {Nature}\ }\textbf {\bibinfo {volume} {536}},\ \bibinfo {pages} {441}
  (\bibinfo {year} {2016})}\BibitemShut {NoStop}%
\bibitem [{\citenamefont {Gertler}\ \emph {et~al.}(2021)\citenamefont {Gertler}, \citenamefont {Baker}, \citenamefont {Li}, \citenamefont {Shirol}, \citenamefont {Koch},\ and\ \citenamefont {Wang}}]{gertler2021}%
  \BibitemOpen
  \bibfield  {author} {\bibinfo {author} {\bibfnamefont {J.~M.}\ \bibnamefont {Gertler}}, \bibinfo {author} {\bibfnamefont {B.}~\bibnamefont {Baker}}, \bibinfo {author} {\bibfnamefont {J.}~\bibnamefont {Li}}, \bibinfo {author} {\bibfnamefont {S.}~\bibnamefont {Shirol}}, \bibinfo {author} {\bibfnamefont {J.}~\bibnamefont {Koch}},\ and\ \bibinfo {author} {\bibfnamefont {C.}~\bibnamefont {Wang}},\ }\href {https://doi.org/10.1038/s41586-021-03257-0} {\bibfield  {journal} {\bibinfo  {journal} {Nature}\ }\textbf {\bibinfo {volume} {590}},\ \bibinfo {pages} {243} (\bibinfo {year} {2021})}\BibitemShut {NoStop}%
\bibitem [{\citenamefont {Braunstein}\ and\ \citenamefont {Van~Loock}(2005)}]{braunstein05}%
  \BibitemOpen
  \bibfield  {author} {\bibinfo {author} {\bibfnamefont {S.~L.}\ \bibnamefont {Braunstein}}\ and\ \bibinfo {author} {\bibfnamefont {P.}~\bibnamefont {Van~Loock}},\ }\href {https://doi.org/10.1103/RevModPhys.77.513} {\bibfield  {journal} {\bibinfo  {journal} {Rev. Mod. Phys.}\ }\textbf {\bibinfo {volume} {77}},\ \bibinfo {pages} {513} (\bibinfo {year} {2005})}\BibitemShut {NoStop}%
\bibitem [{\citenamefont {Sakaguchi}\ \emph {et~al.}(2023)\citenamefont {Sakaguchi}, \citenamefont {Konno}, \citenamefont {Hanamura}, \citenamefont {Asavanant}, \citenamefont {Takase}, \citenamefont {Ogawa}, \citenamefont {Marek}, \citenamefont {Filip}, \citenamefont {Yoshikawa}, \citenamefont {Huntington}, \citenamefont {Yonezawa},\ and\ \citenamefont {Furusawa}}]{sakaguchi2023}%
  \BibitemOpen
  \bibfield  {author} {\bibinfo {author} {\bibfnamefont {A.}~\bibnamefont {Sakaguchi}}, \bibinfo {author} {\bibfnamefont {S.}~\bibnamefont {Konno}}, \bibinfo {author} {\bibfnamefont {F.}~\bibnamefont {Hanamura}}, \bibinfo {author} {\bibfnamefont {W.}~\bibnamefont {Asavanant}}, \bibinfo {author} {\bibfnamefont {K.}~\bibnamefont {Takase}}, \bibinfo {author} {\bibfnamefont {H.}~\bibnamefont {Ogawa}}, \bibinfo {author} {\bibfnamefont {P.}~\bibnamefont {Marek}}, \bibinfo {author} {\bibfnamefont {R.}~\bibnamefont {Filip}}, \bibinfo {author} {\bibfnamefont {J.-i.}\ \bibnamefont {Yoshikawa}}, \bibinfo {author} {\bibfnamefont {E.}~\bibnamefont {Huntington}}, \bibinfo {author} {\bibfnamefont {H.}~\bibnamefont {Yonezawa}},\ and\ \bibinfo {author} {\bibfnamefont {A.}~\bibnamefont {Furusawa}},\ }\href {https://doi.org/10.1038/s41467-023-39195-w} {\bibfield  {journal} {\bibinfo  {journal} {Nat Commun}\ }\textbf {\bibinfo {volume} {14}},\ \bibinfo {pages} {3817} (\bibinfo {year} {2023})}\BibitemShut {NoStop}%
\bibitem [{\citenamefont {Houhou}\ \emph {et~al.}(2022)\citenamefont {Houhou}, \citenamefont {Moore}, \citenamefont {Bose},\ and\ \citenamefont {Ferraro}}]{houhou2022unconditional}%
  \BibitemOpen
  \bibfield  {author} {\bibinfo {author} {\bibfnamefont {O.}~\bibnamefont {Houhou}}, \bibinfo {author} {\bibfnamefont {D.~W.}\ \bibnamefont {Moore}}, \bibinfo {author} {\bibfnamefont {S.}~\bibnamefont {Bose}},\ and\ \bibinfo {author} {\bibfnamefont {A.}~\bibnamefont {Ferraro}},\ }\href {https://doi.org/10.1103/PhysRevA.105.012610} {\bibfield  {journal} {\bibinfo  {journal} {Physical Review A}\ }\textbf {\bibinfo {volume} {105}},\ \bibinfo {pages} {012610} (\bibinfo {year} {2022})}\BibitemShut {NoStop}%
\bibitem [{\citenamefont {Zheng}\ \emph {et~al.}(2021)\citenamefont {Zheng}, \citenamefont {Hahn}, \citenamefont {Stadler}, \citenamefont {Holmvall}, \citenamefont {Quijandr{\'\i}a}, \citenamefont {Ferraro},\ and\ \citenamefont {Ferrini}}]{zheng2021gaussian}%
  \BibitemOpen
  \bibfield  {author} {\bibinfo {author} {\bibfnamefont {Y.}~\bibnamefont {Zheng}}, \bibinfo {author} {\bibfnamefont {O.}~\bibnamefont {Hahn}}, \bibinfo {author} {\bibfnamefont {P.}~\bibnamefont {Stadler}}, \bibinfo {author} {\bibfnamefont {P.}~\bibnamefont {Holmvall}}, \bibinfo {author} {\bibfnamefont {F.}~\bibnamefont {Quijandr{\'\i}a}}, \bibinfo {author} {\bibfnamefont {A.}~\bibnamefont {Ferraro}},\ and\ \bibinfo {author} {\bibfnamefont {G.}~\bibnamefont {Ferrini}},\ }\href {https://doi.org/10.1103/PRXQuantum.2.010327} {\bibfield  {journal} {\bibinfo  {journal} {PRX Quantum}\ }\textbf {\bibinfo {volume} {2}},\ \bibinfo {pages} {010327} (\bibinfo {year} {2021})}\BibitemShut {NoStop}%
\bibitem [{\citenamefont {Hahn}\ \emph {et~al.}(2022{\natexlab{b}})\citenamefont {Hahn}, \citenamefont {Holmvall}, \citenamefont {Stadler}, \citenamefont {Ferrini},\ and\ \citenamefont {Ferraro}}]{hahn2022deterministic}%
  \BibitemOpen
  \bibfield  {author} {\bibinfo {author} {\bibfnamefont {O.}~\bibnamefont {Hahn}}, \bibinfo {author} {\bibfnamefont {P.}~\bibnamefont {Holmvall}}, \bibinfo {author} {\bibfnamefont {P.}~\bibnamefont {Stadler}}, \bibinfo {author} {\bibfnamefont {G.}~\bibnamefont {Ferrini}},\ and\ \bibinfo {author} {\bibfnamefont {A.}~\bibnamefont {Ferraro}},\ }\href {https://doi.org/10.1103/PhysRevA.105.062446} {\bibfield  {journal} {\bibinfo  {journal} {Physical Review A}\ }\textbf {\bibinfo {volume} {105}},\ \bibinfo {pages} {062446} (\bibinfo {year} {2022}{\natexlab{b}})}\BibitemShut {NoStop}%
\bibitem [{\citenamefont {Aharonov}\ \emph {et~al.}(1969)\citenamefont {Aharonov}, \citenamefont {Pendleton},\ and\ \citenamefont {Petersen}}]{aharonov1969}%
  \BibitemOpen
  \bibfield  {author} {\bibinfo {author} {\bibfnamefont {Y.}~\bibnamefont {Aharonov}}, \bibinfo {author} {\bibfnamefont {H.}~\bibnamefont {Pendleton}},\ and\ \bibinfo {author} {\bibfnamefont {A.}~\bibnamefont {Petersen}},\ }\href {https://doi.org/10.1007/BF00670008} {\bibfield  {journal} {\bibinfo  {journal} {Int. J. Theor. Phys}\ }\textbf {\bibinfo {volume} {2}},\ \bibinfo {pages} {213} (\bibinfo {year} {1969})}\BibitemShut {NoStop}%
\bibitem [{\citenamefont {Ketterer}\ \emph {et~al.}(2016)\citenamefont {Ketterer}, \citenamefont {Keller}, \citenamefont {Walborn}, \citenamefont {Coudreau},\ and\ \citenamefont {Milman}}]{ketterer2016}%
  \BibitemOpen
  \bibfield  {author} {\bibinfo {author} {\bibfnamefont {A.}~\bibnamefont {Ketterer}}, \bibinfo {author} {\bibfnamefont {A.}~\bibnamefont {Keller}}, \bibinfo {author} {\bibfnamefont {{\relax SP}.}~\bibnamefont {Walborn}}, \bibinfo {author} {\bibfnamefont {T.}~\bibnamefont {Coudreau}},\ and\ \bibinfo {author} {\bibfnamefont {P.}~\bibnamefont {Milman}},\ }\href {https://doi.org/10.1103/PhysRevA.94.022325} {\bibfield  {journal} {\bibinfo  {journal} {Phys. Rev. A}\ }\textbf {\bibinfo {volume} {94}},\ \bibinfo {pages} {022325} (\bibinfo {year} {2016})}\BibitemShut {NoStop}%
\bibitem [{\citenamefont {Clauser}\ \emph {et~al.}(1969)\citenamefont {Clauser}, \citenamefont {Horne}, \citenamefont {Shimony},\ and\ \citenamefont {Holt}}]{clauser1969}%
  \BibitemOpen
  \bibfield  {author} {\bibinfo {author} {\bibfnamefont {J.~F.}\ \bibnamefont {Clauser}}, \bibinfo {author} {\bibfnamefont {M.~A.}\ \bibnamefont {Horne}}, \bibinfo {author} {\bibfnamefont {A.}~\bibnamefont {Shimony}},\ and\ \bibinfo {author} {\bibfnamefont {R.~A.}\ \bibnamefont {Holt}},\ }\href {https://doi.org/10.1103/PhysRevLett.23.880} {\bibfield  {journal} {\bibinfo  {journal} {Phys. Rev. Lett.}\ }\textbf {\bibinfo {volume} {23}},\ \bibinfo {pages} {880} (\bibinfo {year} {1969})}\BibitemShut {NoStop}%
\bibitem [{\citenamefont {Freedman}\ and\ \citenamefont {Clauser}(1972)}]{freedman1972}%
  \BibitemOpen
  \bibfield  {author} {\bibinfo {author} {\bibfnamefont {S.~J.}\ \bibnamefont {Freedman}}\ and\ \bibinfo {author} {\bibfnamefont {J.~F.}\ \bibnamefont {Clauser}},\ }\href {https://doi.org/10.1103/PhysRevLett.28.938} {\bibfield  {journal} {\bibinfo  {journal} {Phys. Rev. Lett.}\ }\textbf {\bibinfo {volume} {28}},\ \bibinfo {pages} {938} (\bibinfo {year} {1972})}\BibitemShut {NoStop}%
\bibitem [{\citenamefont {Aspect}\ \emph {et~al.}(1982)\citenamefont {Aspect}, \citenamefont {Grangier},\ and\ \citenamefont {Roger}}]{aspect1982}%
  \BibitemOpen
  \bibfield  {author} {\bibinfo {author} {\bibfnamefont {A.}~\bibnamefont {Aspect}}, \bibinfo {author} {\bibfnamefont {P.}~\bibnamefont {Grangier}},\ and\ \bibinfo {author} {\bibfnamefont {G.}~\bibnamefont {Roger}},\ }\href {https://doi.org/10.1103/PhysRevLett.49.91} {\bibfield  {journal} {\bibinfo  {journal} {Phys. Rev. Lett.}\ }\textbf {\bibinfo {volume} {49}},\ \bibinfo {pages} {91} (\bibinfo {year} {1982})}\BibitemShut {NoStop}%
\bibitem [{\citenamefont {Pantaleoni}\ \emph {et~al.}(2023)\citenamefont {Pantaleoni}, \citenamefont {Baragiola},\ and\ \citenamefont {Menicucci}}]{pantaleoni2022}%
  \BibitemOpen
  \bibfield  {author} {\bibinfo {author} {\bibfnamefont {G.}~\bibnamefont {Pantaleoni}}, \bibinfo {author} {\bibfnamefont {B.~Q.}\ \bibnamefont {Baragiola}},\ and\ \bibinfo {author} {\bibfnamefont {N.~C.}\ \bibnamefont {Menicucci}},\ }\href {https://doi.org/10.1103/PhysRevA.107.062611} {\bibfield  {journal} {\bibinfo  {journal} {Phys. Rev. A}\ }\textbf {\bibinfo {volume} {107}},\ \bibinfo {pages} {062611} (\bibinfo {year} {2023})}\BibitemShut {NoStop}%
\bibitem [{Note4()}]{Note4}%
  \BibitemOpen
  \bibinfo {note} {Note that the vacuum state is the optimal Gaussian state for $H$ state distillation~\cite {baragiola2019}.}\BibitemShut {Stop}%
\bibitem [{\citenamefont {Reichardt}(2005)}]{reichardt2005}%
  \BibitemOpen
  \bibfield  {author} {\bibinfo {author} {\bibfnamefont {B.~W.}\ \bibnamefont {Reichardt}},\ }\href {https://doi.org/10.1007/s11128-005-7654-8} {\bibfield  {journal} {\bibinfo  {journal} {Quant. Info. Proc}\ }\textbf {\bibinfo {volume} {4}},\ \bibinfo {pages} {251} (\bibinfo {year} {2005})}\BibitemShut {NoStop}%
\bibitem [{\citenamefont {Inc.}()}]{Mathematica}%
  \BibitemOpen
  \bibfield  {author} {\bibinfo {author} {\bibfnamefont {W.~R.}\ \bibnamefont {Inc.}},\ }\href@noop {} {\bibinfo {title} {Mathematica, {{Version}} 13.2}},\ \bibinfo {note} {champaign, IL, 2022}\BibitemShut {NoStop}%
\bibitem [{\citenamefont {Noh}\ \emph {et~al.}(2022{\natexlab{b}})\citenamefont {Noh}, \citenamefont {Chamberland},\ and\ \citenamefont {Brand{\~a}o}}]{noh2022}%
  \BibitemOpen
  \bibfield  {author} {\bibinfo {author} {\bibfnamefont {K.}~\bibnamefont {Noh}}, \bibinfo {author} {\bibfnamefont {C.}~\bibnamefont {Chamberland}},\ and\ \bibinfo {author} {\bibfnamefont {F.~G.}\ \bibnamefont {Brand{\~a}o}},\ }\href {https://doi.org/10.1103/PRXQuantum.3.010315} {\bibfield  {journal} {\bibinfo  {journal} {PRX Quantum}\ }\textbf {\bibinfo {volume} {3}},\ \bibinfo {pages} {010315} (\bibinfo {year} {2022}{\natexlab{b}})}\BibitemShut {NoStop}%
\bibitem [{\citenamefont {Yamasaki}\ \emph {et~al.}(2020)\citenamefont {Yamasaki}, \citenamefont {Matsuura},\ and\ \citenamefont {Koashi}}]{yamasaki2020}%
  \BibitemOpen
  \bibfield  {author} {\bibinfo {author} {\bibfnamefont {H.}~\bibnamefont {Yamasaki}}, \bibinfo {author} {\bibfnamefont {T.}~\bibnamefont {Matsuura}},\ and\ \bibinfo {author} {\bibfnamefont {M.}~\bibnamefont {Koashi}},\ }\href {https://doi.org/10.1103/PhysRevResearch.2.023270} {\bibfield  {journal} {\bibinfo  {journal} {Phys. Rev. Research}\ }\textbf {\bibinfo {volume} {2}},\ \bibinfo {pages} {023270} (\bibinfo {year} {2020})}\BibitemShut {NoStop}%
\bibitem [{\citenamefont {{Garc{\'\i}a-{\'A}lvarez}}\ \emph {et~al.}(2021)\citenamefont {{Garc{\'\i}a-{\'A}lvarez}}, \citenamefont {Ferraro},\ and\ \citenamefont {Ferrini}}]{garcia-alvarez2019}%
  \BibitemOpen
  \bibfield  {author} {\bibinfo {author} {\bibfnamefont {L.}~\bibnamefont {{Garc{\'\i}a-{\'A}lvarez}}}, \bibinfo {author} {\bibfnamefont {A.}~\bibnamefont {Ferraro}},\ and\ \bibinfo {author} {\bibfnamefont {G.}~\bibnamefont {Ferrini}},\ }in\ \href {https://doi.org/10.1007/978-981-15-5191-8_9} {\emph {\bibinfo {booktitle} {International Symposium on Mathematics, Quantum Theory, and Cryptography}}},\ \bibinfo {editor} {edited by\ \bibinfo {editor} {\bibfnamefont {T.}~\bibnamefont {Takagi}}, \bibinfo {editor} {\bibfnamefont {M.}~\bibnamefont {Wakayama}}, \bibinfo {editor} {\bibfnamefont {K.}~\bibnamefont {Tanaka}}, \bibinfo {editor} {\bibfnamefont {N.}~\bibnamefont {Kunihiro}}, \bibinfo {editor} {\bibfnamefont {K.}~\bibnamefont {Kimoto}},\ and\ \bibinfo {editor} {\bibfnamefont {Y.}~\bibnamefont {Ikematsu}}}\ (\bibinfo  {publisher} {{Springer Singapore}},\ \bibinfo {address} {{Singapore}},\ \bibinfo {year} {2021})\ pp.\ \bibinfo {pages} {79--92}\BibitemShut {NoStop}%
\end{thebibliography}%
\end{document}